%% file: main_Final_M174823.tex
\documentclass[final,onefignum,onetabnum]{siamonline250211}

\input{ex_shared}

\begin{document}
\graphicspath{{./Figures/}}
\maketitle

%
\begin{abstract}
Full Waveform Inversion (FWI) is a powerful wave-based imaging technique, but its inherent ill-posedness and non-convexity make it prone to local minima and poor convergence. Regularization techniques are commonly employed to stabilize FWI by incorporating prior information that enforces structural constraints, such as smooth variations or piecewise-constant behavior. Among them, Tikhonov regularization promotes smoothness, while total variation (TV) regularization preserves sharp boundaries--both widely used in solving ill-posed inverse problems. However, in the context of FWI, we highlight two key shortcomings of these regularization methods. First, subsurface model parameters (P- and S-wave velocities, density) often exhibit complex geological formations with sharp discontinuities separating distinct layers, while parameters within each layer vary smoothly. Neither Tikhonov nor TV regularization alone can effectively constrain such piecewise-smooth structures. Second, and more critically, when the initial model is far from the true model, these regularization assumptions can lead to convergence toward a local minimum. To address these limitations, we propose an adaptive Tikhonov-TV (TT) regularization method that decomposes the model into smooth and blocky components, enabling robust recovery of piecewise-smooth structures. The method is implemented within the Alternating Direction Method of Multipliers (ADMM) framework and incorporates an automated balancing strategy based on robust statistical analysis. Extensive numerical experiments on both acoustic and elastic FWI are conducted using challenging benchmark geological models. The results demonstrate that TT regularization significantly improves convergence and reconstruction accuracy compared to Tikhonov and TV regularization when applied separately. We show that for complex models and remote initial models, both Tikhonov and TV regularization tend to converge to local minima, whereas the TT regularization effectively mitigates cycle skipping through its adaptive combination of the two regularization strategies.
\end{abstract}


%
\section{Introduction.}
Wave-based imaging techniques aim to infer the spatially varying physical properties of an unknown medium, such as seismic wave velocity and density, from a set of observational data. 
This study specifically focuses on subsurface Earth imaging, where the physical properties
are denoted as model and are represented by $ \m \!:= \!\m(\bm{x}) \in \mathbb{R}^n $, with $\bm{x}$ denoting the spatial coordinates and $n$ the number of discrete model parameters. The estimation of $\m$ is indirect, relying on the analysis of seismic measurements or data $\d$ by solving a parameter identification problem of the form:
\begin{equation}\label{eq:F(m)=d}
    \mathcal{F}(\m) = \d,
\end{equation}
where $ \mathcal{F}: \mathbb{M} \rightarrow \mathbb{D}$ is a known mapping that encapsulates the governing physics of the seismic wave propagation; mapping the model parameters to the observed data recorded at specific receiver locations. 
Among various wave-based imaging techniques, full waveform inversion (FWI) has emerged as a high-resolution method, capable of recovering $\m$ with wavelength-scale accuracy \cite{Virieux_2009_OFW}.  Initially introduced in the 1980s \cite{Tarantola_1984_ISR}, FWI has since evolved into a robust inversion technique with broad applications (see e.g., \cite{Tromp_2020_SWI} and references therein). It is typically formulated in the time \cite{Tarantola_1984_ISR}, frequency \cite{Pratt_1998_GNF}, or complex-frequency domains \cite{Shin_2009_WIL} within a multiscale data-fitting local optimization framework \cite{Bunks_1995_MSW}, where medium parameters are iteratively refined by minimizing an appropriate discrepancy metric. 

\subsection{Challenges of FWI: Necessity of regularization.}
Before FWI can be regarded as a powerful imaging technique, several challenges must be addressed. Starting from equation \cref{eq:F(m)=d}, it can be shown that the problem is at least locally ill-posed \cite{Kirsch_2014_STL}, which means that considering $\mathcal{F}(\m)$ as a forward mapping, for any neighborhood of $\m^{*}$ satisfying $\mathcal{F}(\m^{*}) = \d$, there exists a sequence $\{\m^k\}$ such that:
 \begin{equation}
     \lim_{k \rightarrow \infty} \| \mathcal{F}(\m^{*})-\mathcal{F}(\m^{k})\| \rightarrow 0, ~~~~\text{but}~~\lim_{k \rightarrow \infty} \|\m^{*}-\m^{k}\| \nrightarrow 0.
 \end{equation}
 This indicates that the forward map $\mathcal{F}$ is stable: the sequence $\{\m^k\}$ produces data that converge to the actual data $\mathcal{F}(\m^{*}) = \d$, but the inputs $\m^{k}$ do not converge to the true solution $\m^{*}$. Several factors contribute to this ill-posedness:
\begin{itemize}
    \item[I.] {Nonlinearity of the forward mapping:} 
    The complex relationship between model parameters and wave propagation creates multiple local minima, making the inversion sensitive to the initial model and prone to cycle skipping when measured and predicted wavefields differ by over half a cycle \cite{Virieux_2009_OFW}. 
    \item[II.] {Limited low-frequency content:}
    Low-frequency components are essential for mitigating nonlinearity and recovering large-scale variations in $\m$ \cite{Bunks_1995_MSW}. However, practical data often lack sufficient low frequencies, thereby impeding the recovery of a smooth background model.
    \item[III.]  {Acquisition limitations and poor illumination:}
    Seismic sources and receivers are typically deployed only on one side of the medium (e.g., at or near the surface),  which primarily captures backscattered rather than transmitted waves.  Sparse spatial sampling and incomplete illumination further increase the non-convexity of the objective function \cite{Vigh_2021_impact,Bian_2023_IGS}. 
    \item[IV.] {Multiple scattering effects and reflections:} 
    Waves reflect at discontinuities, generating complex multi-arrival wavefields and increasing the possibility of cycle skipping, especially at intermediate to long offsets and in the presence of high-contrast anomalies such as salt bodies \cite{Jones_2014_SIS,Ramirez_2020_LOC}.
    \item[V.] {Assumptions within the modeling process and numerical errors:} 
    The forward model often employs simplifying assumptions (e.g., acoustic vs. elastic, isotropic vs. anisotropic), which lead to model discrepancies. Additionally, numerical errors (e.g., discretization, boundary effects) introduce systematic biases in data matching. 
    \item[VI.] {Presence of noise:}
    Noise in the data leads to instability, potentially creating false minima or bias as the inversion fits both the true signal and the noise. 
    \item[VII.] {Complexity of multiparameter imaging such as elastic media:} 
    More parameters increase the solution space and ambiguity, as different parameter combinations can produce similar wavefield responses. Information can leak between parameters (crosstalk), and elastic FWI involves complex wave physics, including mode conversions, increasing nonlinearity \cite{Operto_2013_GTM,Keating_2020_PCL}. 
\end{itemize}
\vspace{0.5cm}

These challenges contribute to the uncertainty in FWI resolution, making it an ill-posed and non-convex problem. Addressing these issues requires incorporating additional information about $\m$ through {regularization}, denoted as $\mathcal{R}(\m)$. 
The choice of $\mathcal{R}(\m)$ is critical for model recovery, as it biases the solution towards models that possess desirable characteristics or encode prior knowledge about $\m$ \cite{Guitton_2012_BRS,Guitton_2012_CFW,Gholami_2013_BCT,asnaashari2013regularized}.
In this study, we are particularly interested in models that exhibit {piecewise smooth} behavior, which can be represented as $\m = (\m_1, \m_2)$, where:
\begin{itemize}
\item[I.] $\m_1$: Encodes discontinuities and rapid variations, capturing sharp boundaries and structural interfaces.
\item[II.] $\m_2$: Represents the smoothly varying background component. 
\end{itemize}
Without loss of generality, we present our formulation in the frequency domain. For the sake of brevity, the formulations are presented for a single frequency. However, extending them to multiple frequencies is straightforward and can be achieved by summing over frequencies in the objective function. 
\subsection{Related works.}
We focus on regularization techniques based on $\ell_2$-and $\ell_1$-norms, widely used in FWI. 
{Tikhonov regularization} in its isotropic form promotes smoothness by penalizing the $\ell_2$-norm of the model or its derivatives, thereby suppressing high-frequency variations  \cite{Tikhonov_1977_SIP,asnaashari2013regularized}. Its anisotropic variant further refines this approach by incorporating geological tilt information
 \cite{Gholami_2025_OSV}, improving structural alignment with subsurface features. Despite being straightforward and computationally efficient, Tikhonov regularization tends to oversmooth sharp boundaries and is unsuitable for capturing piecewise constant features.
{Sparsity-promoting techniques} enforce sparsity in transform domains (e.g., wavelet, curvelet, seislet) \cite{Loris_2007_TIR,Yang_2019_MFW,Xue_2017_Seislet} using $\ell_1$-norm penalties to enhance resolution and suppress noise. However, their effectiveness depends on the transform’s ability to represent geological structures accurately.
The family of TV regularization \cite{Rudin_1992_NTV}, preserves discontinuities by penalizing the $\ell_1$-norm of the model gradient. Variants include isotropic \cite{Esser_2018_TVFWI,Aghamiry_2019_IBC}, anisotropic  (with directional operators, \cite{Qu_2019_FWI}), and asymmetric TV \cite{Esser_2016_CWI}, offer tailored edge preservation under different geological assumptions. In addition, the use of TV regularization for complex-valued imaging (visco-acoustic FWI in frequency domain) is studied in \cite{Aghamiry_2020_CIT}. TV regularization typically experiences staircase effects, motivating the use of higher-order TV approaches \cite{Du_2021_HTV}.

The previously mentioned $\ell_1$-and $\ell_2$-based regularization methods are effective for enforcing only a single distinct property of the model; either smoothness (by using Tikhonov) or blockiness (by using TV). However, neither approach is well-suited for reconstructing piecewise smooth models, where discontinuities are embedded within a smoothly varying background \cite{Gholami_2013_BCT,Selesnick_2014_SLF,Huska_2021_VAA}.
Instead, promoting solutions that simultaneously exhibit multiple desired properties through
{compound regularization} can be more effective \cite{Gholami_2013_BCT,Aghamiry_2019_CRO}. 
One prominent approach within this framework is the {infimal convolution} method \cite{Chambolle_1997_IRV}, which combines different regularization functionals. A well-known example from this category is Total Generalized Variation (TGV) \cite{Bredies_2010_TGV,Selesnick_2015_GTV}, which integrates  first- and second-order TV functionals to capture both blocky and smooth features. 
An $\ell_p$-norm based TGV formulation has been proposed for imaging piecewise smooth models in FWI \cite{Gao_2019_TGPV}. Additionally, a combination of the Shearlet transform with TGV \cite{Guo_2014_NDP} has been explored  in FWI \cite{Wang_2024_RFW}. 
Following \cite{Huang_2008_FTV}, \cite{Lin_2014_MTV} integrates zero-order Tikhonov regularization with TV for a better balance between smooth and sharp features. A more suitable functional in this form is based on the Tikhonov-TV (TT) regularization \cite{Aghamiry_2018_HTT,Aghamiry_2019_CRO}, where the main challenge is controlling the trade-off between smoothness and blockiness. Another approach for piecewise smooth imaging is Mumford-Shah regularization \cite{Kadu_2018_MSR}, which approximates an image with smooth regions separated by edges. However, its inherent non-convexity complicates optimization and parameter selection.
\subsection{Contributions.}
In general, previous regularization methods for {piecewise smooth} imaging face two primary challenges: (i) selecting appropriate regularization functionals to simultaneously preserve smooth and discontinuous structures, and (ii) determining a proper balance between these terms to ensure accurate model decomposition. To address these issues, we propose the following advancements:

\begin{itemize}
\item[I.] {Balanced Tikhonov-TV regularization:} Unlike prior approaches such as \cite{Gao_2019_TGPV} and \cite{Lin_2014_MTV}, which respectively use second-order TV and zero-order Tikhonov regularization to promote smoothness, we adopt second-order Tikhonov regularization to enforce large-scale smoothness and a first-order TV term to capture blocky features.  This choice, inspired by \cite{Gholami_2013_BCT}, naturally aligns with the structural characteristics of piecewise smooth models. 

\item[II.] Extending TT regularization to multi-parameter elastic FWI: 
Previous applications of TT regularization have been limited to acoustic FWI \cite{Aghamiry_2018_HTT,Aghamiry_2019_CRO}. In this work, we extend the methodology to the elastic case, where multiple parameters with distinct structural properties must be reconstructed simultaneously. The central challenge lies in adaptively tuning the balancing parameters required for decomposing different elastic parameters. We address this by introducing an adaptive parameter selection strategy that ensures a robust and consistent performance across all elastic components.

\item[III.] {Statistical decomposition with automatic parameter selection:} 
A key limitation of prior methods is their reliance on manual or heuristic parameter tuning, making it challenging to accurately decompose the model into its smooth and discontinuous components. Our approach builds on \cite{Gholami_2022_ABP} by formulating an adaptive decomposition scheme in the model gradient domain, allowing for robust separation of the smooth and piecewise constant components. The optimal balancing parameter is automatically determined via a statistical anomaly detection method using median absolute deviation, leveraging gradient variations to dynamically adjust the weighting between regularization terms.
\end{itemize}

\vspace{0.4cm}
The proposed method is rigorously tested in challenging FWI scenarios, demonstrating superior performance in: \begin{itemize} 
\item[I.] Mitigating cycle skipping in FWI for improved seismic imaging. 
\item[II.] Resolving salt and subsalt structures with high accuracy. 
\item[III.] Reconstructing complex models from ultra-long offset data. 
\item[IV.] Ensuring stability against noise, sparse acquisition, and varying balancing parameters ensuring reliable performance for real world conditions. 
\item[V.] Imaging multi-parameter elastic media and evaluating the method’s ability to reconstruct distinct parameters with varying structural properties, highlighting its adaptability across diverse geological environments. 
\end{itemize}
%
%
\section{Full waveform inversion.}

\subsection{Forward problem.}
Let $n$, $n_s$ and $n_r$ denote the number of discrete model parameters, sources, and receivers, respectively.
We assume a set of known source functions in the frequency domain, denote as $\b_i \in \mathbb{C}^{n}$, where $i \in \set{1, \dots, n_s }$ indexes the sources. Given a candidate subsurface model $\m$, the seismic wavefield, $\u_i \in \mathbb{C}^n$, can be obtained by solving the following partial differential equation (PDE) with coefficients $\m$: 
\begin{equation}\label{Au=b}
   \begin{array}{cc}
        \A(\m) \u_i = \b_i, & \text{in}~ \Omega  \\
        ~~~~~~~\u_i=0, & ~~\text{on}~ \partial\Omega
   \end{array}
   \end{equation}
where $\Omega \subset \mathbb{R}^2$ is a bounded domain with boundary $\partial\Omega$. 
  $\A(\m) \in \mathbb{C}^{n \times n}$ is the complex-valued discretized PDE operator and will be defined precisely for acoustic and elastic media in \cref{sec:Sec_Acoustic} and \cref{sec:Sec_Elastic}, respectively. The observed data $\d_i=\P\u_{i}$ is obtained by solving \cref{Au=b} for the wavefield and sampling it at the receiver locations, where the matrix $\P\in \mathbb{R}^{n_r \times n}$ is the sampling operator.
%
\subsubsection{Acoustic media.}\label{sec:Sec_Acoustic}
For two-dimensional (2D) acoustic media, the forward problem is defined by the scalar Helmholtz equation:
\begin{equation}\label{eq:Helmholtz}
\underbrace{[ \Delta + \omega^2 \diag{\m} ]}_{\A(\m)} \u_i = \b_i, 
\end{equation}
where $\omega$ denotes the angular frequency, and $\Delta= \frac{\partial^2}{\partial x^2}+\frac{\partial^2}{\partial z^2}$ represents a discretized Laplace operator using an optimal 9-point finite-difference stencil, with perfectly matched layer (PML) boundary conditions applied to all sides of the computational domain \cite{Chen_2013_OFD}. 
The medium is parameterized by the squared slowness $\m:= \bm{V}(\bm{x})^{-2}$, where  $\bm{V}(\bm{x})$ denotes the P-wave velocity. 
%
\subsubsection{Elastic media.} \label{sec:Sec_Elastic}
For 2D elastic FWI (EFWI), we consider the isotropic form of the frequency-domain elastic wave equation, defined as
\begin{subequations}\label{eq:El_forward}
\begin{align}
&\bm{\rho} \omega^{2}\bold{u}_{i,x}+(\bm{\hat{\lambda}}+2\bm{\hat{\mu}})\partial_{xx}\bold{u}_{i,x}+\bm{\hat{\mu}} \partial_{zz}\bold{u}_{i,x}+(\bm{\hat{\lambda}}+\bm{\hat{\mu}})\partial_{xz}\bold{u}_{i,z} = \bold{b}_{i,x}, \label{EL_bx} \\
&\bm{\rho} \omega^{2}\bold{u}_{i,z}+(\bm {\hat{\lambda}}+2\bm{\hat{\mu}})\partial_{zz}\bold{u}_{i,z}+\bm{\hat{\mu}} \partial_{xx}\bold{u}_{i,z}+(\bm{\hat{\lambda}+\hat{\mu}})\partial_{xz}\bold{u}_{i,x} = \bold{b}_{i,z}, \label{EL_bz}
\end{align}
\end{subequations}
where $\bm{\rho} \in \mathbb{R}^{n \times 1}$ is mass density, $\bm{\hat{\lambda}} \in \mathbb{R}^{n \times 1}$ and $\bm{\hat{\mu}} \in \mathbb{R}^{n \times 1}$ denote the Lamé parameters, $\bold{u}_{i,x} \in \mathbb{C}^{n \times 1}$ and $\bold{u}_{i,z} \in \mathbb{C}^{n \times 1}$ are horizontal and vertical particle displacements, and $\bold{b}_{i,x} \in \mathbb{C}^{n \times 1}$, $\bold{b}_{i,z} \in \mathbb{C}^{n \times 1}$ are the source terms in the respective directions. The P- and
S- wave velocities are defined as
\begin{equation*}
    \text{V}_\text{p} = \sqrt{\frac{\bm{\hat{\lambda}}+2\bm{\hat{\mu}}}{\bm{\rho}}}, ~~~\text{V}_\text{s} = \sqrt{\frac{\bm{\hat{\mu}}}{\bm{\rho}}}.
\end{equation*}
Assuming a constant-density medium, we discretize and rewrite the system of equations in \cref{eq:El_forward} as:
\begin{equation}\label{Aub}
\small
\Amu_{i} = \b_{i}, 
\end{equation} 
where 
\begin{equation}
\noindent 
\bold{A}( \m) = \begin{bmatrix}
\omega^2 +\diag{\m_\text{p}}\partial_{xx}+\diag{\m_\text{s}}\partial_{zz} ~~~~ \diag{\m_\text{p}-\m_\text{s}}\partial_{xz} \\
\diag{\m_\text{p}-\m_\text{s}}\partial_{xz} ~~~~ \omega^2+\diag{\m_\text{p}}\partial_{zz}+\diag{\m_\text{s}}\partial_{xx}
\end{bmatrix} \in \mathbb{C}^{2n \times 2n},
\end{equation}
denotes the PDE operator, and is discretized using the optimal 9-point finite difference stencil proposed by \cite{chen_2016_MFE}, and
\begin{equation}
 \m = 
\begin{bmatrix} 
 \m_\text{p}:=\text{V}_\text{p}^2\\
\m_\text{s}:=\text{V}_\text{s}^2
\end{bmatrix}\in \mathbb{R}^{2n \times 1},
\quad
\bold{u}_i = 
\begin{bmatrix} 
\u_{i,x}\\
\u_{i,z}
\end{bmatrix}\in \mathbb{C}^{2n \times 1},
\quad
\b_i = 
\begin{bmatrix} 
\b_{i,x}\\
\b_{i,z}
\end{bmatrix} \in \mathbb{C}^{2n \times 1}.
\end{equation}
\subsection{Inverse problem.}

 Problem \cref{eq:F(m)=d} involves solving the following $n_s$ coupled systems for ($\u,\m$):
  \begin{equation}\label{eq:coupled}
      \begin{cases}
         \A(\m)\u_{i} = \b_{i}, \\
          \P\u_{i} = \d_{i},
      \end{cases},\quad  i =1,\ldots,n_s.
  \end{equation}
Solving \cref{eq:coupled} can be formulated as the following equality-constrained minimization problem \cite{Haber_2000_OTS,Aghamiry_2019_CRO}:
\begin{mini}
{\m \in \mathcal{M},\u}{\mathcal{R}(\m) + \frac{\alpha}{2}  \sum_{i=1}^{n_s} \|\P\u_{i}-\d_{i} \|_2^2}{}{},
 \addConstraint{\A(\m)\u_{i}}{=\b_{i}, \quad}{ i =1,\ldots,n_s},
\label{eq:FWI_const1}
\end{mini}
where the box $\mathcal{M}=\set{ \m \in \mathbb{R}^{n}, \underline{\m} \le \m \le \overline{\m} }$ is determined by the lower bound $\underline{\m}$ and the upper bound $\overline{\m}$. 
$\mathcal{R}(\m)$ is the regularization function, and $\alpha \!>\!0$ is the regularization parameter, acting as a scaling factor that controls the influence of the data fidelity term and regularization term.
We use the Augmented Lagrangian (AL) method \cite{Powell_1969_NLC}  
in the framework of the the alternating direction method of multipliers (ADMM) for solving \cref{eq:FWI_const1} \cite{Boyd_2011_DOS}. It breaks the problem into three smaller subproblems which are solved in sequence for the wavefield, model, and Lagrange multipliers \cite{Aghamiry_2019_IWR}, as detailed in the following.
 \subsubsection{Wavefield subproblem.} For an initial model $\m$ and an initial multiplier $\dual$, the wavefields are reconstructed by minimizing the AL function for given Lagrange multiplier estimates. This leads to the following linear system of equations for each wavefield: 
     \begin{equation}\label{Ue_pdate}
    \left(\P^{\top} \P + \mu \A(\m^{k-1})^{\top} \A(\m^{k-1}) \right) \u^{k}_i = \P^{\top} 
    \d_i + \mu \A(\m^{k-1})^{\top}\left( \b_i + {\dual}^{k-1}_i \right),
\end{equation}
where $\{\dual_i\}$ denotes the Lagrange multiplier, $\mu\!>\!0$ is the penalty parameter, $k$ represents the iteration number, and superscript $\top$ is the complex-conjugate
(Hermitian) transpose. 
 \subsubsection{Model subproblem.} 
 The update of the model parameters by optimizing the AL function for a given set of wavefields $\set{\u_i}$ and multipliers $\set{\dual_i}$ requires solving the following problem:
\begin{equation}\label{eq:AL_m_update}
 \m^{k} = \argmin_{\m \in \mathcal{M}}~  \mathcal{R}(\m) + \frac{\mu}{2} \| \bm{L}^k\m-\bm{y}^k\|_2^2,
\end{equation} 
where $\bm{L}^k$ and $\bm{y}^k$ are defined differently for acoustic and elastic media \cite{Aghamiry_2019_IWR,Aghazade_2024_REF}. For acoustic case, we have
\begin{equation}
  \bm{L}^k =
  \begin{pmatrix}
      \omega^2 \diag{\u_1^k}\\
      \vdots\\
      \omega^2 \diag{\u_{n_s}^k}
  \end{pmatrix}
   \in \mathbb{C} ^{(n_s n) \times n},~~~~~
   \bm{y}^k = 
   \begin{pmatrix}
       \b_1 + \dual_1^{k-1} -\Delta \u_1^k\\
       \vdots\\
       \b_{n_s} + \dual_{n_s}^{k-1} -\Delta \u_{n_s}^k
   \end{pmatrix}
    \in \mathbb{C}^{n_sn \times 1}.
\end{equation}
For elastic formulation, the structure of $\bm{L}^k$ and $\bm{y}^k$ is more complex as
\begin{equation}
  \bm{L}^k =
  \begin{pmatrix}
      \bm{L}_1^k\\
      \vdots\\
     \bm{L}_{n_s}^k
  \end{pmatrix}
   \in \mathbb{C} ^{(2n_s n) \times 2n},~~~~~
   \bm{y}^k = 
   \begin{pmatrix}
       \bm{y}_1^k\\
       \vdots\\
       \bm{y}_{n_s}^k
   \end{pmatrix}
    \in \mathbb{C}^{2n_sn \times 1},
\end{equation}
where
\begin{subequations}
\begin{align}
& \bm{L}_i^k = \begin{bmatrix}
\diag{\partial_{xx}{\u}_{i,x}^k}+\diag{\partial_{xz}{\u}_{i,z}^k} &  \diag{\partial_{zz}{\u}_{i,x}^k}-\diag{\partial_{xz}{\u}_{i,z}^k} \\
\diag{\partial_{zz}{\u}_{i,z}^k}+\diag{\partial_{xz}{\u}_{i,x}^k} & \diag{\partial_{xx}{\u}_{i,z}^k}-\diag{\partial_{xz}{\u}_{i,x}^k}
\end{bmatrix} \in \mathbb{C} ^{2n \times 2n}, \\
& \bm{y}_i^k = \begin{bmatrix}
{\b}_{i,x}+\dual_{i,x}^{k-1}-\omega^2 \u_{i,x}^k \\
{\b}_{i,z}+\dual_{i,z}^{k-1}-\omega^2 \u_{i,z}^k
\end{bmatrix} \in \mathbb{C}^{2n \times 1}.
\end{align}
\end{subequations}

 \subsubsection{Lagrange multiplier subproblem.} The Lagrange multipliers are updated simply by a gradient ascent step  \cite{Powell_1969_NLC}
\begin{equation}\label{Dual_update}
    {\dual}^{k}_i = {\dual}^{k-1}_i +\b_i - \A(\m^{k})\u^{k}_i.
\end{equation}
\section{Regularization by smooth-blocky decomposition.}

A subsurface model, $\m$, can be represented using a Fourier basis to define its wavenumber content \cite{Clement_2001_MTW,Alkhalifah_2016_FMW}.
FWI relies on constructing the model’s low-wavenumber component first,  followed by refining higher-wavenumber details through  frequency continuation \cite{Bunks_1995_MSW}. 
To present the methodology, we consider a simple 2D Gaussian void model as an example of the subsurface candidate model $\m$ (\cref{fig:nabla_m_a}). 
We assume that $\m$ can be decomposed into two distinct components, each describing specific structural features with characteristic spectral behavior:
\begin{itemize}
    \item[I.] A blocky component, $ \m_1 \in \mathbb{R}^n $, capturing sharp boundaries and discontinuities. The gradient of this component follows a non-Gaussian distribution that is mainly represented in the Fourier domain by high-wavenumber coefficients (\cref{fig:nabla_m_b}).
        \item[II.] A smooth background component, $ \m_2 \in \mathbb{R}^n $, representing the smoothly varying structures (or long-wavelength features). It is characterized by Gaussian-distributed second-order derivatives (\cref{fig:nabla_m_c}).
\end{itemize}

Accordingly, we assume that the desired medium is piecewise smooth, defined as
\begin{equation}\label{m_dec}
  \m = {\m}_{1} + {\m}_{2}. 
\end{equation}

Accurately separating the smooth and blocky components provides significant flexibility in FWI. It enables the stable recovery of the background model through the term $\m_2$, which is crucial for ensuring good convergence of the optimization algorithm in the early iterations. The blocky component is then gradually reconstructed as the solution approaches convergence. This flexibility is achieved through a well-balanced combination of these two terms in the regularization function.
%
\subsection{Tikhonov-TV (TT) regularization.} 
Following the model decomposition under the piecewise smooth assumption, the smooth component, $\m_2$, and blocky component, $\m_1$,  can be effectively constrained using second-order Tikhonov and first-order TV functionals, respectively.
Accordingly, the balanced TT regularization functional can be defined as \cite{Gholami_2013_BCT}:
\begin{equation}\label{eq:R(m)_TT}
\begin{aligned}
    \mathcal{R}(\m)= \|\nabla \m_1\|_1 + \frac{\beta}{2} \|\nabla^2 \m_2\|_2^2,~~~~~\text{with}~~\m=\m_1+\m_2,    
\end{aligned}
\end{equation}
where $\beta>0$ is a balancing parameter that controls the relative influence of the two regularization terms. $\nabla$ and $\nabla^2$ are first-and second order finite difference operators, respectively (see e.g., \cite{Gholami_2022_ABP} for explicit definitions). 
The behavior of the regularization functional is governed by $\beta$:

\begin{enumerate}
    \item For $\beta \rightarrow 0$, the minimization on the TV term dominates. 
    In this regime, the blocky component $\m_1$ tends to a constant field 
    (minimizing total variation), while the smooth component $\m_2 = \m - \m_1$ 
    absorbs essentially all of the signal structure.
    \item For $\beta \rightarrow \infty$, the smooth component is strongly 
    constrained to lie in the null space of the curvature operator 
    (i.e., to a constant or affine trend). Consequently, the formulation 
    asymptotically reduces to TV regularization, where $\m_1$ captures 
    the remaining variability through edge-preserving, piecewise-constant features.
\end{enumerate}

Regularization of the form \cref{eq:R(m)_TT} was initially introduced in  \cite{Gholami_2013_BCT} for signal and image processing applications. More recently, \cite{Aghamiry_2019_CRO} extended this approach to acoustic FWI within an ADMM framework. Specifically, the model update step in \cref{eq:AL_m_update} requires solving the following constrained optimization: 
\begin{equation}\label{eq:TT_AL_1}
\begin{aligned}
&\minimize_{\m_1,\m_2} ~\|\nabla \m_1\|_1 + \frac{\beta}{2} \|\nabla^2 \m_2\|_2^2 +  \frac{\mu}{2} \| \bm{L}[\m_1+\m_2]-\bm{y}\|_2^2, \\
&\text{subject to}~~~\m_1+\m_2=\m \in \mathcal{M}.     
\end{aligned}
\end{equation}
We have omitted the dependence of $\bm{L}$ and $\bm{y}$ on the iteration number $k$ for simplicity and clarity. Despite the successful application of \cref{eq:TT_AL_1}, the choice of $\beta$ significantly influences the solution, and determining an optimal value is not straightforward. In the following, we introduce an equivalent but straightforward version of the TT functional to address this challenge.
%
\subsubsection{Adaptive TT regularization.}
Given the linearity of the gradient operator,  the total model gradient can be decomposed as:
\begin{equation}
\begin{aligned}
 \nabla \m & = \nabla \m_1 + \nabla \m_2, \\
& = \g_1+\g_2,   
\end{aligned}
\end{equation}
where $\g_1= \nabla \m_1$ and $\g_2= \nabla \m_2$  are auxiliary variables representing the gradients of the blocky and smooth components, respectively. Additionally, using the identity $\nabla^2 = \overline{\nabla} \nabla$, the constrained optimization problem in \cref{eq:TT_AL_1} can be reformulated as \cite{Gholami_2022_ABP}:
\begin{equation}\label{TT_AL_2}
  \minimize_{\m \in \mathcal{M},\g_1,\g_2}  \|\g_1\|_1 + \frac{\beta}{2} \|\overline{\nabla}\g_2\|_2^2 +  \frac{\mu}{2} \| \bm{L}\m-\bm{y}\|_2^2,~~ \text{subject to}~~
   \nabla \m = \g_1+\g_2, 
\end{equation}
which can be solved by ADMM (appearing as an inner loop within the FWI iteration). 
The ADMM iteration is defined as
\begin{subequations} \label{ADMM_2}
    \begin{align}
        &  \m^{k} = \argmin_{\m}~\mathcal{L}(\m,\g_1^{k-1},\g_2^{k-1},\p^{k-1},\Nu_1^{k-1},\Nu_2^{k-1}) , \label{ADMM_mTT}\\
         & \g_1^{k} =\argmin_{\g_1}~\mathcal{L}(\m^{k},\g_1,\g_2^{k-1},\p^{k-1},\Nu_1^{k-1},\Nu_2^{k-1})  , \label{ADMM_gb} \\
         & \g_2^{k} =\argmin_{\g_2}~\mathcal{L}(\m^{k},\g_1^{k},\g_2,\p^{k-1},\Nu_1^{k-1},\Nu_2^{k-1})  , \label{ADMM_gs} \\
         & \p^{k} = \Pi_\mathcal{M}(\m^{k}-\Nu_2^{k-1}), \\
         & \Nu_1^{k} = \Nu_1^{k-1} + \g_1^{k}+\g_2^{k}-\nabla \m^{k},\\
         & \Nu_2^{k} = \Nu_2^{k-1}+\p^{k}-\m^{k},  
    \end{align}
\end{subequations}
where 
\begin{multline}\label{AL_TT_new}
   \mathcal{L}(\m,\g_1,\g_2,\p,\Nu_1,\Nu_2)  = \|\g_1\|_1 + \frac{\beta}{2} \|\overline{\nabla} \g_2\|_2^2 +  \frac{\mu}{2} \| \bm{L}\m-\bm{y}\|_2^2 + I_{\mathcal{M}}(\p) \\
   + \frac{\tau_1}{2} \|\nabla \m -\g_1-\g_2-\Nu_1\|_2^2  + \frac{\tau_2}{2} \|\m-\p-\Nu_2\|_2^2.
\end{multline}
In these formulations $\Nu_1$ and $\Nu_2$ are Lagrange multipliers, and $\tau_1, \tau_2 >0$ are the penalty parameters controlling the new constraints enforcement,
$I_{\mathcal{M}}(\p)$ is the indicator function on $\mathcal{M}$ (i.e., $I_{\mathcal{M}}(\p) = 0$ for $\p \in \mathcal{M}$, and $I_\mathcal{M}(\p) = \infty$ for $\p \notin \mathcal{M}$).
 $\Pi_\mathcal{M}(\bm{x})=\text{max}(\text{min}(\bm{x},\overline{\bm{x}}),\underline{\bm{x}})$ is the Euclidean projection of $\bm{x}$ onto the feasible region $\mathcal{M} = \set{ \bm{x}| \underline{\bm{x}} \le \bm{x} \le \overline{\bm{x}}}$. In the following, we provide a detailed analysis for solving the subproblems \cref{ADMM_mTT}–\cref{ADMM_gs}.
%
%
%
\paragraph{Subproblem \cref{ADMM_mTT}.}
The minimization subproblem over $\m$ in \cref{ADMM_mTT} has a closed-form solution given by:
\begin{equation}\label{m_update}
    \m^{k} =   \left( \mu\bm{L}^{\top}\bm{L}+{\tau}_1 \nabla^{\top}\nabla + {\tau}_2  \bm{I}\right)^{-1} \left( \mu\bm{L}^{\top}\bm{y}+{\tau}_1 \nabla^{\top}(\g_1^{k-1}+\g_2^{k-1}+\Nu_1^{k-1}) + {\tau}_2 (\p^{k-1}+\Nu_2^{k-1}) \right),
\end{equation}
where $\bm{I}$ is the identity matrix. 
\paragraph{Subproblem \cref{ADMM_gb}.}

Regarding the update of $\g_1$, we need to solve the following subproblem at iteration $k$:
\begin{equation}
    \min_{\g} ~ \|\g\|_1 + \frac{1}{2\gamma} \|\g - \bold{z}\|_2^2,
    \quad \text{with}\quad\gamma=\frac{1}{\tau_1}\quad\text{and}\quad \bold{z} = \nabla \m^{\,k} - \g_2^{\,k-1} - \Nu_1^{\,k-1}.
\end{equation}
Although this formulation involves the non-differentiable $\ell_1$ norm, it admits a closed-form solution given by the proximal operator of the $\ell_1$ norm \cite{Goldstein_2009_SBM}:  
\begin{equation}\label{eq:g1-prox}
    \g_1^{k}
    =   \text{prox}_{\gamma\ell_l}(\bold{z})=
    \text{prox}_{\frac{1}{\tau_1}\ell_l}
    \left(\nabla \m^{\,k} - \g_2^{\,k-1} - \Nu_1^{\,k-1}\right),
\end{equation}
where $\text{prox}_{\gamma\ell_l}(\bold{z})$ is the well-known soft-thresholding operator, defined as
\begin{equation}\label{eq:soft-thresh}
    \left[\text{prox}_{\gamma\ell_l}(\bold{z})\right]_i
    \;=\; \text{sign}(z_i)\,\max\bigl(|z_i|-\gamma,\,0\bigr).
\end{equation}
This useful property highlights the power of proximal operators in handling non-differentiable optimization problems. To better understand this, consider the Moreau envelope of the $\ell_1$ norm, defined as 
\begin{equation}
   M_{\gamma \ell_1}(\bold{z}) = \min_{\g} ~ \|\g\|_1 + \frac{1}{2\gamma} \|\g - \bold{z}\|_2^2\leq \|\bold{z}\|_1 .
\end{equation}
Intuitively, $M_{\gamma \ell_1}$ can be interpreted as a smooth, regularized version of the $\ell_1$ norm.  It is continuously differentiable, shares the same minimizers as $\|\cdot\|_1$, and is therefore much easier to handle in optimization.  Importantly, the gradient of the Moreau envelope is \cite{Parikh_2014_proximal} 
\begin{equation}
   \frac{d}{d\bold{z}} M_{\gamma \ell_1}(\bold{z}) = 
   \frac{d}{d\bold{z}}\left(\min_{\g} ~ \|\g\|_1 + \frac{1}{2\gamma} \|\g - \bold{z}\|_2^2\right)
   =\frac{1}{\gamma}\left(\bold{z}- \text{prox}_{\gamma\ell_l}(\bold{z})\right).
\end{equation}
Equivalently,
\begin{equation}
    \text{prox}_{\gamma\ell_l}(\bold{z})=\bold{z} - \gamma \frac{d}{d\bold{z}} M_{\gamma \ell_1}(\bold{z}).
\end{equation}
Thus, the update $\g_1^{k}$ via the proximal operator can be viewed as performing a gradient-descent step on the smooth Moreau envelope of the $\ell_1$ norm, with $\gamma$ acting as the step size.
\paragraph{Subproblem \cref{ADMM_gs}.}
The update of $\g_2$ associated with Tikhonov term, is given by:
\begin{equation}\label{g2_update}
    \g_2^{k}=\left(\bm{I}+ \frac{\beta^{k-1} }{\tau_1} \overline{\nabla}^{\top}\overline{\nabla}\right)^{-1}\left( \nabla \m^{k} -\g_1^{k}-\Nu_1^{k-1}\right).
\end{equation}
The ADMM iterations for model updates in \cref{ADMM_2}, utilizing TT regularization with bound constraint (BC), are detailed in \cref{alg:alg1,alg:alg2}. \Cref{alg:alg1} addresses single-parameter acoustic inversion, while \cref{alg:alg2} focuses on two-parameter elastic inversion.
\begin{figure}\label{1D_decom}
\centering
\includegraphics[width=0.75\textwidth]{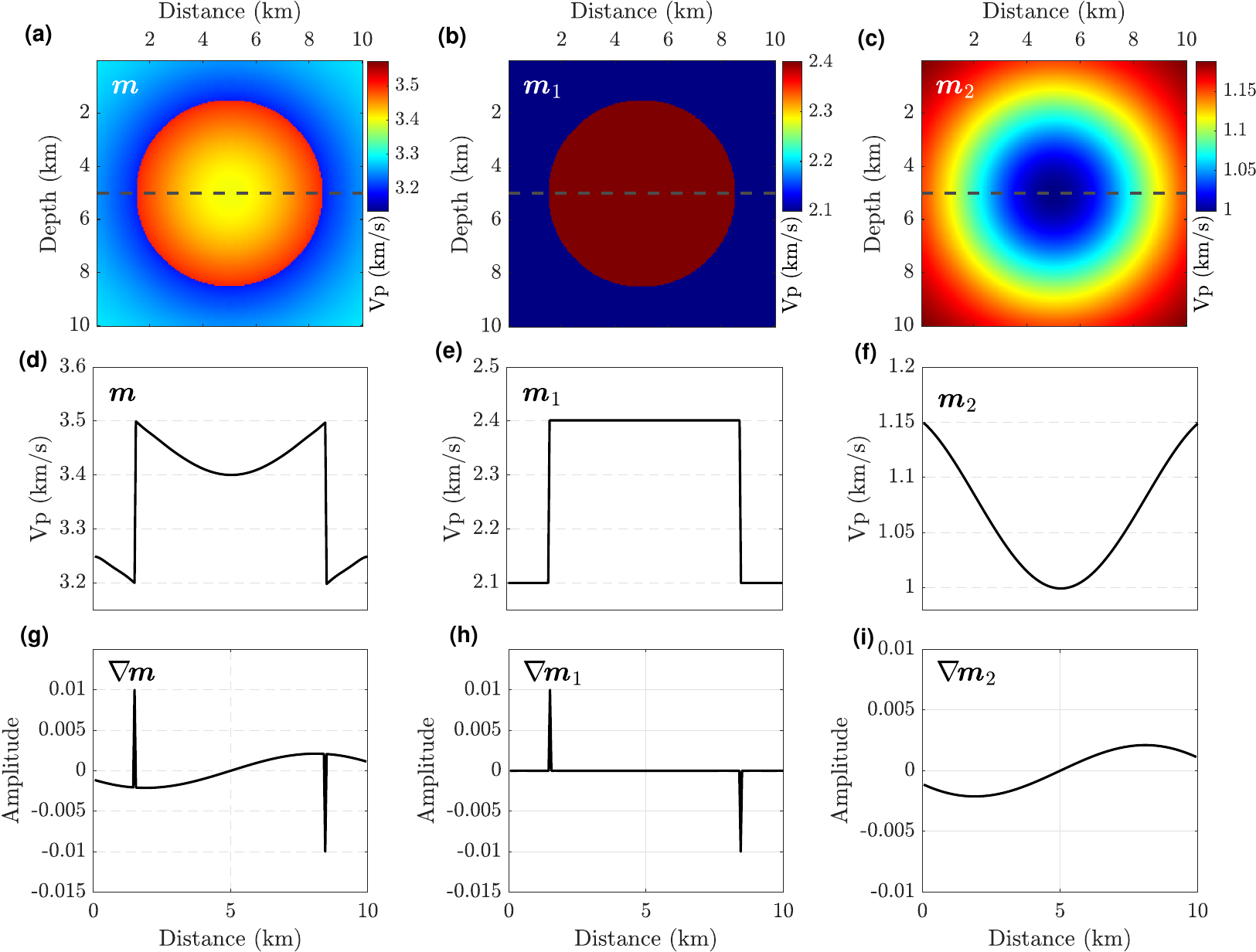}
\caption{An example of a piecewise smooth model. (a) The desired Gaussian void  model, $\m$, that can be decomposed into a piecewise constant component, $\m_1$, (b) and a smooth component, $\m_2$, (c); $\m=\m_1+\m_2$. (d,e,f) extracted horizontal profiles along the dashed lines in (a,b,c) for each model. (g) The gradient of (d) which contains non-Gaussian distributed jumps embedded in a smooth Gaussian distributed trend that is a linear summation of:  (h) the gradient of (e) with sparse non-Gaussian distributed nature and (i), the gradient of (f) with smooth Gaussian nature; $\nabla \m = \nabla \m_1 + \nabla \m_2$. In (g) and (h), amplitudes are clipped to $\pm0.01$ for display.}
\label{fig:nabla_m}
\ps{fig:nabla_m_a}
\ps{fig:nabla_m_b}
\ps{fig:nabla_m_c}
\ps{fig:nabla_m_d}
\ps{fig:nabla_m_e}
\ps{fig:nabla_m_f}
\ps{fig:nabla_m_g}
\ps{fig:nabla_m_h}
\ps{fig:nabla_m_i}
\end{figure}
%
\paragraph{Automatic balancing parameter selection.}
Consider the piecewise smooth model shown in \cref{fig:nabla_m_d}. The TT functional in \cref{eq:TT_AL_1} aims to separately recover the model components ($\m_1$, $\m_2$) shown in \cref{fig:nabla_m_e,fig:nabla_m_f}, while the alternative TT functional in \cref{AL_TT_new} jointly updates $\m$ together with its gradient components, $\g_1 \!=\! \nabla \m_1$ and $\g_2 \!=\! \nabla \m_2$. In both cases, the accuracy of blocky-smooth decomposition is governed by the balancing parameter $\beta$. However, the model's gradient domain offers a more intuitive and adaptive approach to decomposition.
As shown in \cref{fig:nabla_m_g,fig:nabla_m_h,fig:nabla_m_i}, the overall model gradient ${\g}\! =\! \nabla {\m}$ (\cref{fig:nabla_m_g}) can be viewed as a mixture of two distinct components: a non-Gaussian distributed component (sparse part, $\g_1$, \cref{fig:nabla_m_h}) and a Gaussian distributed component (smooth part, $\g_2$, \cref{fig:nabla_m_i}).
The parameter $\beta$ is determined by treating the nonzero entries of $\g_1$ (representing discontinuities)  as anomalies in $\g$, whose magnitudes can be characterized using robust statistics \cite{Rousseeuw_2018_ADR}. 
The procedure consists of three main steps:
\begin{itemize}
 \item[I.] Robust z-score computation: To identify anomalous components in the model gradient, robust z-score is used. The robust z-score for each component of the model gradient is computed as:
\begin{equation}
z_i=\frac{[\g]_i-\operatorname{median}({\g})}{\operatorname{MAD}(\g)},
\end{equation}
where $[\g]_i$ represents the $i^{\text {th}}$ sample of $\g$, and MAD denotes the median absolute deviation, defined as:
\begin{equation}
\text{MAD}(\g) = 1.4826~\text{median}(\g-\text{median}(\g)),
\end{equation}
which measures how far the values in $\g$ typically are from the median. Here, the constant scaling factor $1.4826$ makes MAD a consistent estimator of the standard deviation under the Gaussian distribution (see e.g, \cite{Rousseeuw_1993_MAD}).
\item[II.] Anomaly detection: In the anomaly detection step, elements satisfying $\{ \bm{z}_i |~|\bm{z}_i|>\tau_\text{nrm}\}$ are classified as anomalies, where $\tau_\texttt{nrm} \in [2.5,4]$ is a predefined threshold. Here, $\tau_\texttt{nrm}$ plays the role of the standard deviation cut-off and helps to determine what is considered as anomaly or outlier in the model's gradient.
 In practice, $\tau_\text{nrm}$ is chosen in the range $[2.5,4]$ depending on e.g., the level of sensitivity in detecting outliers. For example, using a higher threshold (e.g., 3-4) will flag fewer data points, making the detection more conservative, while lower thresholds like $2.5$ or even $2$ make the detection more sensitive and are useful to catch even subtle outliers.

The non-anomalous (smooth) components of $\g$, are then extracted by:
\begin{equation}\label{nrm_g}
\texttt{nrm}(\g)=\set{[\g]_i:\left|z_i\right| \leq \tau_{\texttt{nrm}}}.
\end{equation}
\item[III.]  Balancing parameter adaptation: Since $\g \!=\! \g(\beta)$ and $\g_2=\g_2(\beta)$, the balancing parameter $\beta$ is adapted to ensure that the normally distributed elements of $\g$ identified in \cref{nrm_g} match the smooth component of the gradient, $\g_2$, updated in \cref{ADMM_gs}. This is formulated as a root-finding problem:
\begin{equation}\label{eq:phi(beta)}
\text{Find}\; \beta^{*} ~~~ \text{such that}~~\phi(\beta^{*})=0, ~~\text{where}~~ \phi(\beta)=\|\g_2(\beta)\|_{\infty}-\|\texttt{nrm}(\g(\beta))\|_{\infty}.
\end{equation}
Starting from an initial value $\beta^{0}$, the optimal $\beta$ at iteration $k$ is derived through the following fixed-point iteration:
\begin{equation}\label{beta_update}
\beta^{k}=\left(  \frac{2 \|\g_2^{k} \|_\infty}{\|\g_2^{k}\|_\infty + \|
   \texttt{nrm}(\g(\beta))\|_\infty} \right) \beta^{k-1}.
\end{equation}
\end{itemize}
%
\begin{algorithm}[tbhp]
 \begin{algorithmic}[1]
 \caption{Inner ADMM for single parameter acoustic FWI by TT} \label{alg:alg1}
 \STATE $\m^{k} = \left(\mu \bm{L}^{\top}\bm{L}+{\tau}_1\nabla^{\top}\nabla + {\tau}_2  \bm{I}\right)^{-1} \left(\mu \bm{L}^{\top}\bm{y}+ {\tau}_1 \nabla^{\top}(\g_1^{k-1}+\g_2^{k-1}+\Nu_1^{k-1}) +{\tau}_2(\p^{k-1}+\Nu_2^{k-1})\right)$,
   \STATE  $\g_1^{k} = \text{prox}_{\frac{1}{\tau_1}\ell_l}(\nabla \m^{k}-\g_2^{k-1}-\Nu_{1}^{k-1})$,
   \STATE $\g_2^{k}=(\bm{I}+\frac{\beta^{k-1} }{\tau_1}\overline{\nabla}^{\top}\overline{\nabla})^{-1}(\nabla \m^{k} -\g_1^{k}-\Nu_1^{k-1})$,
   \STATE $\p^{k} = \Pi_\mathcal{M}(\m^{k}-\Nu_2^{k-1}) $,
   \STATE $\Nu_1^{k}=\Nu_1^{k-1}+\g_1^{k}+\g_2^{k}-\nabla \m^{k}$,
   \STATE $\Nu_2^{k} = \Nu_2^{k-1}+\bm{p}^{k}-\m^{k} $,
   \STATE $\beta^{k}=\left(  \frac{2 \|\g_2^{k} \|_\infty}{\|\g_2^{k}\|_\infty + \|
   \texttt{nrm}(\nabla \m^{k})\|_\infty} \right) \beta^{k-1}$,
\end{algorithmic}
\end{algorithm}
%
\begin{algorithm}[tbhp]
 \begin{algorithmic}[1]
 \caption{Inner ADMM for two-parameter elastic FWI by TT}\label{alg:alg2}
 \STATE\label{line1}{$\m^{k}=  \begin{bmatrix}
        \m_\text{p}^k \\
        \m_\text{s}^k
    \end{bmatrix}=
    \left(\mu\bm{L}^{\top}\bm{L}+\mathscr{\bm{D}}^{\top}\bm{\Gamma}\mathscr{\bm{D}} + \bm{\Upsilon} \right) ^{-1} \left(\mu\bm{L}^{\top}\bm{y} + \mathscr{\bm{D}}^{\top} \bm{\Gamma} (\g_1^{k-1}+\g_2^{k-1}+\Nu_1^{k-1}) + \bm{\Upsilon} (\bm{p}^{k-1}+\Nu_2^{k-1}) \right)$,}
   \STATE  $\g_1^k = \begin{bmatrix}
         \g_{1,p}^{k} \\
          \g_{1,s}^{k}
     \end{bmatrix} =
     \begin{bmatrix}
          \text{prox}_{\frac{1}{\tau_{1,p}}\ell_l} (\nabla \m_p^{k}-\g_{2,p}^{k-1}-\Nu_{1,p}^{k-1}) \\
           \text{prox}_{\frac{1}{\tau_{1,s}}\ell_l} (\nabla \m_s^{k}-\g_{2,s}^{k-1}-\Nu_{1,s}^{k-1})
     \end{bmatrix}$,
   \STATE $ \g_2^k = \begin{bmatrix}
         \g_{2,p}^{k} \\
          \g_{2,s}^{k}
     \end{bmatrix} = 
     \begin{bmatrix}
         (\bm{I}+\frac{\beta^{k-1}_p}{\tau_{1,p}} \overline{\nabla}^{\top}\overline{\nabla})^{-1}(\nabla \m^{k}_p -\g_{1,p}^{k}-\Nu_{1,p}^{k-1}) \\
         (\bm{I}+\frac{\beta^{k-1}_s}{\tau_{1,s}} \overline{\nabla}^{\top}\overline{\nabla})^{-1}(\nabla \m^{k}_s -\g_{1,s}^{k}-\Nu_{1,s}^{k-1})
     \end{bmatrix}$,
   \STATE $ \p^{k} = \begin{bmatrix}
        \p_{p}^{k} \\
        \p_{s}^{k}
    \end{bmatrix} = 
\begin{bmatrix}
    \Pi_{\mathcal{M}_p}(\m_p^{k}-\Nu_{2,p}^{k-1}) \\
    \Pi_{\mathcal{M}_s}(\m_s^{k}-\Nu_{2,s}^{k-1})
\end{bmatrix}$,
   \STATE $\Nu_1^{k}= \begin{bmatrix}
       \Nu_{1,p}^{k} \\
       \Nu_{1,s}^{k}
   \end{bmatrix} =  
   \begin{bmatrix}
   \Nu_{1,p}^{k-1}+\g_{1,p}^{k}+\g_{2,p}^{k}-\nabla \m_p^{k} \\
   \Nu_{1,s}^{k-1}+\g_{1,s}^{k}+\g_{2,s}^{k}-\nabla \m_s^{k}
   \end{bmatrix}$,
   \STATE $\Nu_2^{k} = \begin{bmatrix}
       \Nu_{2,p}^{k} \\
       \Nu_{2,s}^{k}
   \end{bmatrix} = 
   \begin{bmatrix}
       \Nu_{2,p}^{k-1}+\bm{p}_p^{k}-\m_p^{k}  \\
        \Nu_{2,s}^{k-1}+\bm{p}_s^{k}-\m_s^{k}
   \end{bmatrix}$,
   \STATE $ \beta_p^k=\left(\frac{2\left\|{\g}_{2,p}\left(\beta_p^{k-1}\right)\right\|_{\infty}}{\left\|{\g}_{2,p}\left(\beta_p^{k-1}\right)\right\|_{\infty}+\left\|\texttt{nrm}\left( \nabla \m_p^{k}\right)\right\|_{\infty}}\right) \beta_p^{k-1}$,
   \STATE{$\beta_s^k=\left(\frac{2\left\|{\g}_{2,s}\left(\beta_s^{k-1}\right)\right\|_{\infty}}{\left\|{\g}_{2,s}\left(\beta_s^{k-1}\right)\right\|_{\infty}+\left\|\texttt{nrm}\left(\nabla \m_s^{k}\right)\right\|_{\infty}}\right) \beta_s^{k-1}$.}
\end{algorithmic}
\end{algorithm}
%
\paragraph{Considerations for elastic FWI.}
The AL formulation for elastic media is analyzed in \cite{Aghazade_2024_REF}. In two-parameter elastic FWI, the inner ADMM iteration for TT regularization follows the same principle as in the acoustic case, as summarized in \cref{alg:alg2}. However, due to the joint update of ($\m_\text{p}, \m_\text{s}$), each auxiliary and dual variable consists of two components:
\begin{equation*}
\g_1 = 
    \begin{bmatrix}
        \g_{1,p}\\ \g_{1,s}
    \end{bmatrix}, \quad
\g_2 = 
    \begin{bmatrix}
        \g_{2,p}\\ \g_{2,s}
    \end{bmatrix}, \quad
 \Nu_1 = 
     \begin{bmatrix}
        \Nu_{1,p}\\\Nu_{1,s}
    \end{bmatrix}, \quad
\Nu_2 =
     \begin{bmatrix}
        \Nu_{2,p}\\ \Nu_{2,s}
    \end{bmatrix}, \quad
\bm{p} = 
     \begin{bmatrix}
        \bm{p}_{p}\\ \bm{p}_{s}
    \end{bmatrix}.
\end{equation*}
In multiparameter media, different physical parameters exhibit distinct characteristics, requiring specific tuning for each model. The feasible sets for  ($\m_{p},\m_{s}$) are defined as:
 \begin{align}
   & \mathcal{M}_p = \set{ \m_p \in \mathbb{R}^{n}, \underline{\m}_p \le \m_p \le \overline{\m}_p },\\
    &\mathcal{M}_s = \set{ \m_s \in \mathbb{R}^{n}, \underline{\m}_s \le \m_s \le \overline{\m}_s }.
\end{align}
Moreover, to account for parameter-specific tuning, the diagonal structures of the tuning matrices in line 1 of \cref{alg:alg2} are defined as:
 \begin{equation}\label{EFWI_pars}
     \bm{\Gamma} = 
     \begin{bmatrix}
    {{\tau}}_{1,p} \bm{I}& \bm{0} \\
    \bm{0} & {\tau}_{1,s}\bm{I}
    \end{bmatrix},
    \quad \bm{\Upsilon} = 
    \begin{bmatrix}
    {\tau}_{2,p} \bm{I}& \bm{0} \\
    \bm{0} & {\tau}_{2,s}\bm{I}
\end{bmatrix}.
 \end{equation}
where $\bm{I} \in \mathbb{R}^{n \times n}$ is the identity matrix, and $ {\tau}_{1,p},  {\tau}_{1,s},  {\tau}_{2,p},  {\tau}_{2,s}>0$ are relative weights assigned to the bound constraints and regularization applied to each parameter class. 
The block-diagonal gradient operator (line 1) for the elastic inversion framework is given by:
\begin{equation}
 \mathscr{\bm{D}} =
 \begin{bmatrix}
    \nabla & \bm{0} \\
    \bm{0} & \nabla
\end{bmatrix}.
\end{equation}
Finally, similar to acoustic FWI, to control the balance between regularization terms, separate balancing parameters ($\beta_p,\beta_s$) are determined for each parameter class (lines 7-8).
%
%
\subsection{Practical considerations.}\label{sec:practical_consideration}
 
\subsubsection{On the selection of the free parameters.}\label{sec:params}
The strategy for tuning free parameters follows that of \cite{Aghamiry_2019_IBC} with some modifications to enhance generality and robustness across both acoustic and elastic FWI settings.  
 
The wavefield update in \cref{Ue_pdate} leverages the robustness of the AL formulation with respect to the penalty parameter ${\mu}$. As demonstrated in \cite{Aghamiry_2019_IWR} for acoustic FWI and extended in \cite{Aghazade_2024_REF} to elastic FWI, the algorithm performs reliably with a fixed ${\mu}$ across all iterations. Consequently, we maintain ${\mu}$ as a constant for both problem types.

Regarding the model subproblem for acoustic FWI,  the ratios $(\frac{\tau_{1}}{\mu},\frac{\tau_{2}}{\mu})$  associated with regularization and bound constraints are selected based on the diagonal of the Hessian matrix, $\bm{L}^{\top}\bm{L}$. Their values decrease over iterations to reduce the impact of regularization and constraints near the convergence point:
\begin{equation}\label{tau1_tau2}
\begin{aligned}
& \frac{\tau_{1}}{\mu} = \frac{c_{1}}{k} \times \text{max}(|\text{diag}(\bm{L}^{\top}\bm{L})|),\quad 0<c_{1}<1,  \\
& \frac{\tau_{2}}{\mu}= \frac{c_{2}}{k} \times \text{max}(|\text{diag}(\bm{L}^{\top}\bm{L})|), \quad 0<c_{2}<1,
\end{aligned}
\end{equation}
 with $c_{1} \geq c_{2}$. For the case of elastic FWI, the ratios $(\frac{\tau_{1,j}}{\mu},\frac{\tau_{2,j}}{\mu})$ with $j \in \set{p,s}$ indexing the parameter class; are set similar to \cref{tau1_tau2}. However, for elastic inversion $\bm{L}^{\top}\bm{L}$ is a 2 by 2 block diagonal matrix. Consequently, we use the diagonal of the first and second block for the case of P-and S-wave parameters, respectively.  
Regarding the subproblems \cref{eq:g1-prox} and \cref{g2_update}, for each parameter class ($\{p\}$ for acoustic and $\{p,s\}$ for elastic) we set:
\begin{equation*}
 \frac{1}{\tau_{1,j}} = c_{3,j} \times \text{max}\left(\sqrt{ \left|\nabla\m^{k}_j-\g_{2,j}^{k-1}-\Nu_{1,j}^{k-1}\right|^2} \right), \quad 0<c_{3,j}<1, \quad j \in \set{p,s}  
\end{equation*}
 where the optimal selection for $c_{3,j}$ depends on the properties of the parameter class $j$. Our analysis suggests that starting with a high $c_3$ value during low frequency inversion and gradually decreasing it as the frequency increases leads to improved results.
Finally, for updating the balancing parameter $\beta$, a predefined threshold distance $\tau_\text{nrm}$, is required to separate anomalies in the model gradient. In FWI, the optimal choice of $\tau_\text{nrm}$ depends on the problem, however, its value typically ranges from 2.5 to 4. Higher $\tau_\text{nrm}$ promotes smoother model updates, suitable for early iterations with low-frequency data, while lower $\tau_\text{nrm}$ allows for sharper features as frequency increases.
\subsubsection{Implementation considerations.}

For all numerical examples the variables are initialized as
\begin{itemize}
    \item[I.] The dual variables $\dual$ and ($\Nu_1,\Nu_2$) are initialized to zero. 
    \item[II.] The variables $\g_1$, $\g_2$ are set to zero at the beginning of the inversion process.
    \item[III.] The dual variable $\dual$ is reset to zero whenever we change the data frequency for inversion.
\end{itemize}

We compare the performance of TT regularization with the Tikhonov and TV regularizations, all implemented with the same algorithm described above. 
The Tikhonov regularization is achieved by disabling the update of $\g_1$, i.e. keeping $\g_1=\bm{0}$ throughout the process. The TV regularization is implemented by skipping $\g_2$ to focus solely on the TV term. 

Finally, we use the relative model error (RME) defined as
\begin{equation*}
  \text{RME}=   \frac{\|\m^{k}-\m^{*}\|_2}{\|\m^{*}\|_2},
\end{equation*}
where $\m^{*}$ represents the true model to show the convergence and reconstruction accuracy different methods. Computations were conducted on a dual Intel Xeon Platinum 8176 processor system with 56 cores operating at 2.10 GHz.
%
%
\section{Numerical examples.}\label{sec:experiments}
This section assesses the performance and characteristics of the adaptive TT regularization method using several 2D seismic imaging benchmark models under both constant-density acoustic and elastic approximations. 

\subsection{Acoustic examples.}
We implement the acoustic examples using Gaussian void model, SEAM model, and 2004 BP salt model. \Cref{tab:par_AC} reports the associated user-defined parameters (according to the discussion in section (\cref{sec:params}). 
\begin{table}[tbhp]
\small
\caption{Specified values for the free parameters used for the acoustic examples.}
\label{tab:par_AC}
\centering
\begin{tabular}{l c c c c}
\toprule 
& \multicolumn{3}{c}{Free parameters} \\
\cmidrule(l){2-5}
       & ${\mu}$ & $c_1$ & $c_2$ & $c_3$ \\ 
\midrule 
Gaussian void model (\cref{sec:AC_sec_Gaussian}) & $10^3$ & $0.6$ & $0.1$ & $0.3$ \\ 
SEAM (\cref{sec:AC_sec_SEAM}) & $10^4$ & $0.6$ & $0.1$ & $0.3$ \\ 
2004 BP (\cref{sec:AC_sec_BP}) & $10^4$ (noise-free), $10^7$ (noisy) & $0.7$ & $0.1$ & $0.4$ \\ 
\bottomrule 
\end{tabular}
\end{table}
\subsubsection{Piecewise-smooth reconstruction: Gaussian void model.}\label{sec:AC_sec_Gaussian}

To highlight the challenges of ill-posedness and the need for appropriate regularization, we conducted an inversion test for the Gaussian void model shown in \cref{fig:nabla_m_a}. The model dimension is 10 km by 10 km, discretized with intervals of 50 m.  The acquisition setup consists of 116 sources and receivers distributed along all edges. We invert a single frequency (5~Hz) data by starting from a homogeneous initial model of 3.2 km/s, which suffers from cycle skipping. 
 The inversion results after 50 iterations under different approaches are shown in \cref{fig:inclusion_res_a}, with velocity profiles in \cref{fig:inclusion_res_b}. 
The first column shows the results obtained using bound constraints (without regularization). Despite the overall reconstruction of the model, this strategy cannot fully address the ill-posedness. The bound constraint is thus implemented with regularization.
The second column shows the results obtained after adding the Tikhonov regularization to promote smoothness. The result is improved at the cost of over-smoothed sharp interfaces.
The third column shows the model obtained by replacing the Tikhonov regularization with TV regularization. We observe that, in this case, the model exhibits blocky features and edges are well preserved but the smooth background variations are destroyed by staircase artifacts.
Finally, in the fourth column, we present the inverted model using adaptive TT regularization approach, to balance smooth and blocky features. We observe that the velocity profile closely aligns with the true model, demonstrating effective switching between smoothness and edge preservation. 
The evolution of the RME during iterations, shown in \cref{fig:inclusion_res_error_a}, supports the success of regularization, especially the TT method. For TT regularization, \cref{fig:inclusion_res_error_b} (red curve) shows that after 30 iterations $\phi(\beta)\rightarrow 0$ (equation \cref{eq:phi(beta)}), resulting from reaching the optimal point for $\beta$ (blue curve).
%
%
\begin{figure}[tbhp]
\centering
 \includegraphics[width=.9\textwidth]{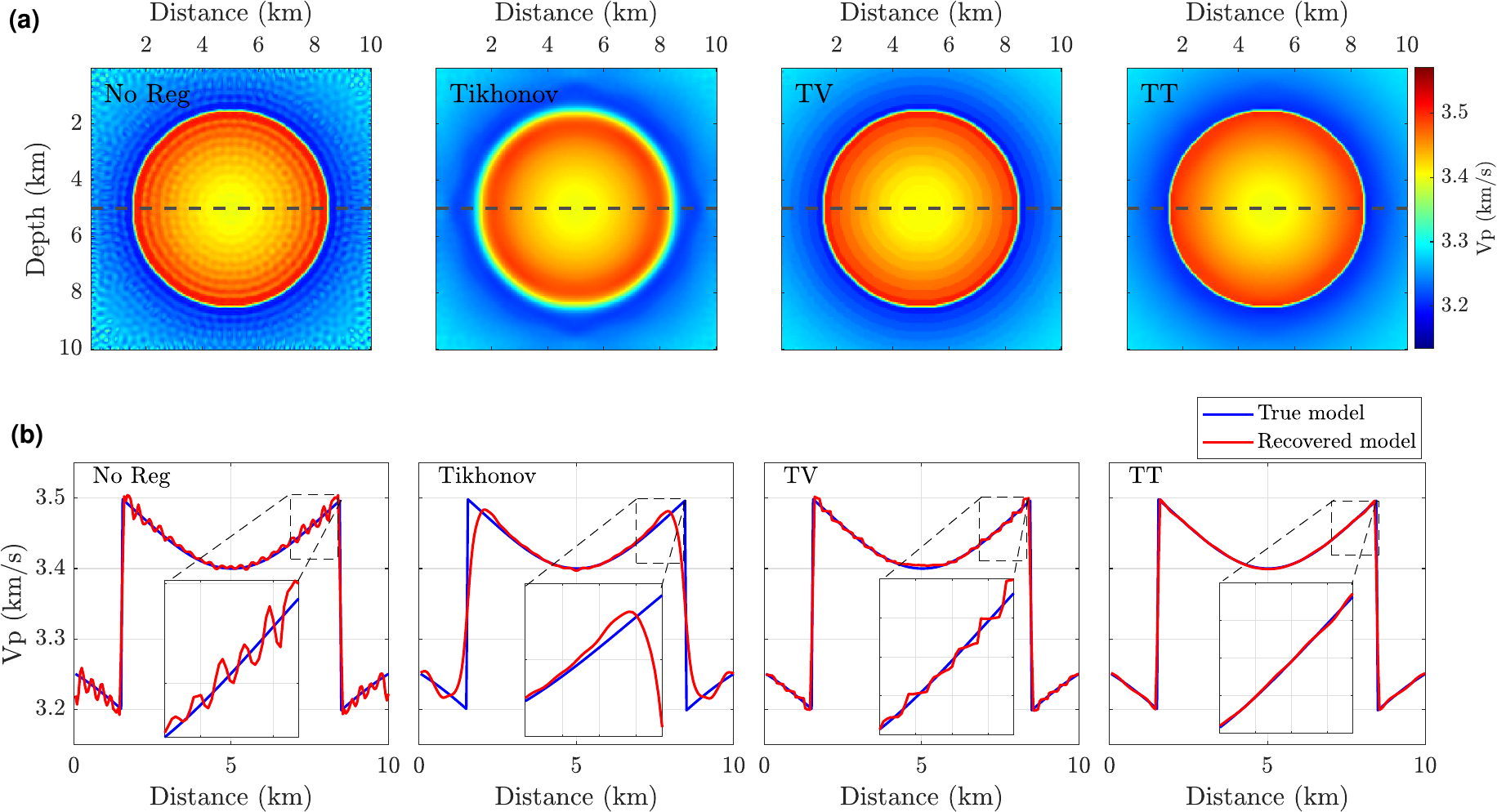}
\caption{Gaussian void test. Recovered velocity models obtained using different approaches. From left to right: without regularization (No Reg), and with  Tikhonov, TV, and TT regularizations.
(a) 2D velocity images.
(b)  1D velocity profiles along the horizontal dashed lines in the top row, comparing the recovered models (red) with the true model (blue).}
\label{fig:inclusion_res}
\ps{fig:inclusion_res_a}
\ps{fig:inclusion_res_b}
 \end{figure}
 %
 %
\begin{figure}[tbhp]
\centering
 \includegraphics[width=0.8\textwidth]{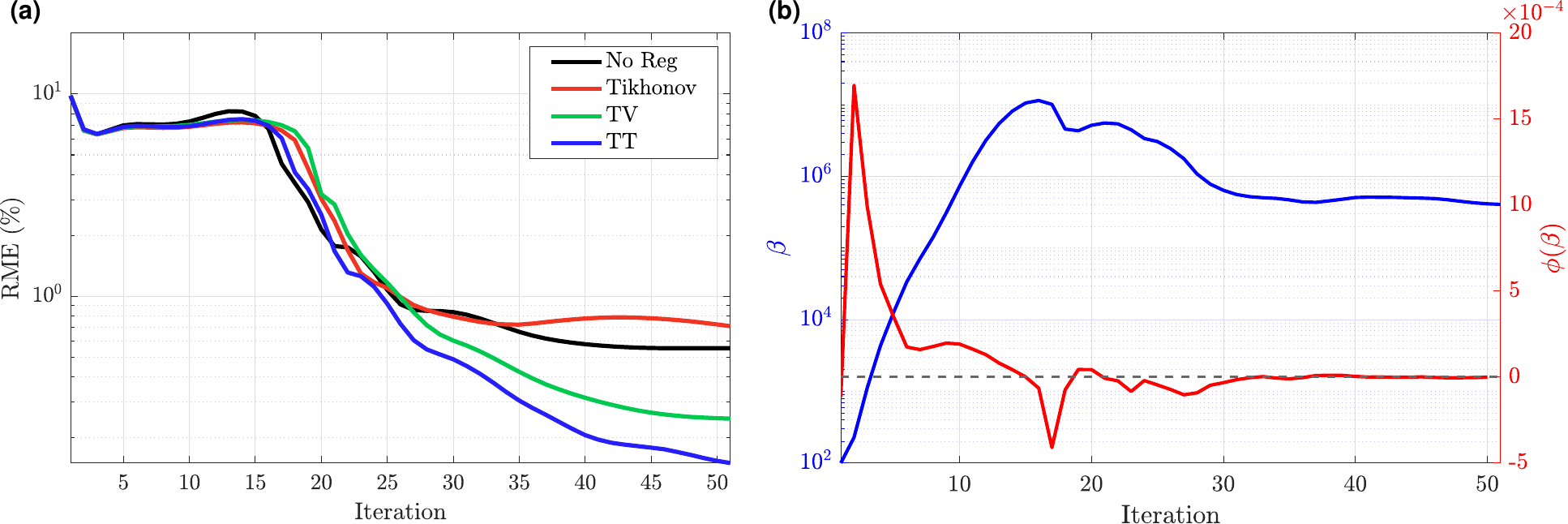}
\caption{Gaussian void test. (a) Evolution of the RME (\%) over iterations for the velocity models shown in \cref{fig:inclusion_res}. (b) The behavior of iteratively adjusted $\beta$ (blue curve) and $\phi(\beta)$ (red curve) for the case of TT regularization.}
\label{fig:inclusion_res_error}
\ps{fig:inclusion_res_error_a}
\ps{fig:inclusion_res_error_b}
 \end{figure}
%
%
\subsubsection{Mitigating cycle skipping: SEAM model.}\label{sec:AC_sec_SEAM}

In this example, we show how the adaptive TT regularization can mitigate the cycle skipping problems and reconstruct a complex model when both the Tikhonov and TV regularizations fail. For this, we examine the imaging of a 2D cross-section of the three-dimensional (3D) SEAM Phase I subsalt Earth model,  representing a deepwater Gulf of Mexico salt domain with fine-scale stratigraphy. The SEAM model poses significant challenges for seismic imaging due to complex high-velocity salt geometry, illumination shadows, distorted wave propagation, and overturned sediments.
The 2D model spans 35 km by 15 km \cite{Fehler_2011_SEAM}, resampled to 251 by 584 points with 60 m intervals (\cref{fig:SEAM_Vel_Shots_init_a}). Ocean bottom seismometer (OBS) acquisition used 35 hydrophones spaced 1 km apart to record seismic signals from 292 pressure sources placed 120 m apart at a depth of 60 m. Of note, spatial reciprocity of Green's functions was utilized to reduce computation costs while maintaining accuracy.
%
%
\begin{figure}[tbhp]
\centering
\includegraphics[width=0.9\textwidth]{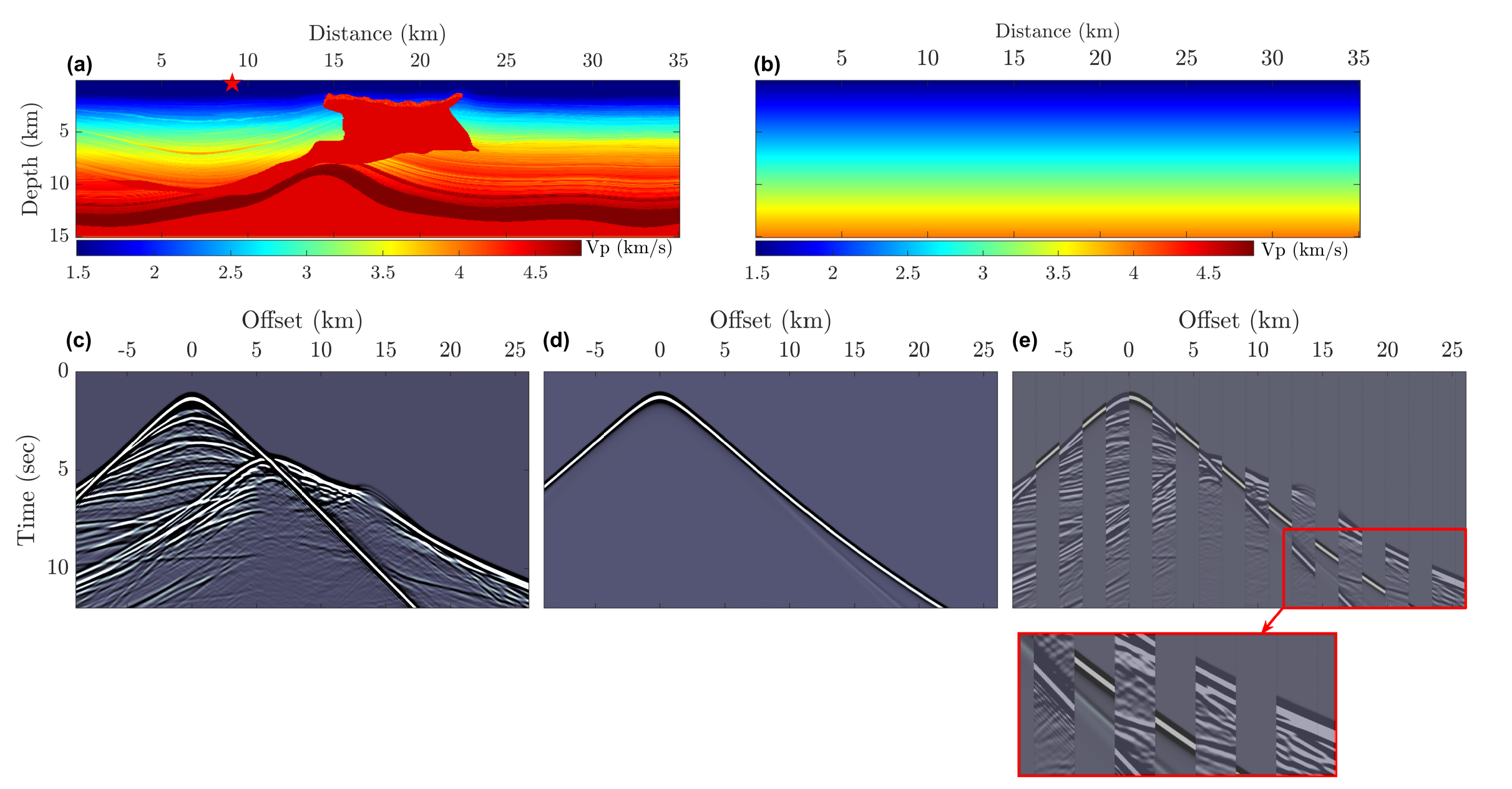}
\caption{SEAM test. (a) The true velocity model. (b) The initial model. Seismograms  computed at the true model (c) and initial model (d). (e) An interleaved display of computed seismograms, showing alternating segments of traces from (c) and (d). A close-up of a segment of the seismograms is also provided for a detailed comparison.  The source position is shown in panel (a) by a red star at $X=9.3$ km.}
\label{fig:SEAM_Vel_Shots_init}
\ps{fig:SEAM_Vel_Shots_init_a}
\ps{fig:SEAM_Vel_Shots_init_b}
\ps{fig:SEAM_Vel_Shots_init_c}
\ps{fig:SEAM_Vel_Shots_init_d}
\ps{fig:SEAM_Vel_Shots_init_e}
\end{figure}
The starting velocity model for inversion linearly increases with depth from 1.5 km/s to 4 km/s (\cref{fig:SEAM_Vel_Shots_init_b}).
Time-domain seismograms are compared in \cref{fig:SEAM_Vel_Shots_init_c,fig:SEAM_Vel_Shots_init_d} for true and initial models, respectively. For a better comparison, the interleaved seismograms are presented in \cref{fig:SEAM_Vel_Shots_init_e}, where alternating sections of traces from both the observed and predicted seismograms are combined, with a zoomed-in view highlighting waveform alignment discrepancies and severe cycle skipping.

\begin{figure}[tbhp]
\centering
\includegraphics[width=0.9\textwidth]{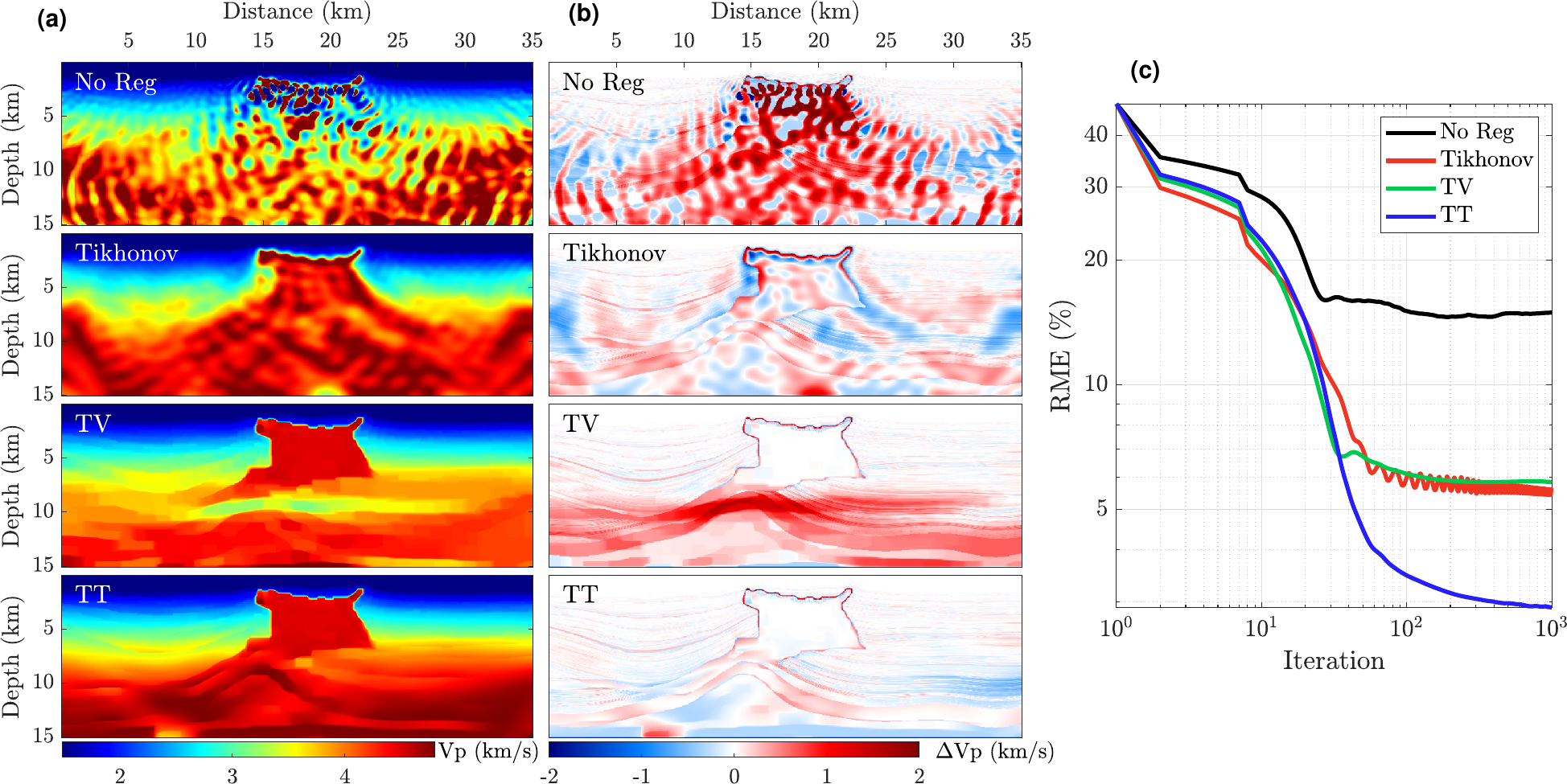}
\caption{SEAM test. (a) Comparison of velocity model reconstructions without regularization (No Reg, first row) and different regularization techniques, Tikhonov (second row) TV (third row) and TT (fourth row) for data frequency of 1.5~Hz after 1000 iterations. (b) The difference between true and estimated velocity model. (c) The evolution of RME (\%) versus iteration for each method.}
\label{fig:SEAM_inv_res}
\ps{fig:SEAM_inv_res_a}
\ps{fig:SEAM_inv_res_b}
\ps{fig:SEAM_inv_res_c}
\end{figure}

\begin{figure}[tbhp]
\centering
\includegraphics[width=.9\textwidth]{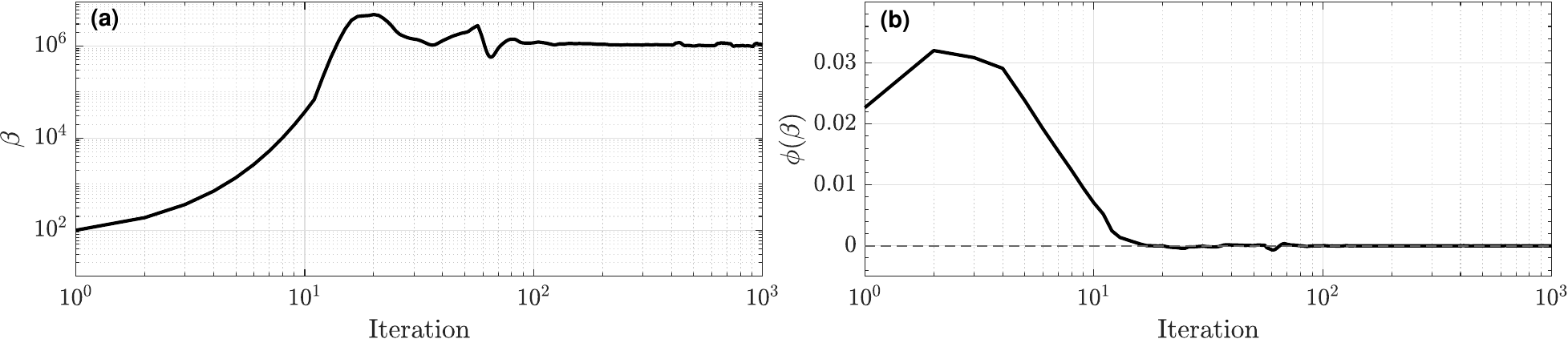}
\caption{SEAM test. (a) Iteratively adjusted $\beta$ using \cref{beta_update}  and (b) variations of function $\phi(\beta)$ \cref{eq:phi(beta)} for the case of TT regularization result shown in \cref{fig:SEAM_inv_res}.}
\label{fig:SEAM_beta}
\end{figure}

\begin{figure}[tbhp]
\centering
 \includegraphics[width=0.8\textwidth,trim={0 2cm 4cm 0},clip]{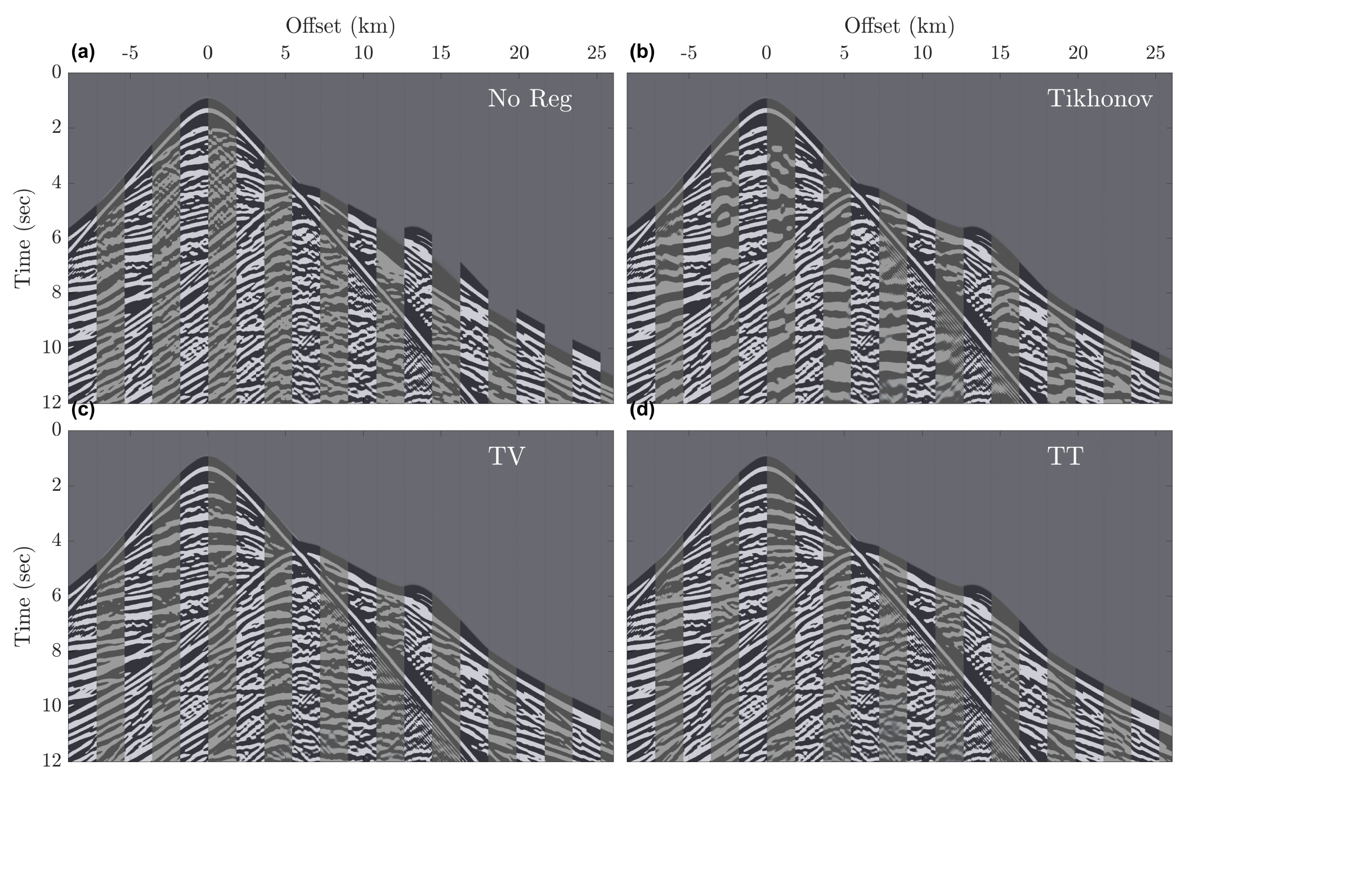}
\caption{SEAM test. Interleaved computed seismograms, combining alternating segments of traces from both the true seismogram and seismograms (highlighted segments) computed at the estimated velocity models shown in \cref{fig:SEAM_inv_res_a}. (a) Without regularization, and with (b) Tikhonov, (c) TV, and (d) TT regularization.}
\label{SEAM_Shots_est}
 \end{figure}

We demonstrate the convergence characteristics of various regularization approaches by inverting a 1.5 Hz single-frequency dataset. Each inversion runs for 1000 iterations to thoroughly evaluate long-term convergence patterns and identify potential local minima issues. For TT regularization, the balancing parameter is initialized as $\beta^{0}=10^2$. \Cref{fig:SEAM_inv_res} shows the inverted results: \cref{fig:SEAM_inv_res_a} depicts the reconstructed velocity models, while \cref{fig:SEAM_inv_res_b} displays the velocity difference map ($V^{*}(\bm{x})\!-\!V(\bm{x})$). 
The first row shows the results without regularization. Clearly, the algorithm gets trapped in a poor local minimum. The second row demonstrates the effect of adding Tikhonov regularization, which improves the result. However, the algorithm still converges to a local minimum—smoother than the previous case, but still suboptimal compared to a case with a proper regularization. The third row presents the results using TV regularization, leading to a different local minimum. While some important features of the velocity model, such as the top of the salt, are recovered, a low-velocity gap is incorrectly introduced in the subsalt region. This issue with TV regularization for subsalt imaging has also been observed by \cite{Esser_2016_CWI}. TT regularization (fourth row) significantly enhances model recovery, overcoming subsalt imaging issues. The RME plot in \cref{fig:SEAM_inv_res_c} confirms these findings, showing that after 100 iterations all methods except TT regularization are trapped in local minima. It also highlights the faster and more stable convergence of the TT method compared with Tikhonov or TV regularization alone. The updated values of the parameter $\beta$ during iterations as well as corresponding objective function $\phi(\beta)$ are shown in \cref{fig:SEAM_beta}. We see that the adjusted parameter converges to the root of $\phi(\beta)$ in a stable manner.
%
%
%
%
To further assess the accuracy of recovered velocity models, \cref{SEAM_Shots_est} presents interleaved computed seismograms, combining traces from both true and estimated models (\cref{fig:SEAM_inv_res_a}). Seismograms from the TT-regularized model exhibit the closest alignment with true data, while models obtained with other regularization methods show misalignment, particularly in later arrivals.
%
%
In \cref{fig:SEAM_inv_res}, we compare the performance of different regularization methods for inverting single-frequency data at 1.5 Hz using 1000 iterations. In practice, however, such a large number of iterations is unnecessary, as the inversion can be more effectively carried out by performing fewer iterations per frequency in a multiscale scheme. To demonstrate this, \cref{fig:SEAM_multi_F_a} shows the result of inverting the same 1.5 Hz data but with only 150 iterations. This model is then used as the starting point for successive inversion of frequencies from 2 to 5 Hz, with a frequency step of 0.5 Hz, each inverted with 20 iterations. The final result at 5 Hz is shown in \cref{fig:SEAM_multi_F_b}, with the corresponding convergence curves displayed in \cref{fig:SEAM_multi_F_c}. We observe that all three methods benefit from the inclusion of higher-frequency information, but TT regularization consistently provides superior results across the entire model, particularly in the deeper regions.
%
%
\begin{figure}[tbhp]
 \centering
  \includegraphics[width=0.9\linewidth]{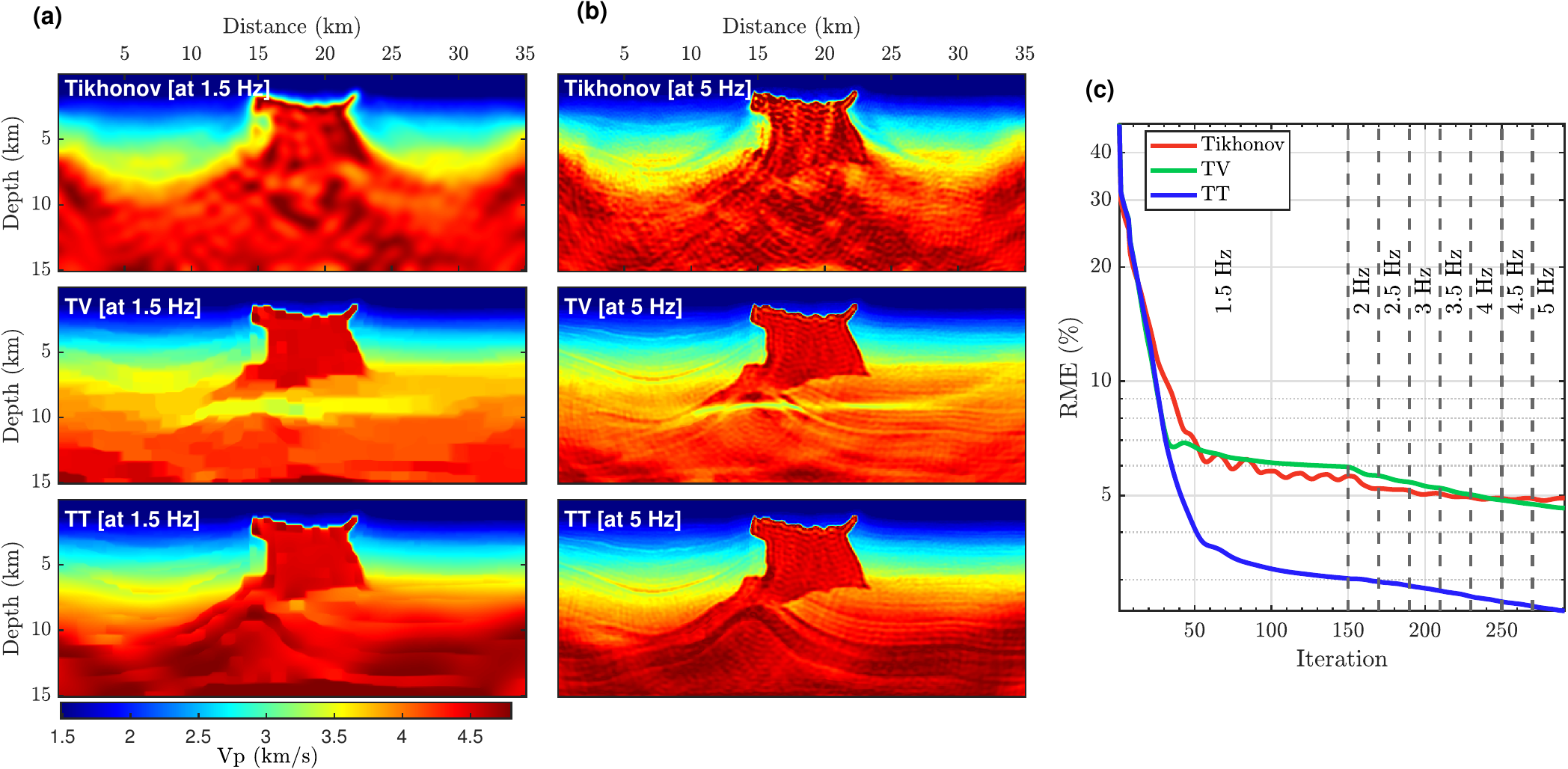}
   \caption{SEAM test. Inversion results comparing Tikhonov, TV, and TT regularization functionals. (a) Inversion results at 1.5 Hz (after 150 iterations), (b) Final results at 5 Hz, and (c) The RME (\%) versus iteration, with frequency transitions indicated by dashed lines.}
\label{fig:SEAM_multi_F}
\ps{fig:SEAM_multi_F_a}
\ps{fig:SEAM_multi_F_b}
\ps{fig:SEAM_multi_F_c}
\end{figure}
The computational performance analysis reported in \cref{tab:SEAM_runtime} demonstrates minimal computational overhead across all regularization methods. All timing measurements represent wall-clock execution time averaged over 10 independent runs to ensure statistical reliability. TV regularization exhibits the fastest execution (0.76 seconds/iteration), while adaptive TT shows comparable efficiency (0.83 seconds/iteration). The cost of non-adaptive TT (fixed $\beta$) is also reported to confirm negligible cost for parameter selection in adaptive TT. Considering TT's higher accuracy, this minimal computational penalty establishes it as a cost-effective choice for large-scale FWI implementations.
\begin{table}[tbhp] 
\small
\caption{Computational performance of regularization methods for the SEAM model: the average runtime per iteration (in seconds) for model regularization (\cref{alg:alg1}) and one FWI iteration (including regularization). Results are averaged over 10 runs.}
\label{tab:SEAM_runtime}
\centering  
\begin{tabular}{l c c c c c} 
\toprule 
& \multicolumn{4}{c}{Runtime (sec)} \\ 
\cmidrule(l){2-5} 
       & Tikhonov & TV & Non-adaptive TT & Adaptive TT \\ 
\midrule 
One FWI iteration  & 5.58 & 5.52 & 5.59 & 5.6\\ 
Regularization part  & 0.80 & 0.76 & 0.81 & 0.83\\ 
\bottomrule 
\end{tabular}
\end{table}
\subsubsection{The role of various parameters: 2004 BP salt model.}\label{sec:AC_sec_BP}

We extend our analysis using the 2004 BP salt model \cite{Billette_2004_BPB}, designed to evaluate seismic imaging techniques in complex geological settings. This model features sedimentary basins with increasing velocity strata and high-velocity, high-contrast salt formations (\cref{fig:BP_Vel_Shots_init_a}).
The original model size of 12 km by 67.5 km is retained, with a grid spacing of 75 m. We employed an ultra-long-offset fixed-spread OBS acquisition with 67 seismometers spaced 1 km apart to capture signals from 450 pressure sources, uniformly distributed 150 m apart at a depth of 75 m below the sea surface. In addition, Green’s function reciprocity was applied to reduce computational costs.
The inversion began with a 1D initial model that linearly increases from 1.5 km/s to 3.5 km/s with depth (\cref{fig:BP_Vel_Shots_init_b}). In this case, conventional FWI started with minimum 1 Hz data frequency converges to a local minimum. This is evident in the computed seismograms shown in \cref{fig:BP_Vel_Shots_init_c,fig:BP_Vel_Shots_init_d,fig:BP_Vel_Shots_init_e}, where panels (c) and (d) correspond to the true and initial models, and panel (e) presents the interleaved traces combining the two.
%
%
\begin{figure}[tbhp]
\centering
\includegraphics[width=1\textwidth]{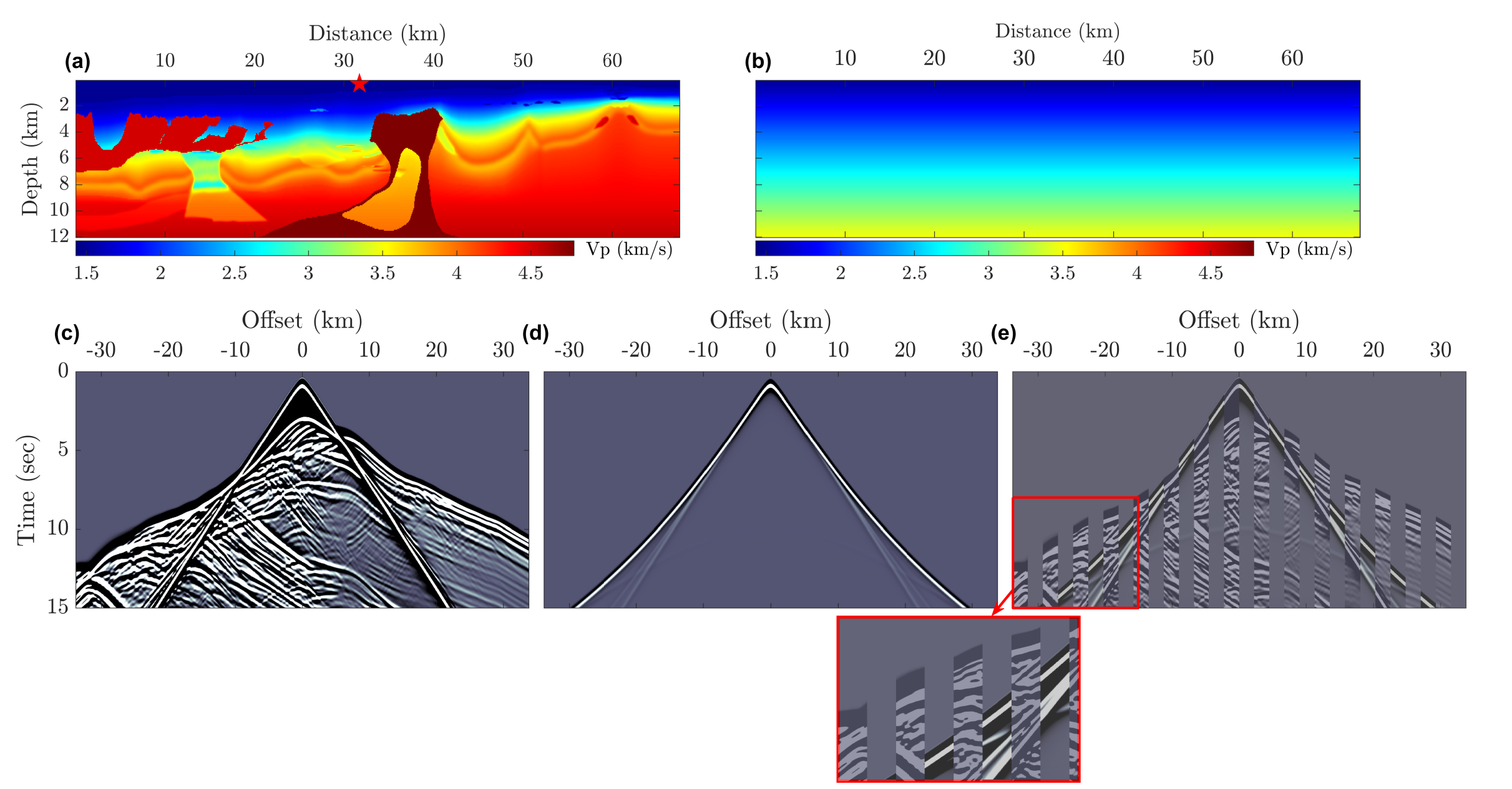}
\caption{2004~BP test. (a) The true velocity model. (b) Initial model. (c) Seismogram computed at true model and (d) initial model. (e) Interleaved computed seismograms; combining alternating segments of traces from (c) and (d). The source position is shown in panel (a) by a red star at $X=31.725$ km.}
\label{fig:BP_Vel_Shots_init}
\ps{fig:BP_Vel_Shots_init_a}
\ps{fig:BP_Vel_Shots_init_b}
\ps{fig:BP_Vel_Shots_init_c}
\ps{fig:BP_Vel_Shots_init_d}
\ps{fig:BP_Vel_Shots_init_e}
 \end{figure}
Various approaches can be employed to design a robust algorithm for FWI and partially mitigate cycle skipping. These include the multipath frequency continuation strategy \cite{Bunks_1995_MSW} and different data conditioning techniques \cite{Gorszczy_k2017_TRW}. The former involves sequentially inverting different frequency bands, starting from low frequencies and gradually progressing to higher ones. This process is repeated multiple times to recover different scales of the model and improve data fitting across frequencies.
For instance, \cite{Gholami_2024_FWI_LM} implemented a frequency continuation strategy with three cycles: (i) 1 Hz to 3.5 Hz, (ii) 1 Hz to 4.5 Hz, and (iii) 1 Hz to 5 Hz, each using a 0.5 Hz frequency interval. This strategy enhances convergence, as demonstrated in \cite{Gholami_2024_FWI_LM} (their Fig. 15).
However, since this study focuses on examining the role of regularization in FWI, we adopt a single-path inversion approach. This makes the inversion process more challenging, thereby providing a clearer assessment of regularization's impact.

\begin{figure}[tbhp]
\centering
\includegraphics[width=0.9\textwidth]{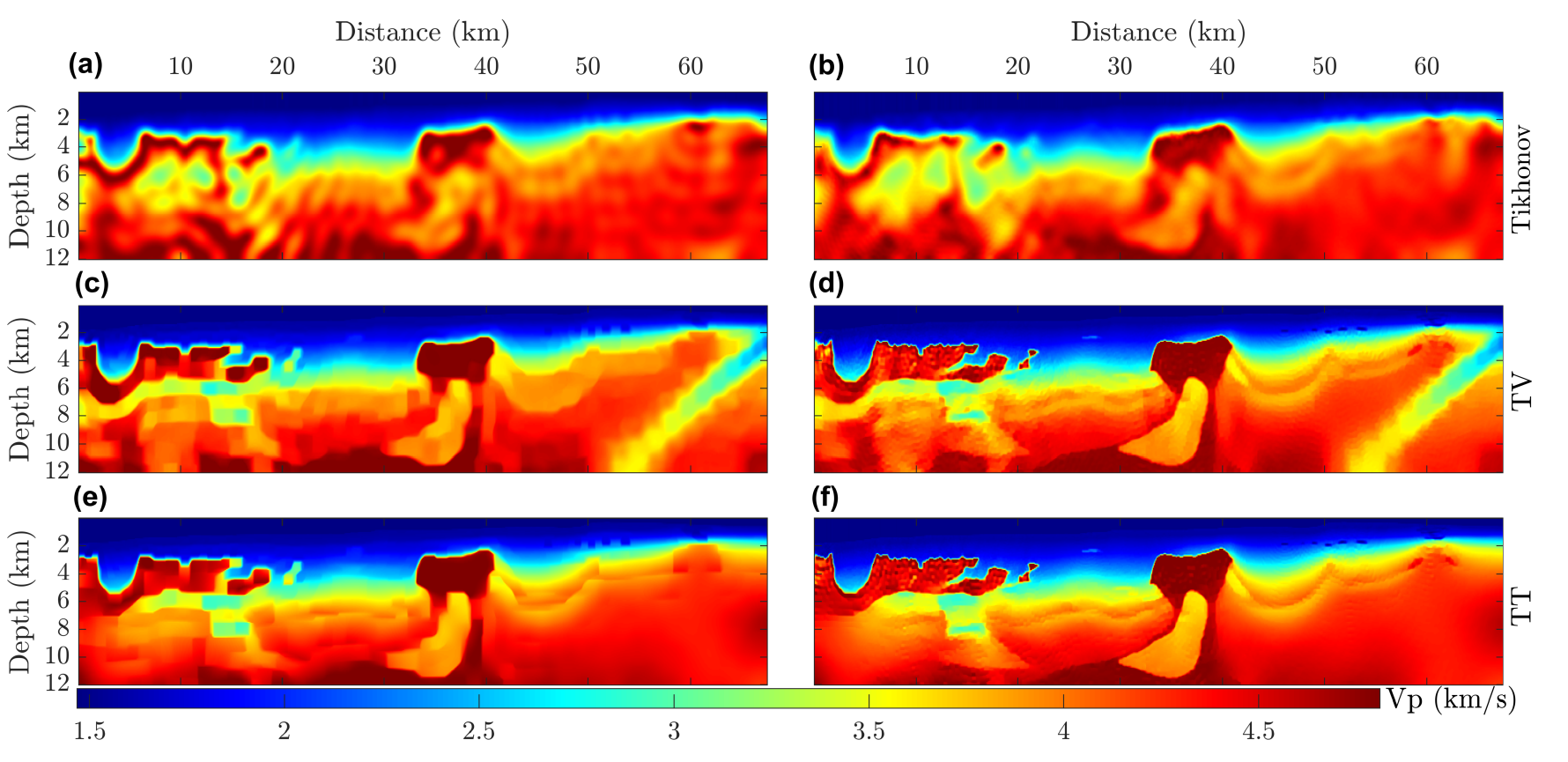}
\caption{2004~BP test. The inversion results obtained by different regularization methods. (a-b) Tikhonov regularization, (c-d) TV regularization, and (e-f) TT regularization. (a,c,e) The inverted results at 1~Hz. (b,d,f) The final results.}
\label{fig:BP_inv_res}
\ps{fig:BP_inv_res_a}
\ps{fig:BP_inv_res_b}
\ps{fig:BP_inv_res_c}
\ps{fig:BP_inv_res_d}
\ps{fig:BP_inv_res_e}
\ps{fig:BP_inv_res_f}
\end{figure}

We perform sequential inversion of four discrete frequencies (1, 2, 3, 5 Hz) over 225 iterations, starting with 150 iterations at 1 Hz to evaluate convergence towards a large-wavelength background model, followed by 25 iterations each for 2 Hz, 3 Hz, and 5 Hz. For the TT method, the initial value of the balancing parameter is set to $\beta^{0}=10^2$. The inversion results shown in \cref{fig:BP_inv_res} compare Tikhonov, TV, and TT regularization techniques. The first column (\cref{fig:BP_inv_res_a,fig:BP_inv_res_c,fig:BP_inv_res_e}) shows models obtained using 1 Hz data, while the second column (\cref{fig:BP_inv_res_b,fig:BP_inv_res_d,fig:BP_inv_res_f}) shows final models after inverting up to 5 Hz. At 1 Hz, all methods capture large-scale features. Tikhonov regularization (\cref{fig:BP_inv_res_a}) produces smooth models but lacks structural detail, especially on the left. TV regularization (\cref{fig:BP_inv_res_c}) enhances sharp contrasts but introduces low-velocity artifacts on the right. TT regularization (\cref{fig:BP_inv_res_e}) provides a balanced approach, yielding sharper features while maintaining continuity. As higher frequencies are incorporated (\cref{fig:BP_inv_res_b,fig:BP_inv_res_d,fig:BP_inv_res_f}), Tikhonov regularization remains overly smooth, failing to recover the left-side salt body and missing finer details. TV regularization enhances detail resolution but introduces a low-velocity artifact on the right. This failure is consistent with the issue observed in \cref{fig:SEAM_inv_res}, because high-velocity anomalies at a distance interval of 59~km-63~km and depth interval of 2.8~km-3.8~km, combined with low illumination beneath them, degrade the inversion result.  In contrast, TT regularization achieves a near-accurate reconstruction of the model. These findings show the importance of selecting an appropriate regularization strategy based on target resolution and model characteristics. TT regularization effectively balances smoothness and blockiness, making it suitable for media with varied structural properties.  
%
%
\noindent
\Cref{fig:BP_inv_res_FT_a,fig:BP_inv_res_FT_c,fig:BP_inv_res_FT_e} compares the final models, \cref{fig:BP_inv_res}, in the Fourier domain. Magnitude spectrum differences between the true and recovered models are shown in \cref{fig:BP_inv_res_FT_b,fig:BP_inv_res_FT_d,fig:BP_inv_res_FT_f}. Tikhonov regularization (\cref{fig:BP_inv_res_FT_a,fig:BP_inv_res_FT_b}) focuses on low-wavenumber content, producing a compact central pattern. TV regularization (\cref{fig:BP_inv_res_FT_c,fig:BP_inv_res_FT_d}  captures finer details at higher frequencies, though inaccuracies are present at low-wavenumber. The TT method (\cref{fig:BP_inv_res_FT_e,fig:BP_inv_res_FT_f} combines the low-frequency strengths of Tikhonov and TV methods with the high-frequency accuracy of the TV method. This balanced recovery of low and high wavenumber is crucial for effective wave-based imaging applications.
%
%
\begin{figure}[tbhp]
\centering
\includegraphics[width=0.7\textwidth]{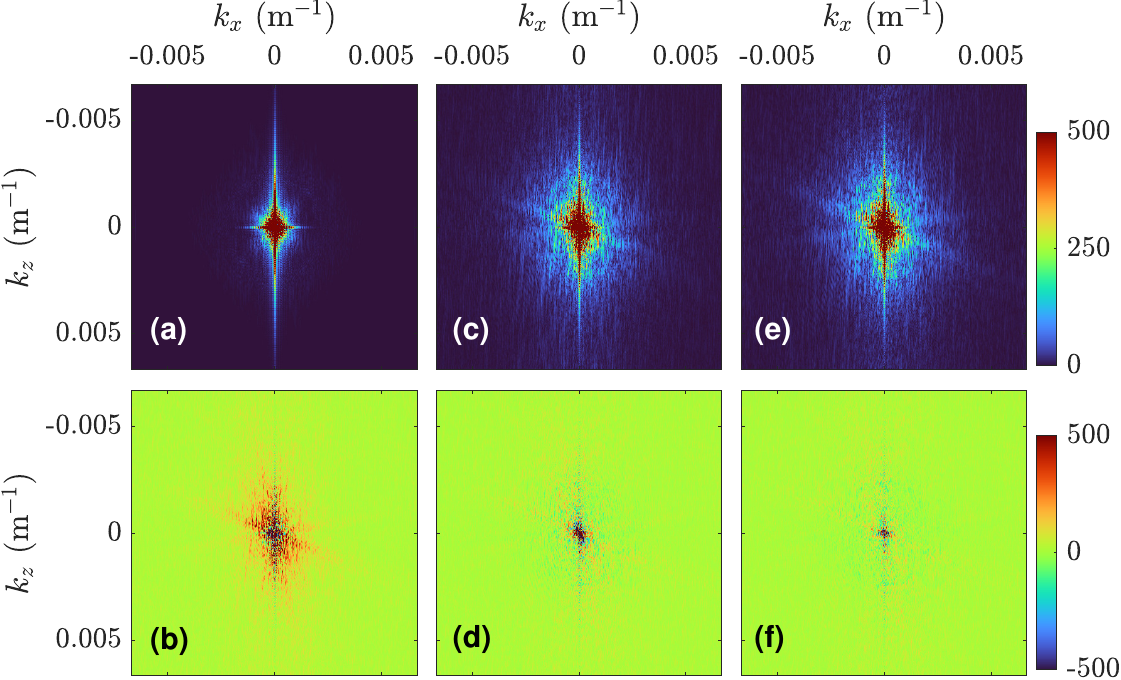}
\caption{2004~BP test. (a,c,e) The Fourier spectra of the final reconstructed velocity models (shown in \cref{fig:BP_inv_res}) using (a) Tikhonov,  (c) TV, and (e) TT regularization. (b,d,f) The difference between true  Fourier spectrum and (a,c,e).}
\label{fig:BP_inv_res_FT}
\ps{fig:BP_inv_res_FT_a}
\ps{fig:BP_inv_res_FT_b}
\ps{fig:BP_inv_res_FT_c}
\ps{fig:BP_inv_res_FT_d}
\ps{fig:BP_inv_res_FT_e}
\ps{fig:BP_inv_res_FT_f}
\end{figure}
Detailed comparisons of velocity profiles (\cref{fig:BP_logs_a}) highlights the superiority of TT regularization, showing a closer alignment with the true model at various depths. In addition, \cref{fig:BP_logs_b} quantitatively compares the efficacy of three regularization methods using RME (\%) throughout iterations, with TT regularization demonstrating faster and more stable convergence than Tikhonov and TV. After 150 iterations, TT achieves a model error below 5\%, while Tikhonov and TV show delayed convergence and higher errors. 
%
%
\begin{figure}[tbhp]
\centering
\includegraphics[scale=0.4]{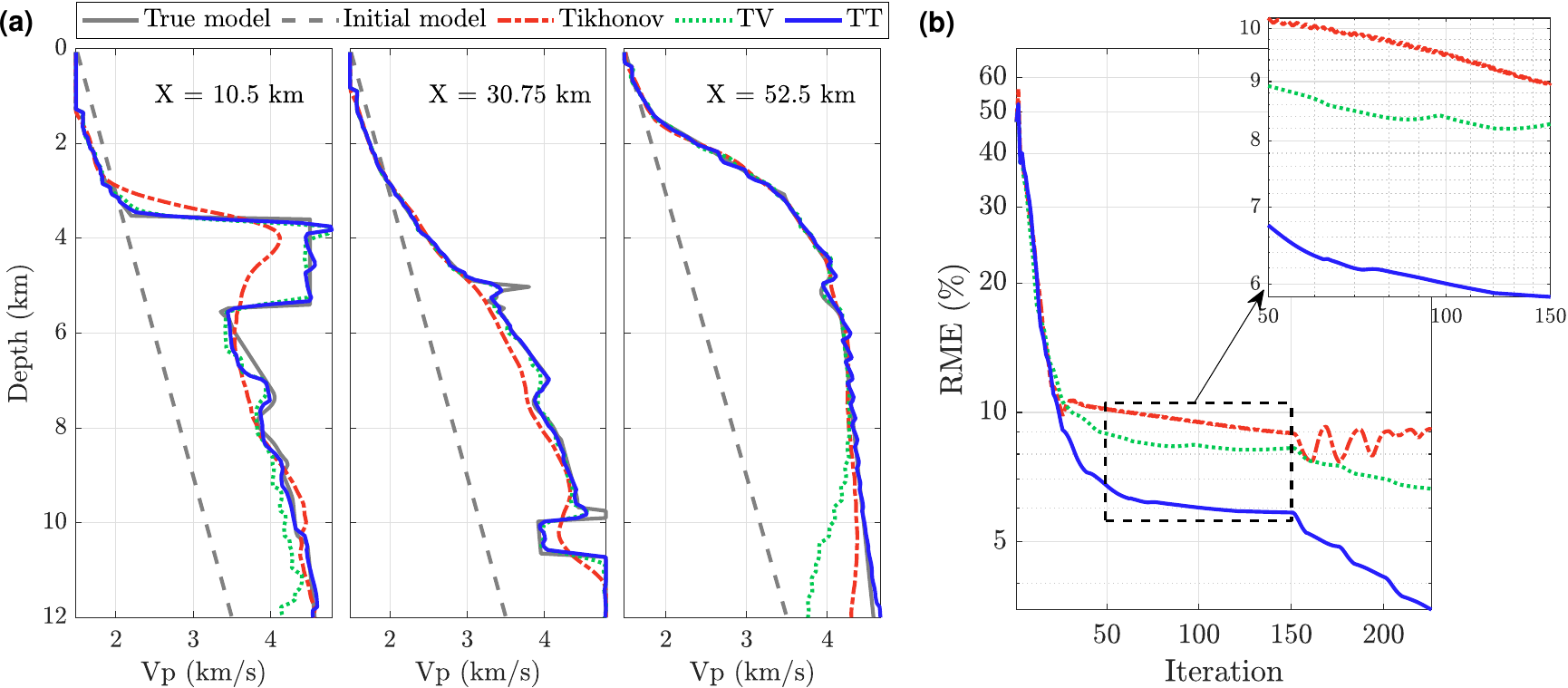}
\caption{2004~BP test. (a) Velocity logs extracted from the models in \cref{fig:BP_inv_res_b,fig:BP_inv_res_d,fig:BP_inv_res_f} at three locations denoted by $X$ in comparison with the true and initial models. (b) The evolution of the RMEs (\%) during iterations.
}
\label{fig:BP_logs}
\ps{fig:BP_logs_a}
\ps{fig:BP_logs_b}
\end{figure}
The smooth and blocky components of the estimated model obtained with TT regularization  are displayed in \cref{fig:BP_m1_m2}, together with their corresponding wavenumber spectra. The decomposition successfully separates the blocky structures ($\m_1$) from the smooth 
background variations ($\m_2$) in the 2004 BP salt model. The Fourier spectra confirm that  $\m_1$ captures the high-wavenumber content associated with sharp geological boundaries,  whereas $\m_2$ retains the low-wavenumber trends and gradual velocity variations. 
This property of TT regularization may be interpreted as an implicit multiscale strategy,  
the method favors smooth models in the early iterations and progressively incorporates 
finer details as the data misfit decreases. This interpretation is supported by the  behavior of the adaptively updated parameter $\beta$ and its objective function $\phi(\beta)$, shown in \cref{fig:beta_phi_BP1}. During the early iterations, 
$\beta$ remains small (\cref{fig:beta_phi_BP1_a}), promoting updates of the background model and enhancing low-wavenumber features. As the iterations advance, $\beta$ increases, thereby shifting the emphasis toward high-wavenumber, blocky updates through the TV term. 
%
%
\begin{figure}[tbhp]
\centering
\includegraphics[width=.8\linewidth]{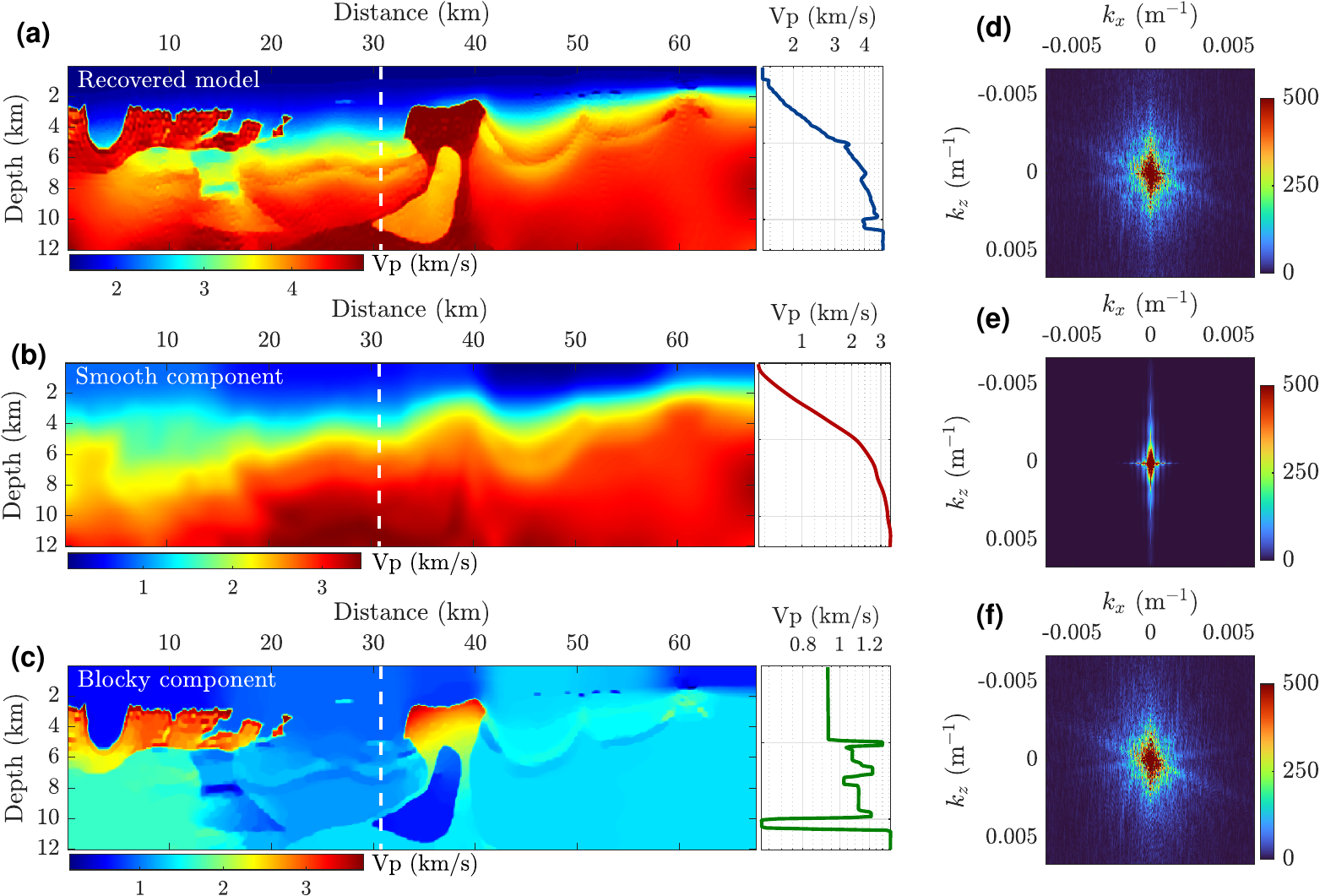}
\caption{2004~BP test. Decomposition of the inversion result by TT regularization (a) into smooth (b) and blocky components (c). For the white dashed line in each model, the corresponding vertical profile is shown. (e-f) The Fourier magnitude spectrum of (a-c).
}
\label{fig:BP_m1_m2}
\end{figure}
%
%
\begin{figure}[tbhp] 
    \centering
    \includegraphics[width=0.9\linewidth]{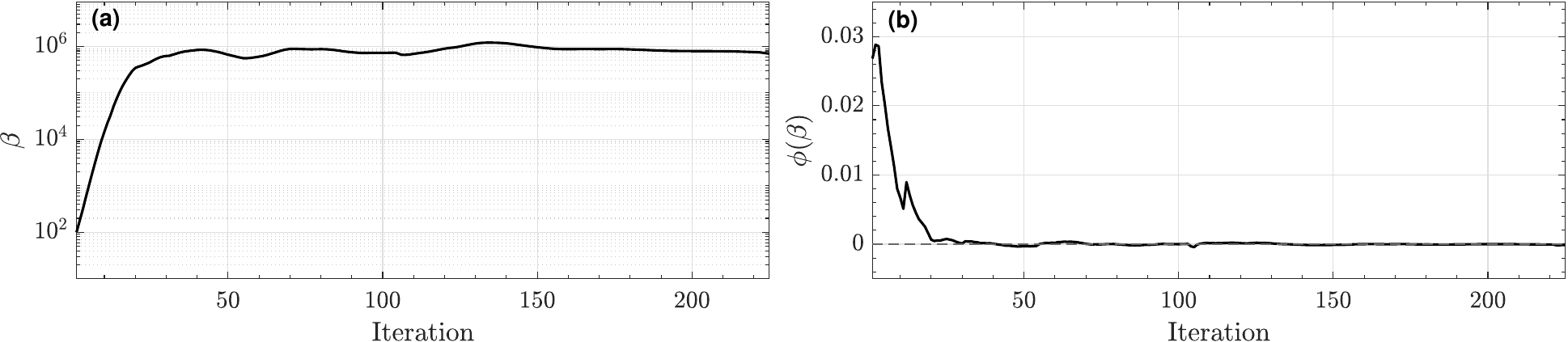}
    \caption{2004~BP test. (a) Iteratively adjusted $\beta$ using \cref{beta_update}  and (b) variations of function $\phi(\beta)$ \cref{eq:phi(beta)} during TT regularization for the result shown in \cref{fig:BP_inv_res_f}.}
 \label{fig:beta_phi_BP1}
\ps{fig:beta_phi_BP1_a}
\ps{fig:beta_phi_BP1_b}    
\end{figure}
\begin{table}[tbhp] 
\caption{Computational performance of regularization methods for 2004 BP model: the average runtime per iteration (in seconds) for model regularization (\cref{alg:alg1}) and one FWI iteration (including regularization). Results are averaged over 10 runs.} 
\label{tab:BP_runtime} 
\centering  
\begin{tabular}{l c c c c c} 
\toprule 
& \multicolumn{4}{c}{Runtime (sec)} \\ 
\cmidrule(l){2-5} 
       & Tikhonov & TV & Non-adaptive TT & Adaptive TT \\ 
\midrule 
One FWI iteration  & 8.47 & 8.45 & 8.55 & 8.64\\ 
Regularization part & 0.80 & 0.78 & 0.82 & 0.84\\ 
\end{tabular}
\end{table}
\Cref{tab:BP_runtime} summarizes the computational performance of each regularization method applied to the 2004 BP model, reporting the average runtime per iteration for both the model regularization step (\cref{alg:alg1}) and the one FWI iteration (including regularization). The difference in total runtime per iteration is marginal (less than 0.2 sec), indicating that adaptive TT regularization adds minimal overhead despite its more sophisticated balancing mechanism.
\paragraph{On the robustness of the balancing parameter selection.}
To evaluate the effectiveness of adaptive balancing parameter selection in TT regularization, the inversion is conducted with different initializations of $\beta^{0} \in [10^{-2}, 10^3]$. The corresponding results are shown in \cref{fig:BP_b_var}. Despite differing initial $\beta$ values, the method achieved similarly accurate results. 
%
%
\begin{figure}[tbhp]
\centering
\includegraphics[width=0.9\textwidth]{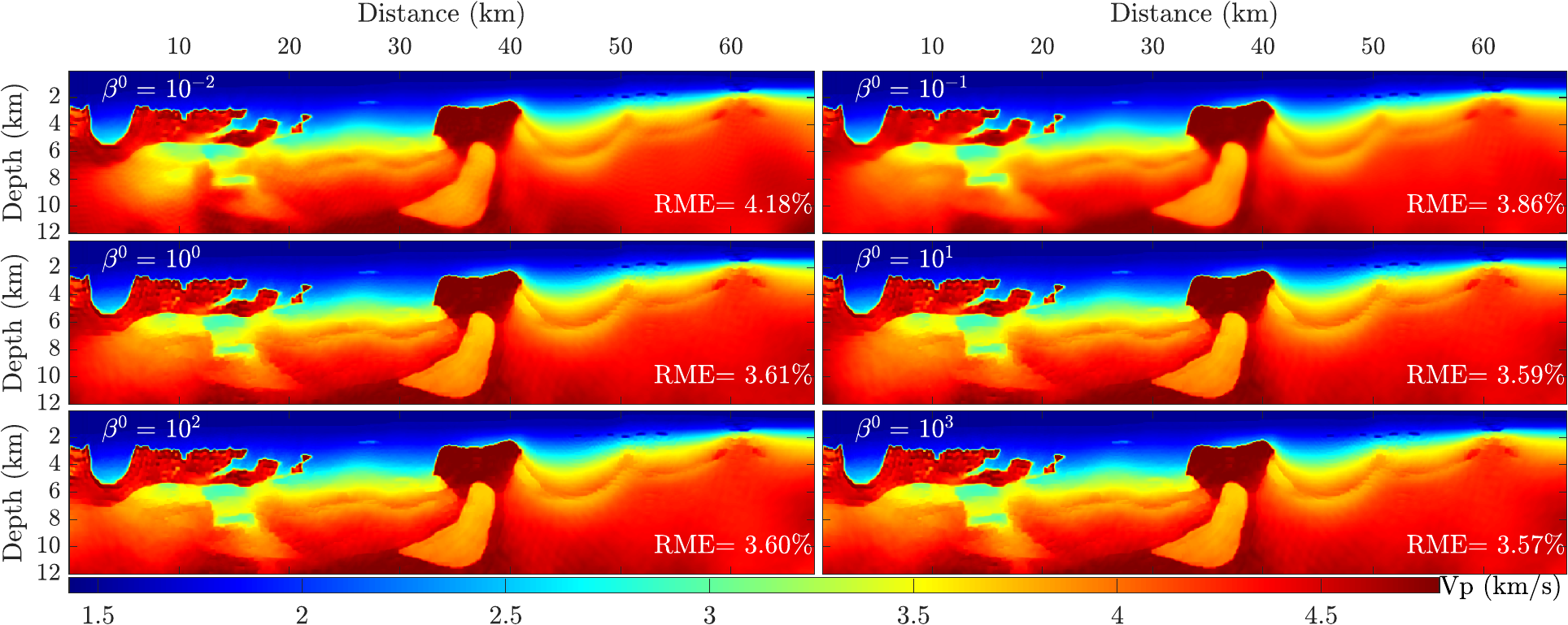}
\caption{2004 BP test. Sensitivity of adaptive TT to $\beta^0$. The inversion results obtained by adaptive TT regularization with different initialization of the  balancing parameter labeled by $\beta^{0}$. The final RME value is also specified for each plot.}
\label{fig:BP_b_var}
\end{figure}
\Cref{fig:BP_mse_beta_a,fig:BP_mse_beta_b} illustrate the evolution of $\beta$ and $\phi(\beta)$ over iterations for different initial values, shown with various colors and line styles. Regardless of the starting $\beta^{0}$, all curves converge to a similar value, though the convergence rate varies.
Model error reduction over iterations is displayed in \cref{fig:BP_mse_beta_c}, where all initial $\beta$ values lead to a comparable final error. This indicates that while the choice of $\beta^{0}$ affects convergence speed, the overall model accuracy remains largely unaffected by the initial parameter selection.
%
%
\begin{figure}[tbhp]
\centering
\includegraphics[width=.5\textwidth]{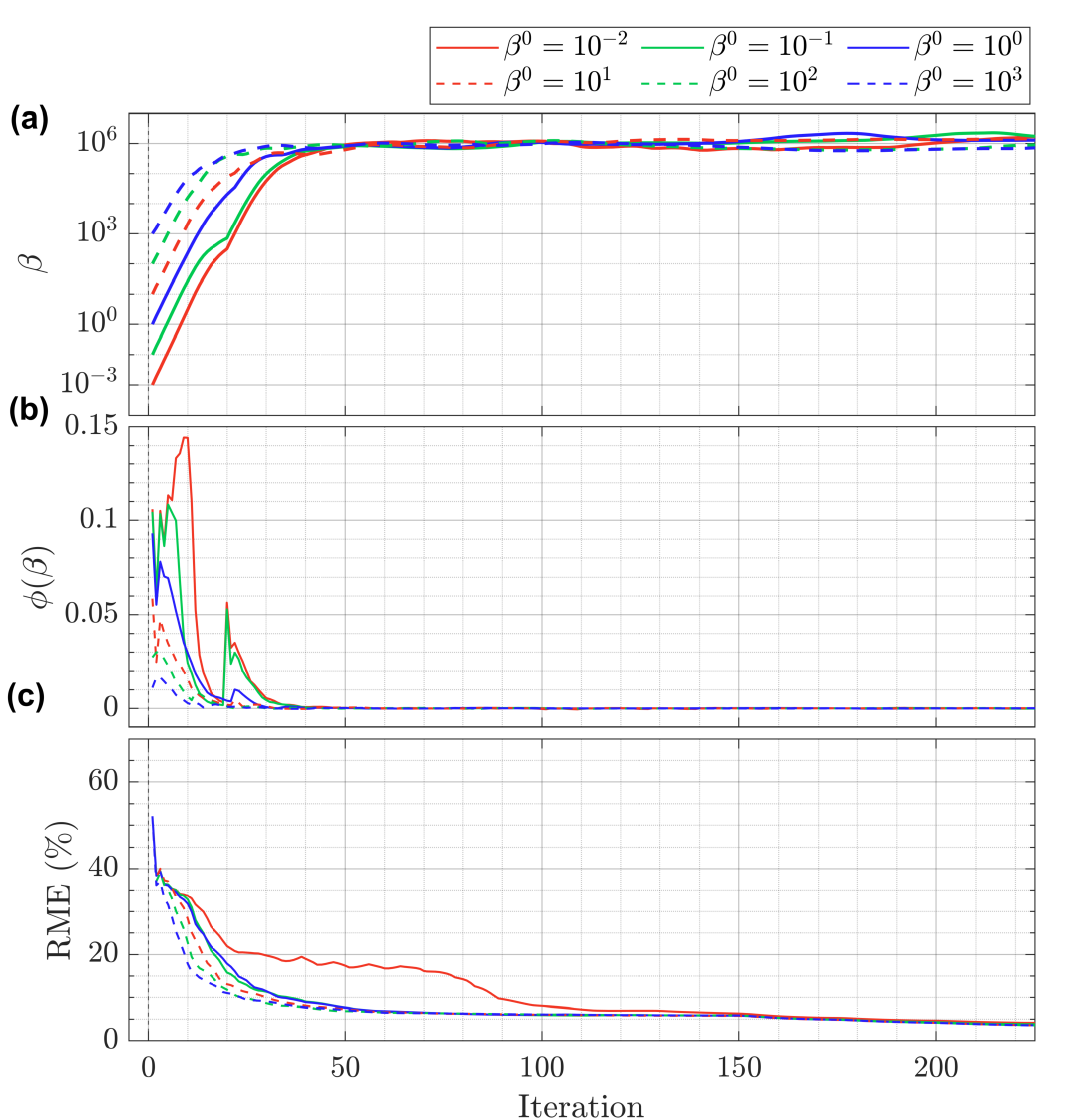}
\caption{2004 BP test. Sensitivity of adaptive TT to $\beta^0$. The variation of (a) $\beta$ and (b) $\phi(\beta)$ versus iteration for the models shown in \cref{fig:BP_b_var} for different initialization specified by $\beta^{0}$. (c) The evolution of the associated RME (\%)  versus iteration number.}
\label{fig:BP_mse_beta}
\ps{fig:BP_mse_beta_a}
\ps{fig:BP_mse_beta_b}
\ps{fig:BP_mse_beta_c}
\end{figure}

\begin{figure}[tbhp]
\centering
\includegraphics[width=1\textwidth]{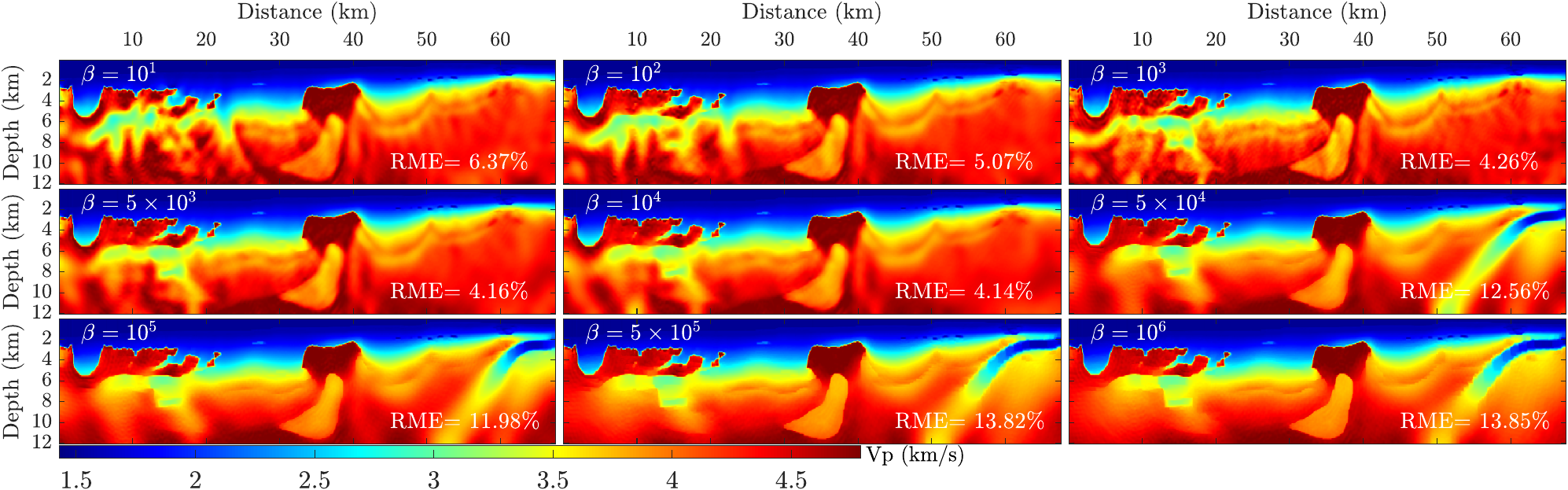}
\caption{2004 BP test. Sensitivity of non-adaptive TT to $\beta$. Inverted velocity models obtained by different $\beta$ values ranging from $10$ to $10^{6}$.}
\label{fig:BP_inv_fixed_beta}
\end{figure}

\begin{figure}[tbhp]
\centering
\includegraphics[scale=0.4]{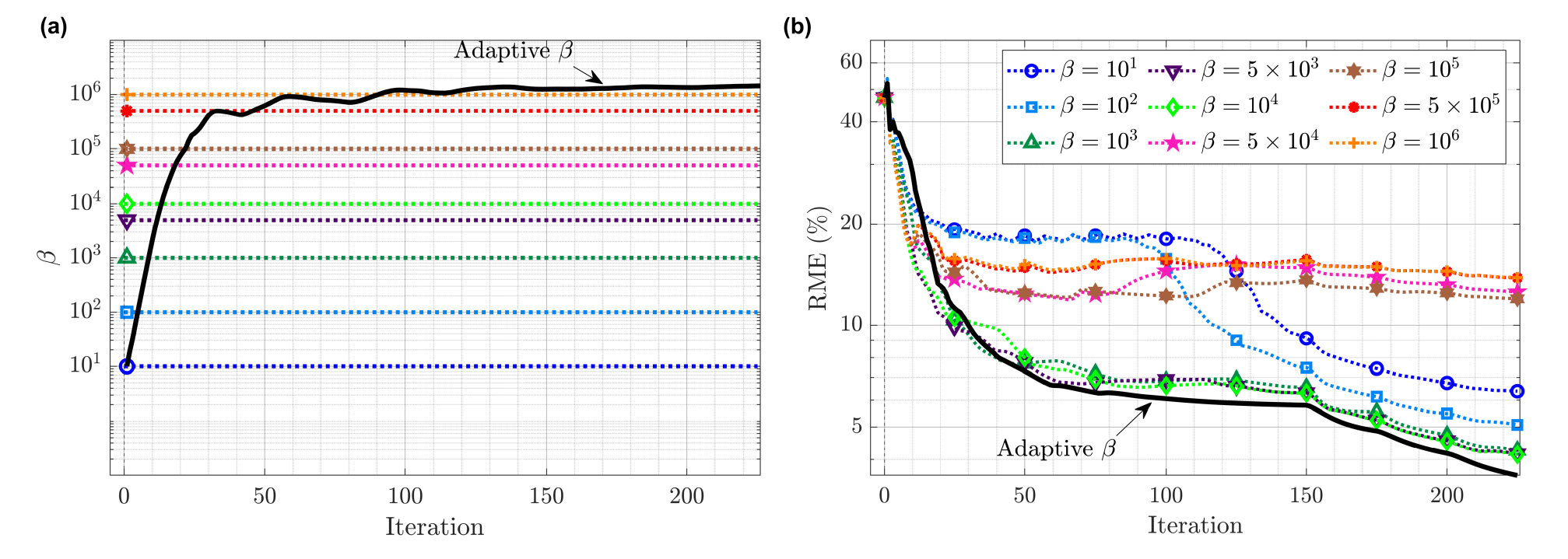}
\caption{2004 BP test. (a) Fixed $\beta$ values (colored curves) used for non-adaptive TT in \cref{fig:BP_inv_fixed_beta}, compared with the adaptive TT strategy (black curve) showing the evolution of $\beta$. (b) Corresponding RME as a function of iteration.}
\label{fig:BP_fixed_beta_mse}
\end{figure}

To further validate the superiority of the adaptive approach, we compare it with non-adaptive TT regularization using fixed $\beta$ values. \Cref{fig:BP_inv_fixed_beta} presents the inversion results for various fixed $\beta$ values ranging from $10$ to $10^6$. We can see that a small value ($\beta = 10$) lead to overly smooth reconstruction (approaching Tikhonov regularization; observed in \cref{fig:BP_inv_res_b}), while a large value ($\beta >  10^4$) behaves similarly to TV regularization as previously observed in \cref{fig:BP_inv_res_d}. Only a narrow range of intermediate values ($\beta \approx 10^3$ to $10^4$) yields acceptable results, but even then the reconstructions exhibit higher RME values compared to the adaptive approach in \cref{fig:BP_b_var}. The corresponding convergence curves are displayed in \cref{fig:BP_fixed_beta_mse}, with different colors for fixed $\beta$ and the adaptive method highlighted in black. These results demonstrate that the adaptive method consistently outperforms the non-adaptive counterpart, achieving lower final error and exhibiting robustness with respect to the initialization.
%
%
%
%

\begin{figure}[tbhp]
\centering
\includegraphics[scale=0.35]{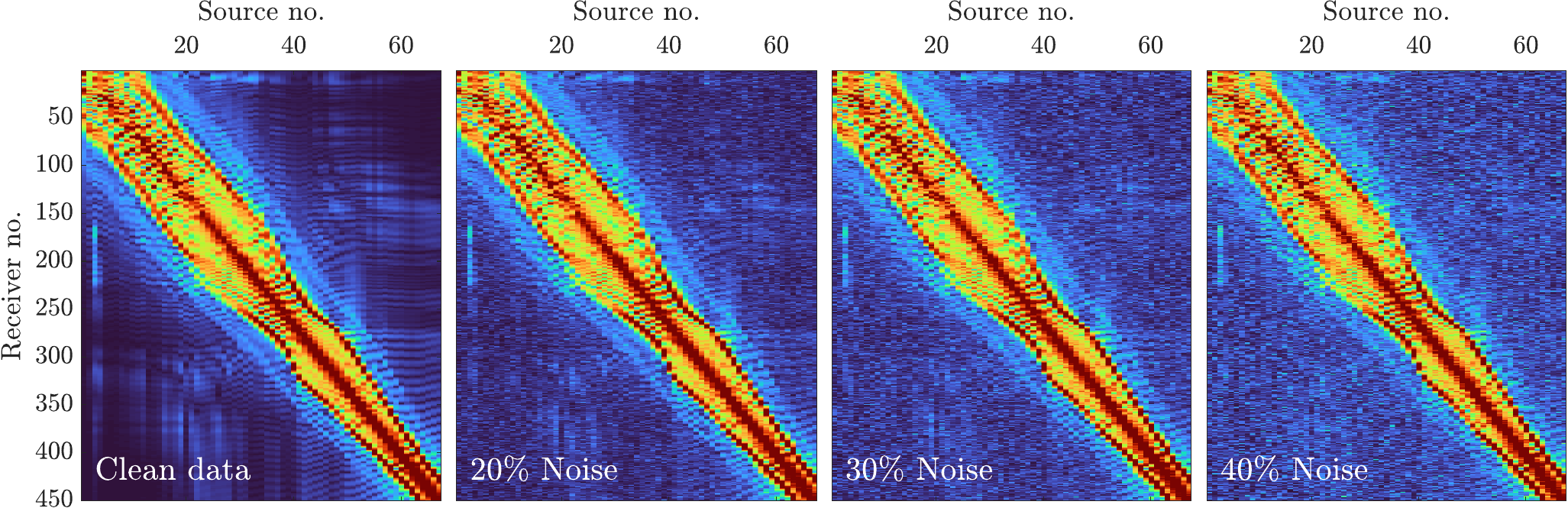}
\caption{Comparison between the monochromatic 2 Hz clean data (in source-receiver coordinates) and the data with added normally distributed Gaussian random noise, where the standard deviation of noise is 20\%, 30\%, and 40\% of the mean value of the clean data.
}
\label{fig:BP_nosie_data}
\end{figure}

\begin{figure}[tbhp]
\centering
\includegraphics[scale=0.4]{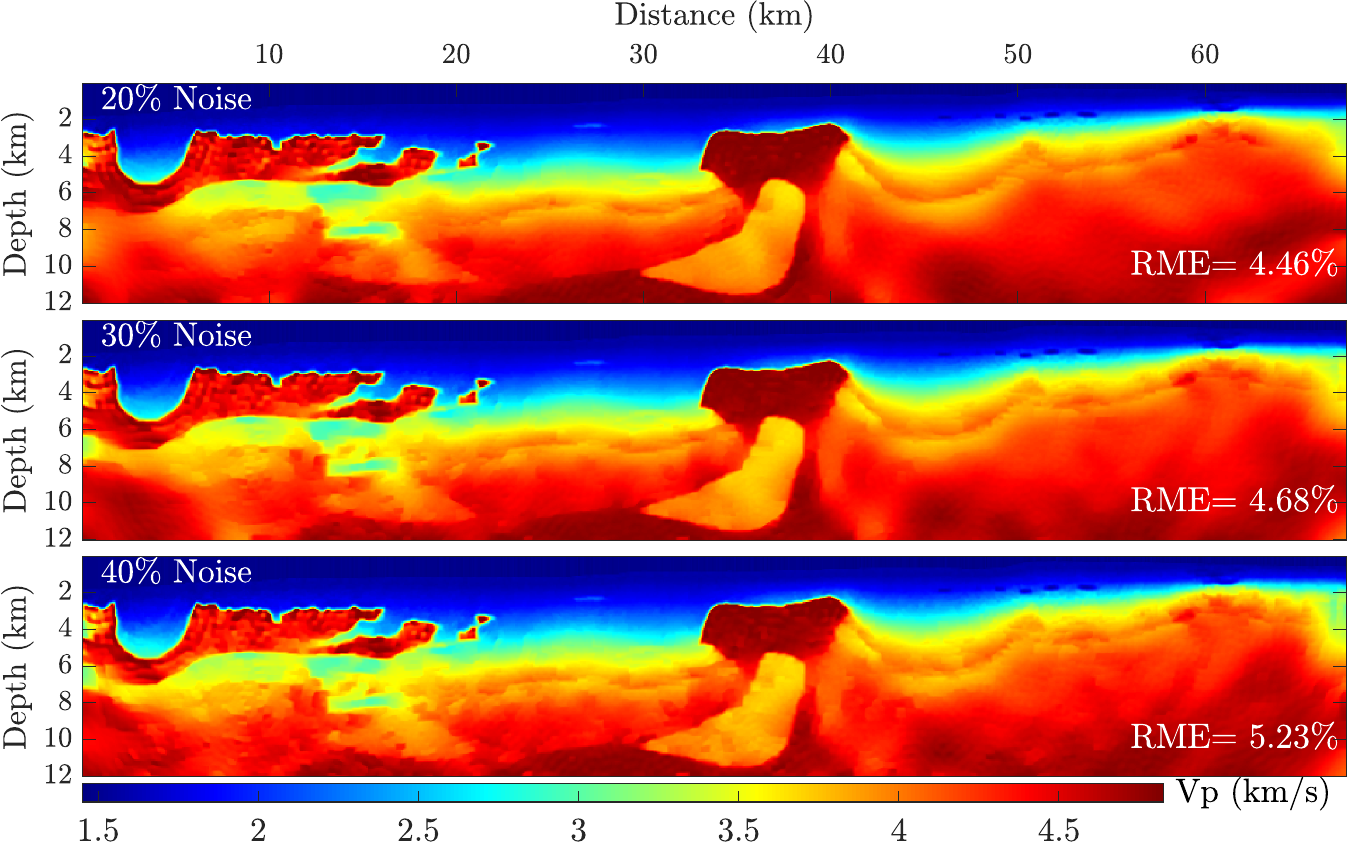}
\caption{2004 BP test. Inversion of noisy data. The final inversion results, using TT regularization, for different noise levels in the data shown in \cref{fig:BP_nosie_data}, with each row corresponding to a certain noise level. }
\label{fig:BP_inv_noise}
\end{figure}

\paragraph{Inversion of noisy data.}
To evaluate the robustness of TT regularization under noisy conditions, we added Gaussian random noise to the recorded data at levels of 20\%, 30\%, and 40\% of the mean value. A comparison between clean and noisy data at varying noise levels for monochromatic 2 Hz data in shot-receiver coordinates is shown in  \cref{fig:BP_nosie_data}. The final inversion results using TT regularization (\cref{fig:BP_inv_noise}) show robust performance across all noise levels, despite reduced accuracy with increasing noise.
\Cref{fig:BP_inv_noise_error_a} shows the energy distribution of added noise across frequencies. The evolution of RME over iterations for each noise level is illustrated in \cref{fig:BP_inv_noise_error_b}, with the noise-free case included as a reference. TT regularization maintained stable convergence, demonstrating resilience against noise. \Cref{fig:BP_inv_noise_error_c} presents the convergence of the data residual for different frequencies.
%
%
%
%
%
%
\begin{figure}[tbhp]
\centering
\includegraphics[scale=0.4]{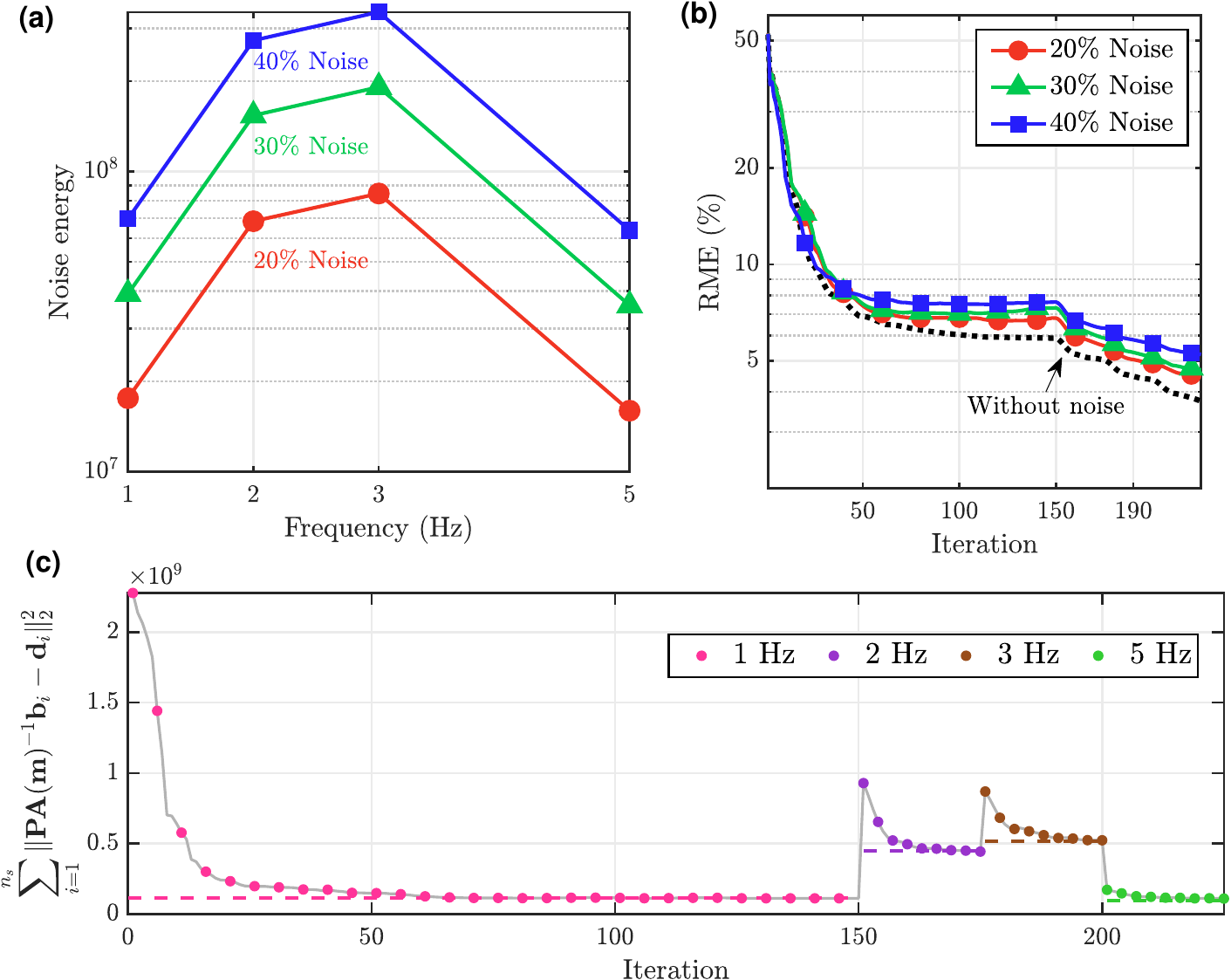}
\caption{2004 BP test. (a) Energy of noise versus frequency for varying noise levels. (b) Evolution of computed RMEs (\%) throughout iterations at varying noise levels. The error curve for the noise-free case is shown as a dotted black curve for reference. (c) Evolution of calculated data residuals with frequency and iteration number for the case of inverting 40\% added noise. Residuals for each frequency component are color-coded, with horizontal dashed lines indicating noise energy for each frequency.}
\label{fig:BP_inv_noise_error}
\ps{fig:BP_inv_noise_error_a}
\ps{fig:BP_inv_noise_error_b}
\ps{fig:BP_inv_noise_error_c}
\end{figure}

\subsubsection{Inversion of sparse data.}
This experiment evaluates TT regularization for velocity model recovery under sparse data acquisition. The inversion uses three setups with varying OBS numbers $n_s=[25, 15, 10]$, corresponding to OBS intervals $\Delta \text{S} \!\approx[2.8, 4.8, 7.5]$ km.
\Cref{fig:BP_sparse_res_a} shows inverted velocity models with increasing sparsity from top to bottom. As the source interval widens, the inversion struggles to recover fine details, particularly in complex regions. \Cref{fig:BP_sparse_res_b} illustrates model error reduction over iterations for each setup. Denser acquisition ($\Delta \text{S}\! \approx \! 2.8$ km) achieves lower final model error and faster convergence. Sparser setups ($\Delta \text{S}\! \approx \! 7.5$ km) still benefit from TT regularization but show slower convergence and higher final error. Interestingly, for a 4.8 km OBS spacing, TT regularization still yields an accurate subsurface estimate. The dotted black line, representing a denser reference case ($\Delta \text{S} \approx 1$ km), emphasizes the trade-off between acquisition density and model accuracy.
%
%
\begin{figure}[tbhp]
\centering
\includegraphics[scale=0.45]{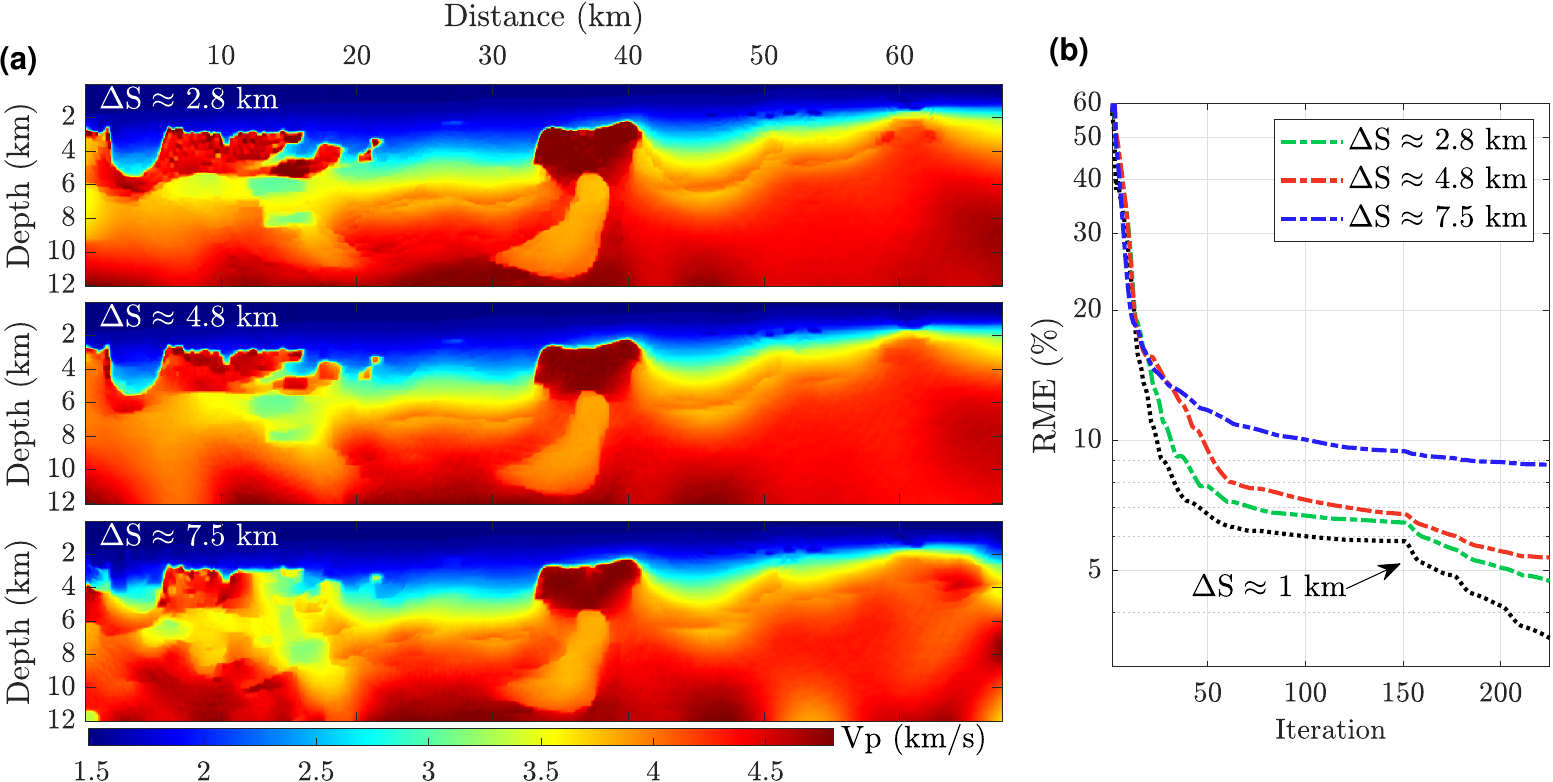}
\caption{2004 BP test. Inversion results for the sparse acquisition test, comparing TT regularization performance under different data sparsity levels. (a) Rows represent varying OBS intervals of $\Delta \text{S}\approx[2.8, 4.8, 7.5]$ km, showing progressively sparser acquisition setups. (b) Evolution of RME (\%) over iterations for each acquisition density, where
the dotted black line represents RME for $\Delta \text{S} \approx 1$ km as a reference.}
\label{fig:BP_sparse_res}
\ps{fig:BP_sparse_res_a}
\ps{fig:BP_sparse_res_b}
\end{figure}
\subsection{Elastic examples.}  
The proposed TT regularization is assessed using two synthetic examples. The associated user-defined parameters are reported in \cref{tab:par_EL}. 

\begin{table}[tbhp]
\caption{Specified values for the free parameters used for the elastic FWI examples.}
\label{tab:par_EL}
\centering
\begin{tabular}{l c c c c c c c}
\toprule 
& \multicolumn{7}{c}{Free parameters} \\
\cmidrule(l){2-8}
Experiment & $\mu$ & $c_{1,p}$ & $c_{1,s}$ & $c_{2,p}$ & $c_{2,s}$ & $c_{3,p}$ & $c_{3,s}$ \\ 
\midrule 
Inclusion (\cref{sec:EL_sec_Inclusion}) & $10^3$ & $0.01$ & $0.05$ & $10^{-3}$ & $10^{-3}$ & $0.2$ & $0.2$ \\ 
Overthrust (\cref{sec:EL_sec_overthrust}) & $10^6$ & $0.015$ & $0.15$ & $10^{-3}$ & $10^{-3}$ & $0.3$ & $0.3$ \\ 
\bottomrule 
\end{tabular}
\end{table}
\subsubsection{Example 1: Inclusion model.}\label{sec:EL_sec_Inclusion}
\begin{figure}[tbhp]
\centering
\includegraphics[scale=0.45]{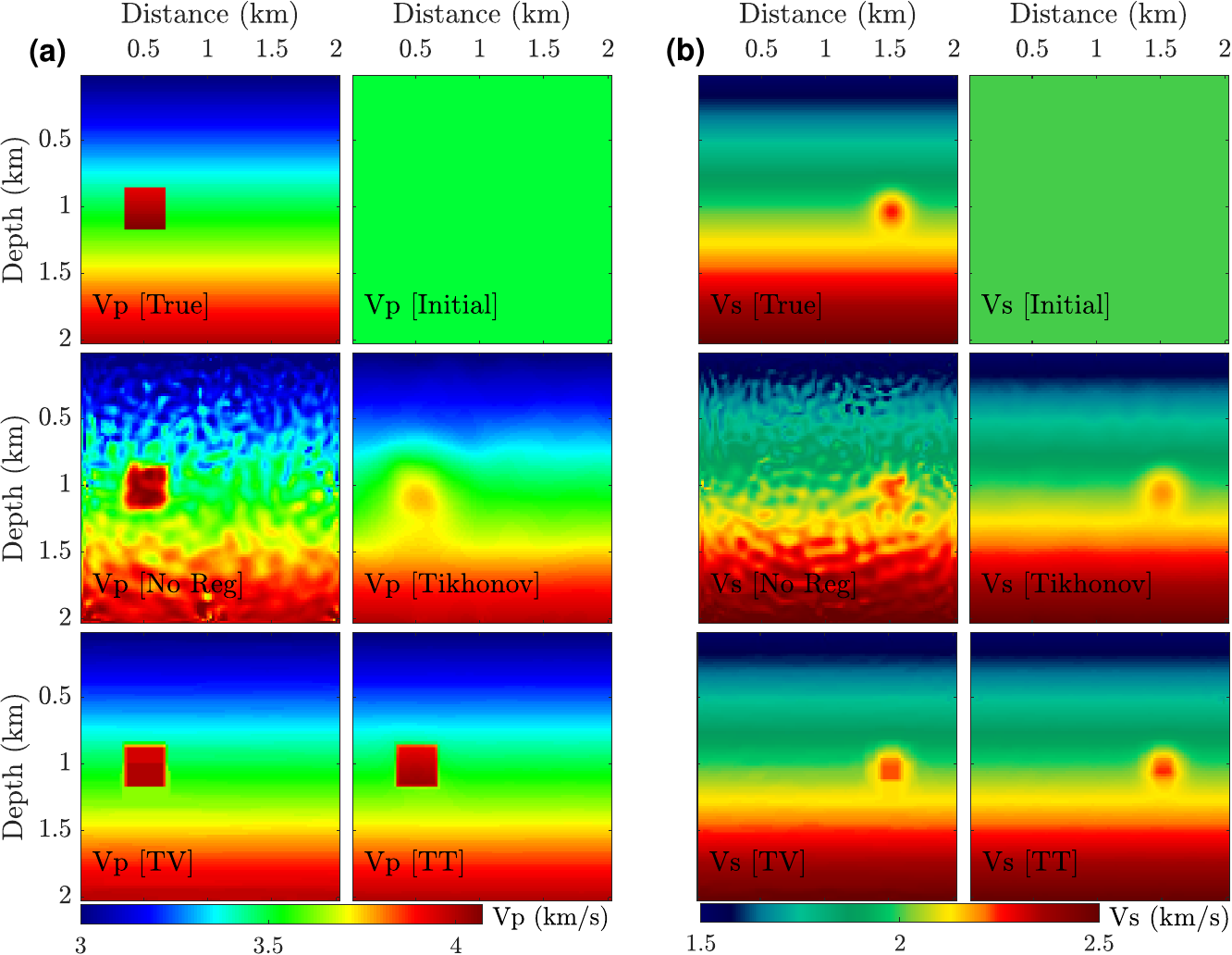}
\caption{Elastic FWI (Inclusion test). Analysis of FWI without regularization (No Reg) and the performance of three regularization techniques-Tikhonov, TV, and TT-on the recovery of the  $\text{V}_\text{P}$ and $\text{V}_\text{S}$ models. (a)  $\text{V}_\text{P}$ model, and (b)  $\text{V}_\text{S}$ model. Each panel consists of six subplots, organized in pairs by regularization type. The first row in each panel shows the True and Initial velocity models. The remaining panels illustrate recovered velocity models without regularization (No Reg), and with Tikhonov, TV, and TT regularizations.}
\label{fig:EL_inclusion}
\ps{fig:EL_inclusion_a}
\ps{fig:EL_inclusion_b}
\end{figure}

To evaluate the performance of different regularization techniques on reconstruction of the $\text{V}_\text{P}$ and $\text{V}_\text{S}$ models in elastic FWI, a toy experiment using an inclusion model was conducted (\cref{fig:EL_inclusion}). \Cref{fig:EL_inclusion_a,fig:EL_inclusion_b} show the true $\text{V}_\text{P}$ and $\text{V}_\text{S}$ models, featuring a blocky anomaly in $\text{V}_\text{P}$ and a smooth Gaussian-shaped anomaly in $\text{V}_\text{S}$ within a linearly increasing depth background.
Forward modeling uses 8 two-component sources (Ricker wavelet, 10 Hz) and 200 two-component receivers evenly distributed along all edges. The joint inversion of 5 and 7 Hz frequencies starts with homogeneous models (3.5 km/s for $\text{V}_\text{P}$ and 2 km/s for $\text{V}_\text{S}$), with initial balancing parameters $\beta_p^{0}=\beta_s^{0}=10^2$.
When regularization is not implemented, the reconstructed models are noisy with high-frequency artifacts and poorly defined boundaries, especially around inclusions. Tikhonov regularization improves smoothness but blurs boundaries, particularly for the blocky anomaly in $\text{V}_\text{P}$.
TV regularization, as expected, effectively captures the blocky inclusion in $\text{V}_\text{P}$ but lacking smoothness in the $\text{V}_\text{S}$ anomaly. TT regularization, however, achieves a balanced result, preserving sharp boundaries while maintaining smooth transitions, offering a hybrid approach that accurately represents both sharp and smooth features.
%
%
\Cref{fig:EL_inclusion_MSE_a,fig:EL_inclusion_MSE_b}  show vertical profiles of $\text{V}_\text{P}$ and $\text{V}_\text{S}$ at specific lateral positions ($X=0.5$ km for $\text{V}_\text{P}$ and $X=1.48$ km for $\text{V}_\text{S}$). TT regularization closely matches the true model, capturing sharp $\text{V}_\text{P}$ boundaries and maintaining smoothness in $\text{V}_\text{S}$. TV regularization achieves sharp transitions but introduces minor oscillations. Tikhonov regularization over-smooths profiles, resulting in less accuracy.
The evolution of the RME curves for $\text{V}_\text{P}$ and $\text{V}_\text{S}$ is shown in \Cref{fig:EL_inclusion_MSE_c,fig:EL_inclusion_MSE_d}, respectively. TT regularization achieves the fastest and most significant error reduction. TV regularization converges more slowly. Tikhonov regularization shows slower error reduction, oscillation in convergence, and higher final error, highlighting its limitations.
\Cref{fig:EL_inclusion_MSE_e} shows the balancing parameter $\beta$ for TT regularization, where $\beta_p$ (brown) controls the influence of TV and Tikhonov regularization in $\text{V}_\text{P}$ and $\beta_s$ (dark purple) does the same for $\text{V}_\text{S}$. The balancing parameter is adjusted based on structural properties, leading to distinct optimal values. $\text{V}_\text{P}$ exhibits a blocky anomaly, while $\text{V}_\text{S}$ contains a smooth anomaly, resulting in $\beta_p^{opt}>\beta_s^{opt}$.
\Cref{fig:EL_inclusion_MSE_f} shows the behavior of function $\phi(\beta)$ for each model, providing insights into adaptive regularization dynamics.

To compare adaptive TT regularization with its non-adaptive counterpart, we applied a common heuristic for selecting optimal regularization parameters in the non-adaptive case. Specifically, we sampled the $(\beta_p,\beta_s)$ plane on a $15 \times 15$ logarithmically spaced grid within the range $[10^{-3},10^5]^2$. \Cref{fig:EL_beta_anal_a,fig:EL_beta_anal_d} show the RME surfaces for $\text{V}_\text{P}$ and $\text{V}_\text{S}$, respectively. The minima of these surfaces, marked by yellow stars, occur at different $(\beta_p,\beta_s)$ values. The corresponding reconstructed velocity models are displayed in the right-hand panels.
Three key observations can be made: i) The optimal values of $\beta_p$ and $\beta_s$ are generally not equal.
ii) The minima of the RME surfaces for $\mathbf{m}_\text{p}$ and $\mathbf{m}_\text{s}$ occur at different points.
iii) Both surfaces exhibit strong irregularity.
Thus, while careful manual tuning of $(\beta_p,\beta_s)$ can yield satisfactory results for non-adaptive TT, this strategy is inherently limited in practice. These findings motivate the need for adaptive regularization that dynamically adjust $\beta_p$ and $\beta_s$ during inversion.
To demonstrate this, we implemented the adaptive TT regularization scheme, initializing $(\beta_p^0,\beta_s^0)$ with the same values used in the non-adaptive case. \Cref{fig:MSE_EL_beta_anal_a,fig:MSE_EL_beta_anal_b}  show the RME evolution of the $\mathbf{m}_p$ and $\mathbf{m}_s$ models for the non-adaptive TT, while \Cref{fig:MSE_EL_beta_anal_c,fig:MSE_EL_beta_anal_d}  present the corresponding results for the adaptive TT. Panels (e–f) display the evolution of $\beta_p$ and $\beta_s$ during adaptive inversion. Compared to the non-adaptive approach, the adaptive scheme yields RME curves with significantly lower variance and consistently achieves lower error bounds, regardless of the initialization—results that align with the observations from the acoustic FWI experiment.
%
\begin{figure}[!ht]
\centering
\includegraphics[scale=0.5]{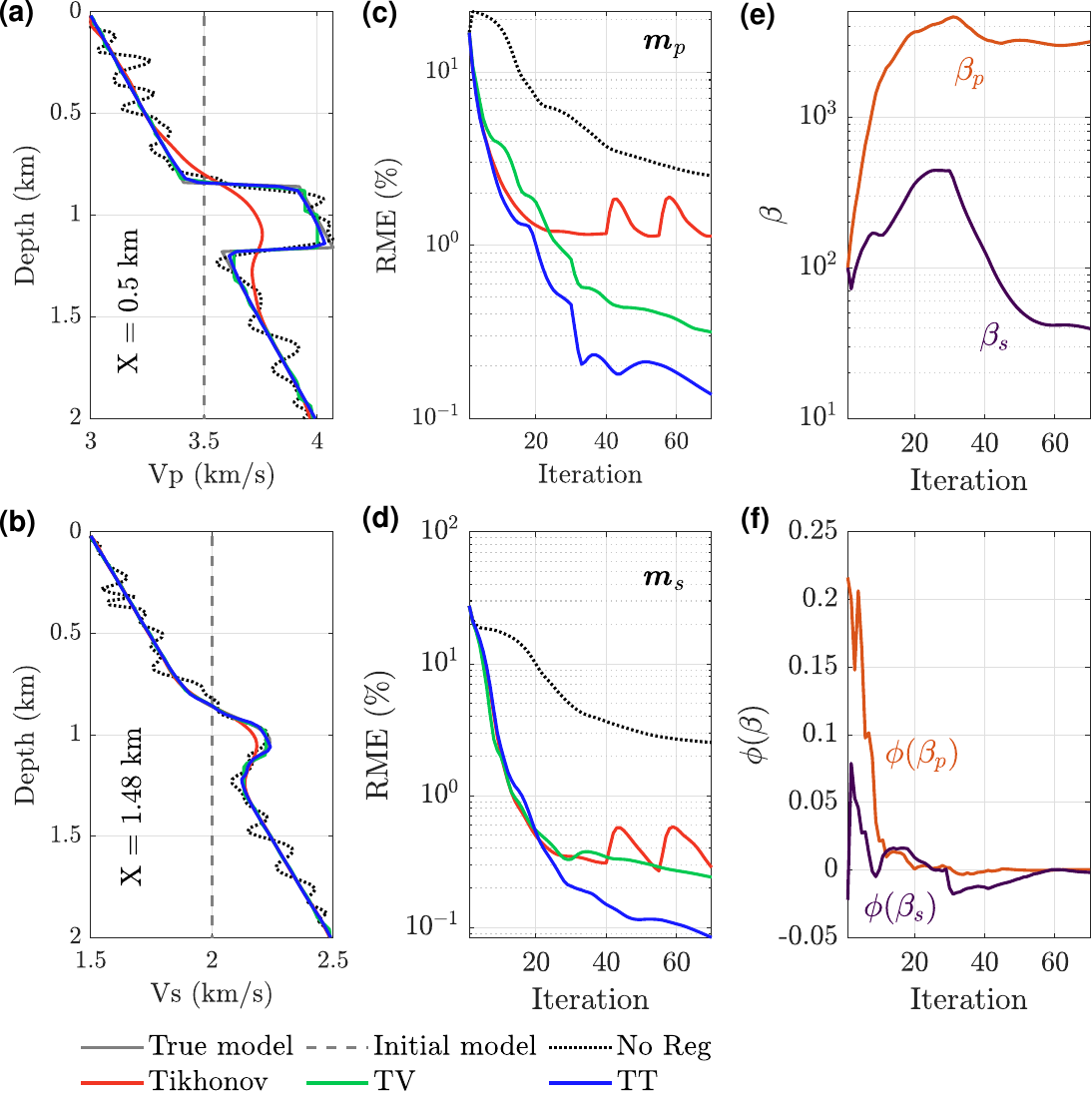}
\caption{Elastic FWI (Inclusion test). Detailed comparison of three regularization methods for elastic FWI based on $\text{V}_\text{P}$ and $\text{V}_\text{S}$ models. Comparison of vertical profiles of $\text{V}_\text{P}$ (a) and $\text{V}_\text{S}$ (b) at two locations ($X = 0.5~km$ for $\text{V}_\text{P}$ and $X = 1.48~km$ for $\text{V}_\text{S}$). RME (\%) convergence over iterations for $\text{V}_\text{P}$ (c) and $\text{V}_\text{S}$ (d). (e-f) The evolution of the balancing parameter 
$\beta$ and variation of function $\phi(\beta)$ for $\text{V}_\text{P}$ and $\text{V}_\text{S}$ models in the TT regularization method.}
\label{fig:EL_inclusion_MSE}
\ps{fig:EL_inclusion_MSE_a}
\ps{fig:EL_inclusion_MSE_b}
\ps{fig:EL_inclusion_MSE_c}
\ps{fig:EL_inclusion_MSE_d}
\ps{fig:EL_inclusion_MSE_e}
\ps{fig:EL_inclusion_MSE_f}
\end{figure}
\begin{figure}[tbhp]
\centering
\includegraphics[scale=0.5]{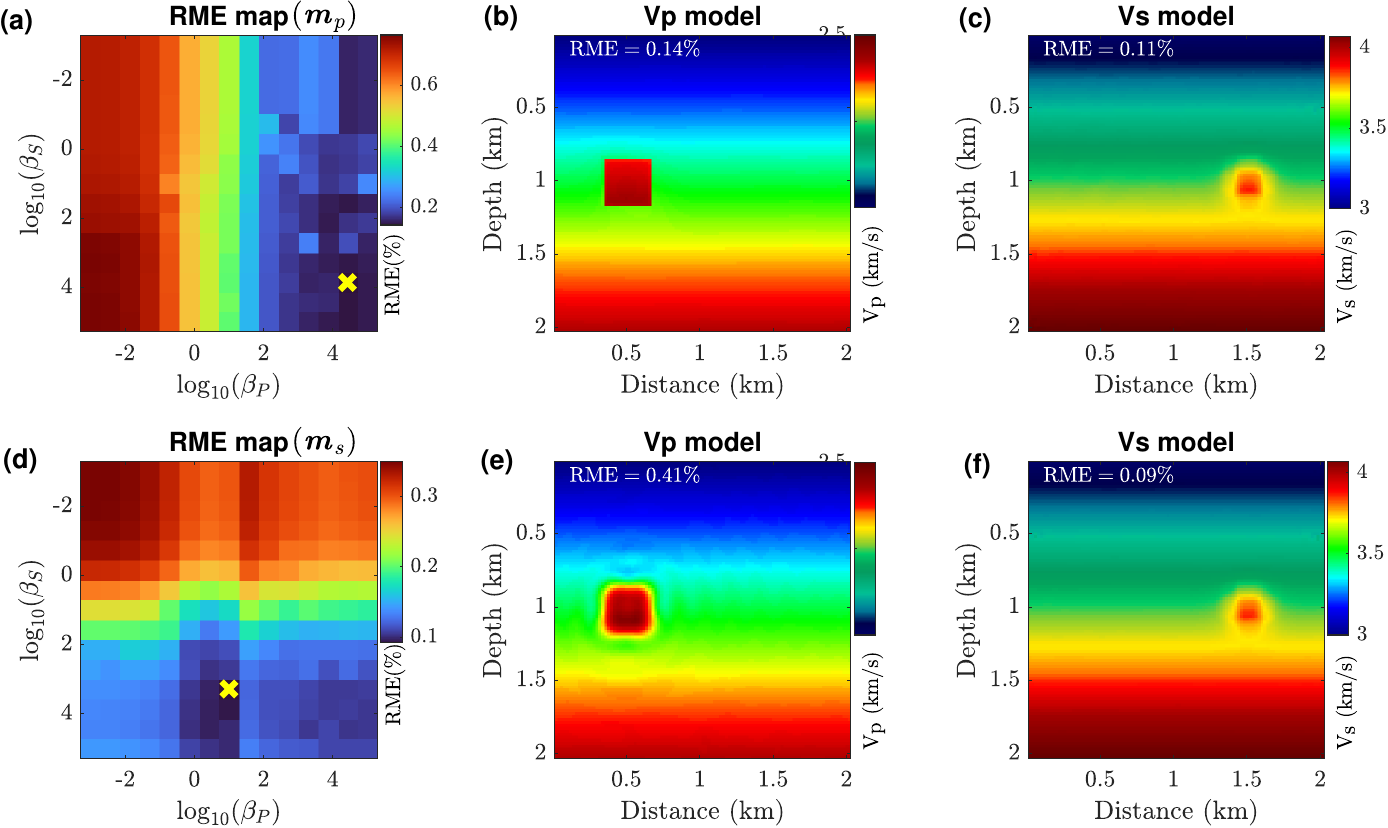}
\caption{Elastic FWI (Inclusion test). Non-adaptive TT regularization for different $(\beta_p,\beta_s)$. (a) RME surface for $\mathbf{m}_\text{p}$. (b) $\text{V}_\text{P}$ and (c) $\text{V}_\text{S}$ models corresponding to the minimum in (a) (yellow star). (d) RME surface for $\mathbf{m}_\text{s}$. (e) $\text{V}_\text{P}$ and (f) $\text{V}_\text{S}$ models corresponding to the minimum in (d) (yellow star).}
\label{fig:EL_beta_anal}
\ps{fig:EL_beta_anal_a}
\ps{fig:EL_beta_anal_b}
\ps{fig:EL_beta_anal_c}
\ps{fig:EL_beta_anal_d}
\ps{fig:EL_beta_anal_e}
\ps{fig:EL_beta_anal_f}
\end{figure}
%
%
\begin{figure}[tbhp]
\centering
\includegraphics[scale=0.4]{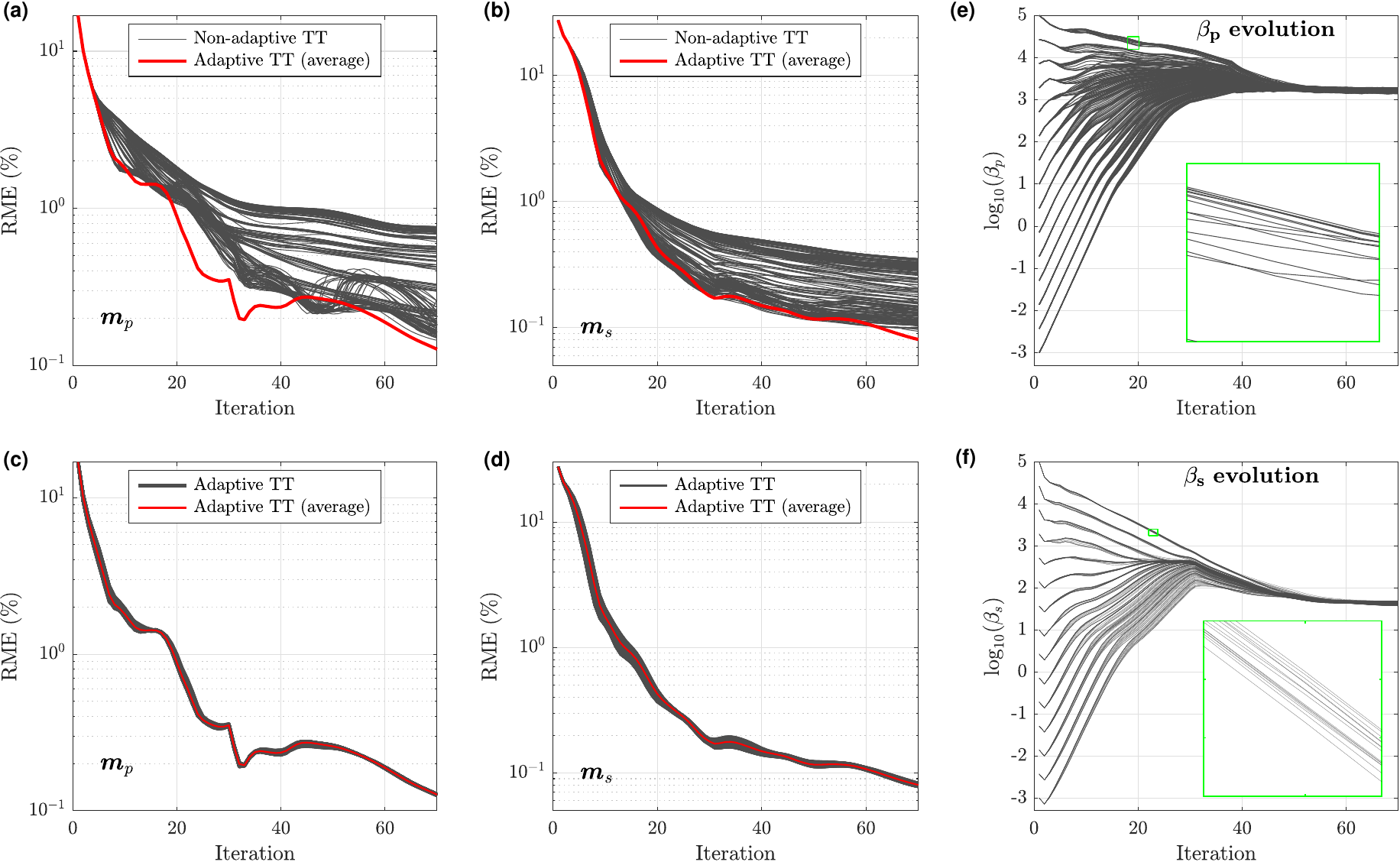}
\caption{Elastic FWI (Inclusion test). Convergence comparison of non-adaptive and adaptive TT regularizations. (a, b) RME evolution curves for $\m_\text{p}$ and $\m_\text{s}$ obtained by non-adaptive approach, where each curve is related to a fixed ($\beta_p$, $\beta_s$). (c,d) RME convergence of adaptive TT initialized with the fixed values used in non-adaptive case. (e,f) Evolution of $\beta_p$ and $\beta_s$ during adaptive inversion.}
\label{fig:MSE_EL_beta_anal}
\ps{fig:MSE_EL_beta_anal_a}
\ps{fig:MSE_EL_beta_anal_b}
\ps{fig:MSE_EL_beta_anal_c}
\ps{fig:MSE_EL_beta_anal_d}
\ps{fig:MSE_EL_beta_anal_e}
\ps{fig:MSE_EL_beta_anal_f}
\end{figure}

\subsubsection{Example 2: SEG/EAGE overthrust model.}\label{sec:EL_sec_overthrust}
As the final example, the analysis of TT regularization is extended using realistic surface recorded data from a 2D section of the 3D SEG/EAGE overthrust $\text{V}_\text{P}$ model (\cref{fig:EL_over_true_a}). The model dimensions are 4.67 km $\times$ 20 km with a grid interval of 25 m. The $\text{V}_\text{S}$ model is inferred from the $\text{V}_\text{P}$ model using an empirical relation \cite{brocher2005empirical}: 
\begin{equation}
\begin{aligned}
\text{V}_\text{S}=&0.7858-1.2344\text{V}_\text{P}+0.7949\text{V}_\text{P}^2 
-0.1238\text{V}_\text{P}^3+0.0044\text{V}_\text{P}^4.
\end{aligned}
\end{equation}
This relationship results in higher Poisson's ratio near the surface, necessitating dense spatial sampling. To avoid this issue, we set the minimum $\text{V}_\text{S}$ to 1.4 km/s (\cref{fig:EL_over_true_b}). The Poisson's ratio field is computed as: 
\begin{equation}
    \sigma = \frac{\text{V}_\text{P}^2-2\text{V}_\text{S}^2}{2(\text{V}_\text{P}^2-\text{V}_\text{S}^2)},
\end{equation}
and shown in \cref{fig:EL_over_true_c}.
The surface acquisition configuration involves 67 sources at 300 m intervals using 8 Hz Ricker wavelets ($\bold{b}_x$ and $\bold{b}_z$). Receivers consist of 401 two-component sensors at 50 m intervals. The inversion starts with models increasing linearly with depth: $\text{V}_\text{P}$ ranges from 2.7 km/s to 6.5 km/s (\cref{fig:EL_over_true_d}), and $\text{V}_\text{S}$ ranges from 1.2 km/s to 3.8 km/s (\cref{fig:EL_over_true_e}). The initial Poisson's ratio is shown in \cref{fig:EL_over_true_f}.
%
%
\begin{figure}[tbhp]
\centering
\includegraphics[width=1\textwidth]{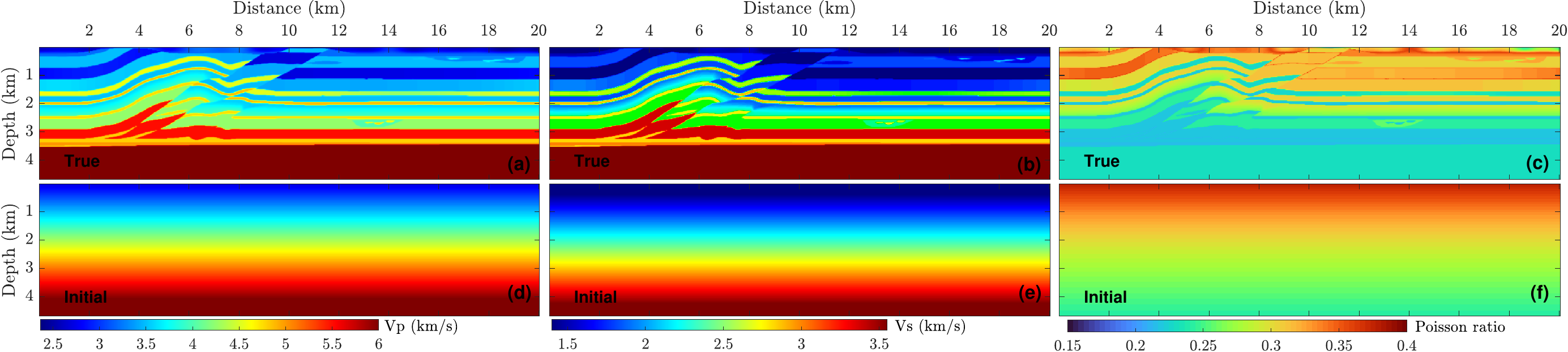}
\caption{Elastic FWI (Overthrust model). (a) True $\text{V}_{\text{P}}$ model, (b) True  $\text{V}_{\text{S}}$ model, (c) Computed Poisson's ratio based on (a,b). (d-f) Corresponding initial $\text{V}_{\text{P}}$, $\text{V}_{\text{S}}$, and Poisson's ratio.}
\label{fig:EL_over_true}
\ps{fig:EL_over_true_a}
\ps{fig:EL_over_true_b}
\ps{fig:EL_over_true_c}
\ps{fig:EL_over_true_d}
\ps{fig:EL_over_true_e}
\ps{fig:EL_over_true_f}

\end{figure}
The inversion process follows a multiscale strategy \cite{Bunks_1995_MSW} with frequencies set between 3 Hz and 12 Hz, in 0.5 Hz increments. The cycles are:  3~Hz to 6~Hz,  3~Hz to 7.5~Hz, and 3~Hz to 12~Hz. Furthermore, for TT regularization $\beta_\text{p}=\beta_\text{s}=10^2$.
Constructing the long wavelength for the models via the inversion of the minimum available frequency is crucial. The initial frequency inversion is executed over 20 iterations, and inversion results are presented \cref{fig:EL_over_F3_a} for the case without regularization, and in \cref{fig:EL_over_F3_b,fig:EL_over_F3_c,fig:EL_over_F3_d} for reconstructions incorporating Tikhonov, TV, and TT regularization, respectively. The results demonstrate that without regularization (\cref{fig:EL_over_F3_a}), the recovered $\text{V}_{\text{P}}$, $\text{V}_{\text{S}}$, and Poisson’s ratio are dominated by high-wavenumber artifacts. Tikhonov and TV regularization (\cref{fig:EL_over_F3_b,fig:EL_over_F3_c}) improve the recovery of low-wavenumber background structures. TT regularization (\cref{fig:EL_over_F3_d}) further enhances the large-scale smooth components, mitigating high-frequency influence and leading to a coherent long-wavelength model, crucial for accurate subsurface imaging.
%
%
\begin{figure}[tbhp]
\centering
\includegraphics[width=1\textwidth]{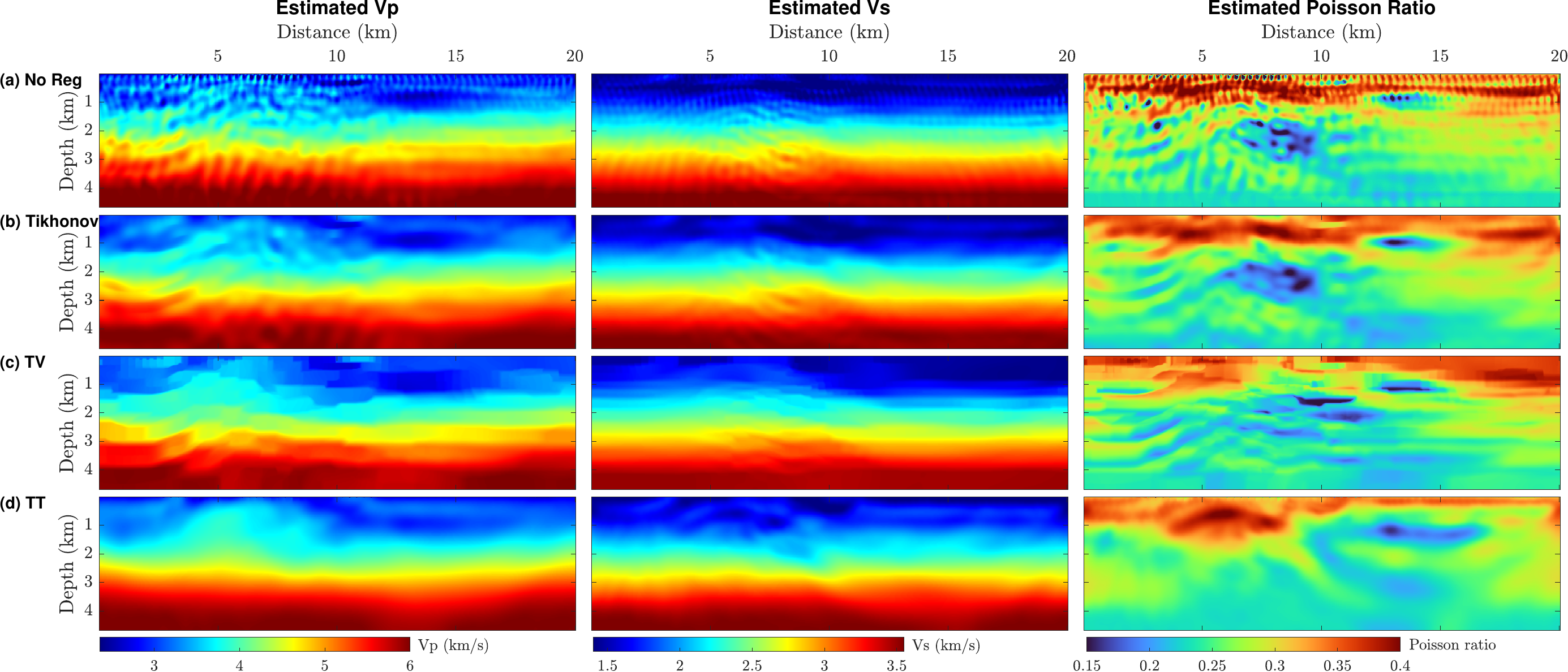}
\caption{Elastic FWI (Overthrust model). Recovered model after inverting data frequency at 3 Hz. The first column shows the recovered \(\text{V}_{\text{P}}\) model, the second column displays the \(\text{V}_{\text{S}}\) model, and the third column presents the computed Poisson's ratio. Panel (a) corresponds to the case without regularization, while panels (b), (c), and (d) show results obtained with Tikhonov, TV, and TT regularization, respectively.}
\label{fig:EL_over_F3}
\ps{fig:EL_over_F3_a}
\ps{fig:EL_over_F3_b}
\ps{fig:EL_over_F3_c}
\ps{fig:EL_over_F3_d}
\ps{fig:EL_over_F3_e}
\ps{fig:EL_over_F3_f}

\end{figure}

\begin{figure}[tbhp]
\centering
\includegraphics[width=1\textwidth]{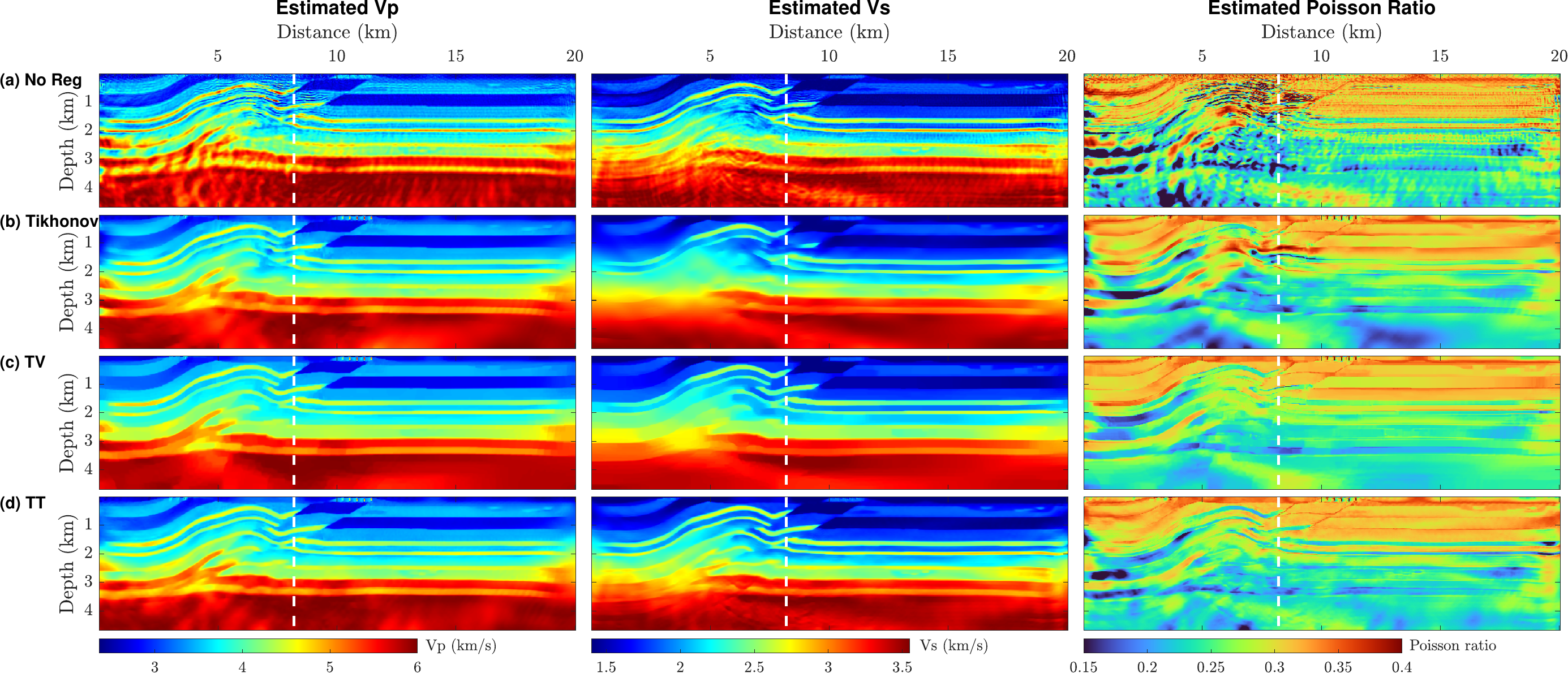}
\caption{Elastic FWI (Overthrust model). Same as \cref{fig:EL_over_F3} at the final frequency data.}
\label{fig:EL_over_final}
\end{figure}

\begin{figure}[tbhp]
\centering
\includegraphics[scale=0.4]{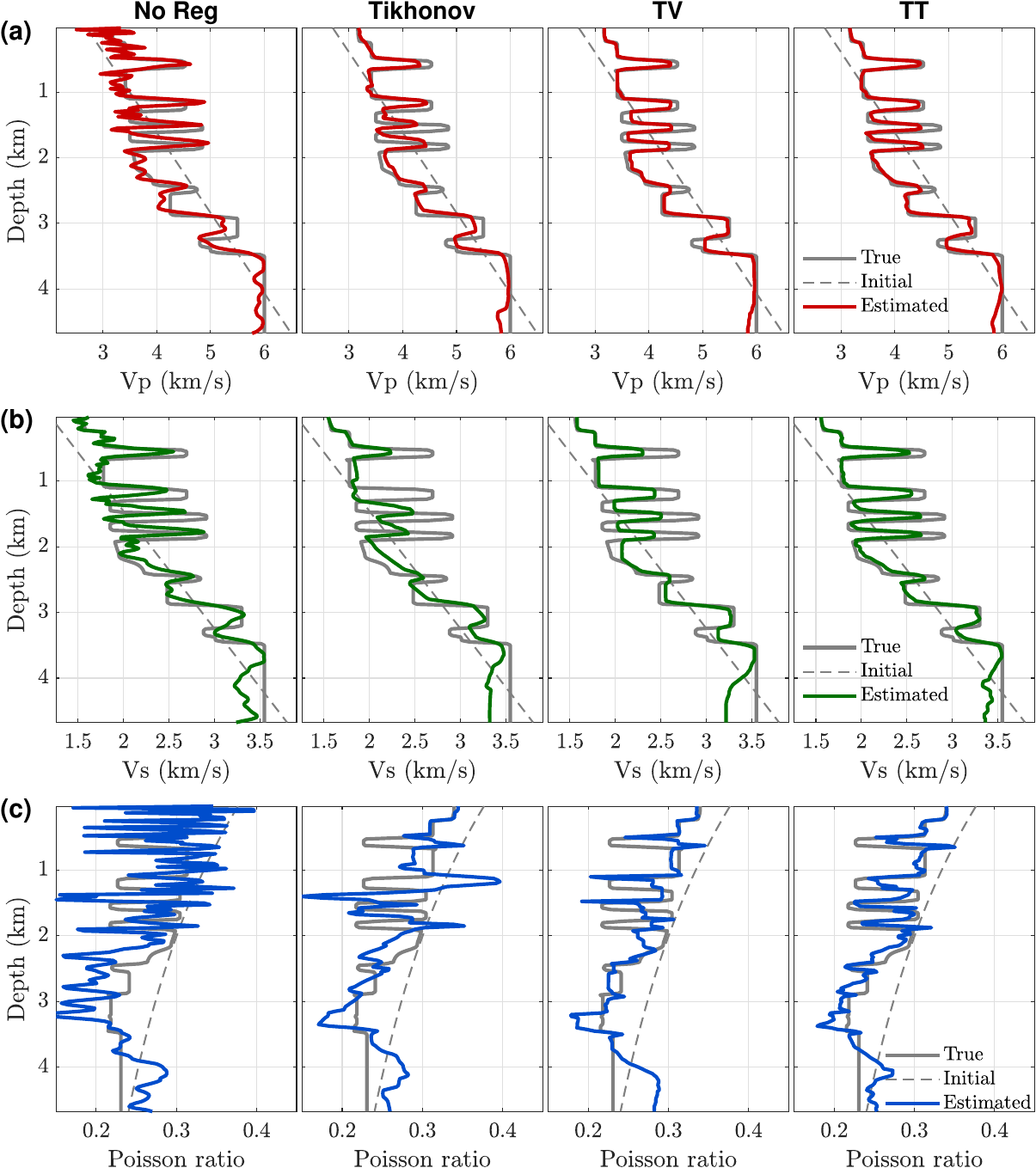}
\caption{Elastic FWI (Overthrust model). Vertical log comparison at distance 8.2 km (indicated by the dashed white line in \cref{fig:EL_over_final}). Each row shows depth profiles for (a) $\text{V}_{\text{P}}$, (b) $\text{V}_{\text{S}}$, and (c) Poisson's ratio.}
\label{fig:EL_over_final_logs}
\end{figure}

Following the initial frequency inversion, the higher frequencies are progressively updated, with each frequency processed over 10 iterations. The final inversion results are illustrated in \cref{fig:EL_over_final}. The effectiveness of TT regularization is evident in the reconstructed models, which is supported by vertical log comparison in \cref{fig:EL_over_final_logs} and the evolution of model error shown in \cref{fig:EL_over_error_a} for $\m_\text{p}$ model and \cref{fig:EL_over_error_b} for $\m_\text{s}$  model. These detailed comparisons demonstrate that TT regularization produces reconstructions that better match the true models across all depth levels, while maintaining faster convergence behavior throughout the inversion process. \Cref{fig:EL_over_error_c,fig:EL_over_error_d} correspond to the update of [$\beta_\text{p},\beta_\text{s}$] and [$\phi_\text{p},\phi_\text{s}$], respectively. 

The computational performance analysis reported in \cref{tab:over_runtime} reveals that all regularization methods exhibit comparable efficiency, with regularization steps requiring between 1.73-1.92 seconds per iteration. While TT-based approaches show slightly higher computational costs (1.86-1.92 sec) compared to Tikhonov (1.83 sec) and TV (1.73 sec), the overhead remains minimal at less than 5\% of the total one FWI iteration time (including regularization).
%
%
%
%
%
%
\begin{figure}[tbhp]
\centering
\includegraphics[width=0.5\textwidth]{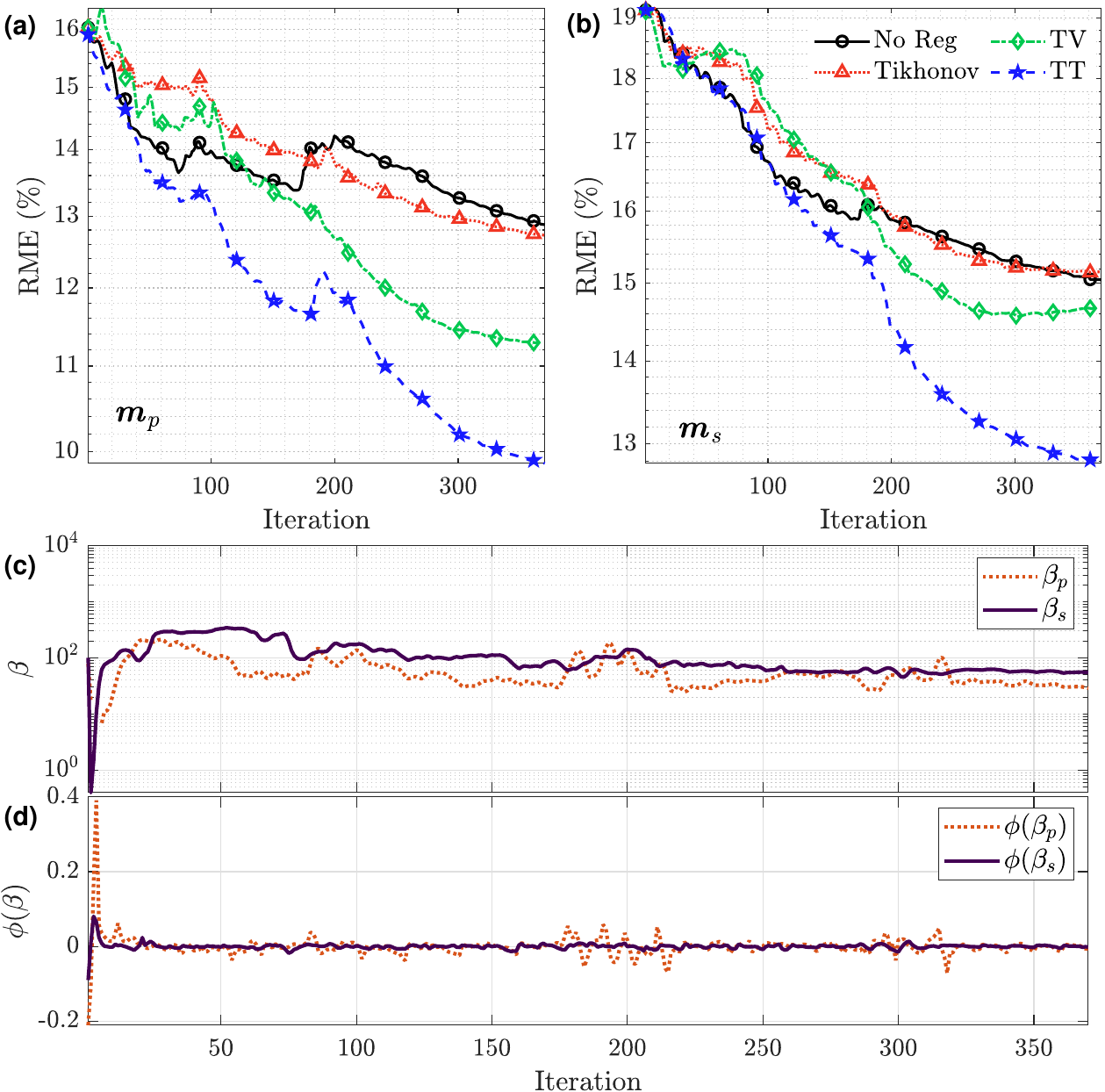}
\caption{Elastic FWI (Overthrust model). (a,b) comparison of the evolution of the RME (\%) for $\text{V}_{\text{p}}$ model ($\m_\text{p}$) and $\text{V}_{\text{s}}$ model ($\m_\text{s}$) in the absence of regularization (No Reg) and with Tikhonov, TV, and TT regularization. (c) Variations of  $\beta_\text{p}$ and $\beta_\text{s}$ during iterations. (d)  variations of  $\phi(\beta_\text{p})$ and $\phi(\beta_\text{s})$ during iterations.}
\label{fig:EL_over_error}
\ps{fig:EL_over_error_a}
\ps{fig:EL_over_error_b}
\ps{fig:EL_over_error_c}
\ps{fig:EL_over_error_d}
\end{figure}
\begin{table}[tbhp] 
\small
\caption{Computational performance of regularization methods for EFWI using the overthrust model: the average runtime per iteration (in seconds) for model regularization (\cref{alg:alg2}) and one EFWI iteration (including regularization). Results are averaged over 10 runs.}
\label{tab:over_runtime} 
\centering  
\begin{tabular}{l c c c c c} 
\toprule 
& \multicolumn{4}{c}{Runtime (sec)} \\ 
\cmidrule(l){2-5} 
       & Tikhonov & TV & Non-adaptive TT & Adaptive TT \\ 
\midrule 
One EFWI iteration  & 52.59 & 51.68 & 52.67 & 52.74\\ 
Regularization step  & 1.83 & 1.73 & 1.86 & 1.92\\ 
\bottomrule 
\end{tabular}
\end{table}
%
\section{Conclusions.}
We proposed an adaptive Tikhonov-TV regularization technique for acoustic and elastic Full Waveform Inversion (FWI), utilizing the efficacy and simplicity of the alternating direction method of multipliers. The proposed method dynamically balances smoothness and edge preservation by adjusting the influence of Tikhonov and TV regularizations, allowing accurate imaging of subsurface models with integrated structural characteristics. Numerical experiments on acoustic and elastic FWI benchmarks demonstrated the method's resilience to local minima, accelerated convergence rate, and superior reconstruction quality compared to Tikhonov and TV regularizations when applied separately. The method demonstrated notable efficacy in addressing the challenges of imaging complex geological media, including subsalt imaging and parameter crosstalk mitigation in elastic FWI. The adaptive technique for adjusting the balancing parameter demonstrated robustness against variations in initial parameter selection and maintained stable performance despite sparse acquisition configurations and noisy data. Importantly, the added computational cost of the adaptive strategy remains minimal, making it a practical and adaptable tool for high-resolution seismic imaging.
\section*{Acknowledgments.}
This research was financially supported by the SONATA BIS grant
(No. 2022/46/E/ST10/00266) of the National Science Center in
Poland.

\pagebreak

\bibliographystyle{siamplain.bst}

\newcommand{\SortNoop}[1]{}

\end{document}

%% file: ex_shared.tex
\usepackage{lipsum}
\usepackage{braket,amsfonts}
\usepackage{mathrsfs}
\usepackage{multirow} 
\usepackage{booktabs} 
\usepackage{graphicx,epstopdf} 
\usepackage{algorithmic}
\usepackage{algorithm}
\usepackage{amsmath, amssymb,bm}

\usepackage{enumitem}
\usepackage{hyperref}
\usepackage{optidef}

\setlist[enumerate]{leftmargin=.5in}
\setlist[itemize]{leftmargin=.5in}

\renewcommand{\bold}[1]{\bm{#1}}
\newcommand{\diag}[1]{\text{diag}(#1)}

\renewcommand{\u}{\bold{u}}

\renewcommand{\b}[0]{\bold{b}}   
\renewcommand{\d}[0]{\bold{d}}
\renewcommand{\P}[0]{\bold{P}}

\newcommand{\g}{\bold{g}} 
\newcommand{\A}{\bold{A}}  
\newcommand{\m}{\bold{m}}
 
\newcommand{\Amu}{\A(\m)\u} 
\newcommand{\p}{\bold{p}}

\newcommand{\dual}[0]{\bm{\lambda}}
\newcommand{\Nu}[0]{\bm{\nu}}

\DeclareMathOperator*{\minimize}{minimize}

\DeclareMathOperator*{\argmin}{argmin}

\makeatletter
\renewcommand{\ALG@name}{\sc Algorithm}
\makeatother

\Crefname{ALC@unique}{Line}{Lines} 

\usepackage[caption=false]{subfig}

\renewcommand{\thesubfigure}{(\alph{subfigure})}

\usepackage{cleveref}

\crefformat{figure}{#2Figure~#1#3}
\Crefformat{figure}{#2Figure~#1#3}

\crefformat{subfigure}{{#2Figure~#1#3}}
\Crefformat{subfigure}{{#2Figure~#1#3}}

\makeatletter
\newcommand{\ps}[1]{%
  \refstepcounter{subfigure}%
  \protected@edef\@currentlabel{\thesubfigure}%
  \label{#1}%
}
\makeatother

\ifpdf
  \DeclareGraphicsExtensions{.eps,.pdf,.png,.jpg}
\else
  \DeclareGraphicsExtensions{.eps}
\fi

\newsiamremark{remark}{Remark}
\newsiamremark{hypothesis}{Hypothesis}
\crefname{hypothesis}{Hypothesis}{Hypotheses}
\newsiamthm{claim}{Claim}

\headers{FWI by adaptive TT regularization}{K. Aghazade, and A. Gholami}

\title{Robust acoustic and elastic full waveform inversion by adaptive Tikhonov-TV regularization 
\thanks{Submitted to the editors DATE
\funding{This research was financially supported by the SONATA BIS grant
(No. 2022/46/E/ST10/00266) of the National Science Center in
Poland. }}}

\author{Kamal Aghazade \thanks{Institute of Geophysics, Polish Academy of Sciences, Warsaw, Poland 
  (\email{aghazade.kamal@igf.edu.pl}, \email{agholami@igf.edu.pl}). }
 \and Ali Gholami \footnotemark[2]}

\usepackage{amsopn}

\usepackage{titlesec}
\titleformat{\paragraph}
  {\normalfont\normalsize\bfseries}{\theparagraph}{1em}{}
\titlespacing{\paragraph}{0pt}{3.25ex plus 1ex minus .2ex}{1.5ex plus .2ex}
